\documentclass{article}
\usepackage[T1]{fontenc}
\usepackage{authblk}
\usepackage{mathtools}
\usepackage{graphicx} 
\usepackage{hyperref}
\usepackage{braket}
\usepackage[table,xcdraw]{xcolor}
\usepackage{physics}
\usepackage{multirow}
\usepackage{natbib}
\usepackage[margin=1.1in]{geometry}
\usepackage{quantikz}
\usepackage{subcaption}
\usepackage{amsthm}

\usepackage{rotating}

\newcommand\blfootnote[1]{%
  \begingroup
  \renewcommand\thefootnote{}\footnote{#1}%
  \addtocounter{footnote}{-1}%
  \endgroup
}

\title{Energy efficiency of quantum computers}

\author[1]{Miquel Carrasco-Codina$^{\dagger}$} 
\author[1]{Pau Escofet}
\author[2]{Paul Hilaire}
\author[3]{Ariane Soret} 
\author[4]{Sam Nerenberg}
\author[4]{Victor Champain}
\author[5]{Gerard Milburn} 
\author[5]{Klara Theophilo}
\author[6]{Sophie H. Li}
\author[7]{Irais Bautista}
\author[7]{Andrés Gómez}
\author[8]{Jose Miralles}
\author[1]{Sergi Abadal}
\author[9]{Carmen G. Almudéver}
\author[1]{Eduard Alarcón}
\author[4]{Raja Yehia$^{\dagger}$}

\affil[1]{Universitat Politècnica de Catalunya, Barcelona, Spain}
\affil[2]{LTCI, Inria, T\'el\'ecom Paris, Institut Polytechnique de Paris, Palaiseau, France}
\affil[3]{Quandela SAS, 7 Rue Léonard de Vinci, 91300 Massy, France}
\affil[4]{ICFO-Institut de Ciències Fotòniques, The Barcelona Institute of Science and Technology, Av. Carl Friedrich Gauss 3, 08860 Castelldefels (Barcelona), Spain.}
\affil[5]{National Quantum Computing Centre, Didcot OX11 0QX, United Kingdom}
\affil[6]{Department of Physics, Harvard University, Cambridge, MA, USA}
\affil[7]{Galicia Supercomputing Center (CESGA), Av. de Vigo S/N, 15705 Santiago de Compostela, Spain}
\affil[8]{Qilimanjaro Quantum Tech, Carrer de Veneçuela 74, 08019 Barcelona, Spain}
\affil[9]{Universitat Politècnica de València, Spain}

\date{\today}
\usepackage{tabularx}
\usepackage{float}
\usepackage{xcolor}
\definecolor{darkgreen}{rgb}{0.0,0.5,0.0}
\begin{document}

\maketitle
\begin{abstract}
    How much energy does a quantum computer consume? Are they more efficient than their classical counterparts? In this work, we make a step towards answering these questions. We define the energy efficiency of a quantum computer as the ratio of the number of algorithms it can perform during a given time over the energy consumed by the hardware during this time. We analyze the most representative physical platforms currently envisioned to be used as building blocks of quantum computers: superconducting qubits, silicon spin qubits, trapped ions, neutral atoms and photonic qubits. Including insights from experts in all these technologies and taking into account algorithm compilation constraints, we discuss the advantages and inconveniences of each platform from an energy standpoint. Beyond providing concrete values of the energy consumption of current quantum computers, we lay the foundation of a framework to benchmark the energy efficiency of any future quantum computing architecture. 
\end{abstract}

\blfootnote{$^{\dagger}$ corresponding authors : \url{miquel.carrasco.codina@upc.edu}, \url{raja.yehia@gmail.com}}
\tableofcontents

\newpage

\section{Introduction}


Quantum computing is emerging as a technology of the near future, with promising applications in various fields ranging from many-body physics simulations for molecule synthesis or nuclear structure analysis~\cite{1st_VQE,All_vqe,Review_VQE}, to solving optimization problems for logistics~\cite{Phillipson2025,Ciacco2025}. It enables a fundamentally different computing paradigm from the so-called \textit{classical} current computing platforms, by encoding information in the degrees of freedom of quantum particles. Several physical systems are envisioned to create these quantum bits, and it is not clear today which of them will be the key components of future quantum computers.

While quantum computing presents us with groundbreaking opportunities, it also poses the challenge of environmental sustainability. As the world struggles with the climate crisis, the construction and energy cost of future technologies has become a critical concern~\cite{QEImanifesto}. More and more stringent resource constraints urge the community to understand the cost of quantum computing. It can be that a particular technology, while currently being very efficient at running quantum algorithms or storing quantum information, is not scalable in a sustainable way and thus has to be discarded. It is important to study this early on, so that \textit{path dependence} effects, where past events or decisions constrain later developments, are avoided. On the other end, the discovery of an energy advantage, where a particular task is solved using less energy than a classical counterpart, could speed up the development of a particular platform energy-wise.

In this paper, we make a first attempt at understanding the energy footprint of current quantum computing platforms. Informed by the diverse expertise of the contributing authors, we established a list of the main platforms envisioned to scale up to a large-scale and fault-tolerant quantum computer, capable of successfully executing quantum algorithms and therefore, run useful applications~\cite{eisert2025mindgapsfraughtroad}. While several paradigms and degrees of freedom can be employed to manipulate quantum information on a particular platform, we provide an estimation of the power consumption of near-future noisy intermediate scale quantum (NISQ) computers. We discuss the possibility of extracting trends and study the scalability of each platform from an energy standpoint. Our goal here is not to rank the different platforms, but rather to understand what the critical components are and identify pros and cons of each platform. 

To this purpose, we define an energy efficiency metric for quantum computer architectures and analyze their performances. For each particular platform, we listed the important hardware components as well as their energy consumption. Our analysis shows that, although executing specific algorithms can increase the classical control overhead, the total energy consumption of the quantum system does not increase significantly when transitioning from an idle (passive) state to active operation. This is largely because major hardware element, such as cryostats and laser systems, remain continuously powered, even when no computation is being performed. In our analysis, instead of focusing on a task-specific energy comparison across platforms, we evaluate their general computational capabilities and compare it to their energy consumption. This requires the derivation of different compilation methods for quantum algorithms running on each of the platforms considered.

We find that, as could be expected, most of the energy expenditure in NISQ quantum computers stems from maintaining a stable qubit environment, through cooling or temperature control. It can be noticed however that, should current architectures scale up without new fundamental discoveries and innovative engineering, the consumption related to qubit control electronics will rapidly outgrow this cost. Beyond establishing a baseline list of components for NISQ computers, our analysis studies the main constraints on the running time of quantum algorithms on each platform: qubit connectivity in the case of superconducting and spin qubits, number of interaction zones and shuttling time in the case of trapped ions and neutral atoms, and probabilistic operations in the case of photonic qubits. Our work shows that operational times of the devices, such as two-qubit gate time or the reset time between different shots of a quantum circuit, are critical parameters in improving the energy efficiency. 

Using our methodology, recent techniques under investigation to improve operational times can be compared to benchmark their impact on the energy efficiency. As examples, we study active reset on superconducting platforms, qubit connectivity in spin qubit platforms, number of interaction zones in trapped ion platforms, continuous reload in neutral atom platforms and different materials for chip-based photonic computing. These results can guide policy-making and engineering decisions in the near future to improve the sustainability of quantum computers.

We note that the study conducted in this paper is only one of the building blocks required to construct a complete view of the future energy cost of quantum technologies. One key cost that is not covered and might prove to be a dominant one, is the manufacturing cost. Materials used in the construction of quantum devices are present in finite quantities, for example rare-earth elements or Helium 3. The direction of current technological development is not concerned with the re-usability of devices and, when the state of the art advances, the previous generation of quantum computers is usually discarded. A full life-cycle assessment of quantum computers might give very different outcomes than the ones we present in this study.

Additional factors can be considered to provide a comprehensive understanding. The rebound effect for example, which refers to the phenomenon whereby improvements in the efficiency of a technology reduce the effective cost of its use, leading to increased consumption that partially offsets the expected resource or energy savings. This response can occur at multiple levels, including direct effects (increased use of the more efficient technology) and indirect effects (reallocation of saved resources to other consumption activities). Consequently, the net environmental or energy benefits of technological innovations may be smaller than anticipated without accounting for these behavioral and systemic responses.

This paper is organized as follows. Our methodology is first introduced in Section~\ref{sec:methdolo} with the definition of the energy efficiency metric and elements of the main elements required to compute it. In Sections~\ref{sec:superco}, \ref{sec:spinqubit}, \ref{sec:trappedions}, \ref{sec:neutralatoms} and \ref{sec:photon}, we study respectively superconducting platforms, spin-qubit platforms, trapped-ions platforms, neutral-atoms platforms and photonic platforms. In each of these sections, a baseline hardware component list and a compilation model are derived, after which the energy efficiency is showcased. Section~\ref{sec:future} touches on error correction codes and future research directions opened by this work, and Section~\ref{sec:conclusion} concludes the main text with general remarks. Finally, Appendix~\ref{app:hardware} contains the hardware component lists for each platform, Appendix~\ref{app:compil} gives additional information on our compilation models and Appendix~\ref{app:chiptech} discusses different chip technologies for photonic computing.


\section{Methodology}
\label{sec:methdolo}
In previous works~\cite{Fellous-Asiani2023,woitzikEnergyEstimationBenchmark2024}, the energetic footprint of quantum computers has been calculated by identifying the energy needed to perform the building block tasks of quantum algorithms: single qubit gates, measurements, heat evacuation etc. This approach can be valuable to understand the theoretical energetic demand of the individual operations taking place during computation, but it should be understood more as a lower bound for resource demand, rather than an accurate estimation that matches real hardware. In this work, we approach the issue from a hardware-aware perspective, by estimating the consumption of quantum computers using power consumption values from existing components and fully working setups. Although this method is less suitable to predict the scaling laws of future quantum computers, it serves our purpose of displaying the energetic demands of hardware that exists or is within reach of current NISQ technologies.

Our energetic model of quantum computing is based on two main parameters: energy consumption of the hardware and number of computations performed in a given time. We first note that the power consumption of state-of-the-art quantum computers does not change significantly whether the device is computing or not. In most architectures, either the hardware elements involved in performing computation cannot be turned on and off between operations, or the passive power demand is orders of magnitude larger compared to the elements that are only involved on the computing process. In addition, the consumption of these systems can not be, in general, broken down into isolated operations or actions that they perform on the qubits (such as gates, or measurements). For these reasons, our model has a more experimental, hardware-aware approach, similar to Yehia et al.'s work~\cite{Yehia2025}, where the consumption of several communication protocols for photonic circuits was estimated based on the power demand of real hardware. 

Since power consumption remains roughly constant regardless of whether a quantum computer is actively performing computations, our figure of merit should account for how efficiently a given platform executes quantum algorithms. For instance, given two quantum computing platforms $\pi_A$ and $\pi_B$, if $\pi_A$ consumes 10 times more but is 100 times faster at performing tasks than $\pi_B$, one might prefer to use platform $\pi_A$. We thus estimate the energy consumed by the devices given a certain period of time rather than for a certain task. This leads us to the following high level definition of the energy efficiency of a quantum computer:

\begin{equation}
\label{eq:EE}
    EE_{\mathcal{A}}^{\pi} = \frac{\textrm{\# Algorithms performed in time t}}{\textrm{Energy consumed in time t} }.
\end{equation}

As demonstrated later in this section, this metric is a time-independent function that quantifies the cost of running algorithms on a quantum computing platform $\pi$, in computation per Joule. Strictly speaking, Equation~\ref{eq:EE} do not represent an efficiency; that would be a dimensionless coefficient that returns the effective output out of a theoretical baseline. However, it can be used to compare the consumption of different platforms performing the same task, by accounting for the time a certain computer takes to perform a targeted task. Precise estimation of how long each platform needs to compute a given function is however a challenging endeavor, and providing exact values of computing times of specific tasks is out of the scope of this work. In the remainder of this section, we elaborate on how the energy consumed and the number of algorithms performed in a given time are estimated.

It should be noted that, while the energy efficiency might allow to compare between quantum computing platforms, this is not the main point of this work. Each platform has their pros and their cons and should not be examined solely with one metric. We acknowledge that the heavy task of benchmarking quantum computers is a global effort, with, for example, DARPA recently launching a Quantum Benchmarking Initiative (QBI)~\cite{DARPA} or the Quantum Energy Initiative (QEI) working group in the IEEE~\cite{QEIWG}. Their approach usually consists either in defining a \textit{price-per-qubit} metric or to find a task that would allow a fair comparison between platforms. This study takes an alternative approach, showcasing a metric that attempts to capture the computational capabilities of a quantum computer, and that can readily be used on NISQ devices to analyze different architecture.

\subsection{Hardware power consumption}

The total power consumption of a given quantum computing platform $\pi$ is simply given by the sum of the power consumption of each hardware element (classical and quantum) that it involves :

\begin{equation}
    P^{\pi} = \sum_{i\in \mathcal{H^{\pi}}}P^{i},
\end{equation}

where $\mathcal{H^{\pi}}$ is the list of hardware elements included in $\pi$ (for example lasers, vacuum chamber etc.) and $P^{i}$ is the power consumption of each of them. By simply multiplying $P^{\pi}$ by a time $t$, we can estimate the energy consumption of the platform during this time.

An interesting quantity of study is $E^{\pi}(0)$, the initial energy required to turn on the quantum computer $\pi$. This cost is usually quite large, since most quantum computing systems require considerable cool-down times to achieve suitable cryogenic and vacuum conditions. Studying $E^{\pi}(0)$ would provide, for example, understanding of how long a quantum computer needs to remain active to compensate its initialization cost. In current implementations of quantum computers, this energy is spent only once. The computer can then be left turned on for months. For this reason, the initialization energy is not included in our energy efficiency metric. 

Ignoring the initialization cost, one can estimate the energy consumption of a computer running for a time $t$ as

\begin{equation}
\label{eq:energycons}
    E^{\pi}(t) = t \cdot P^{\pi}.
\end{equation} 

However, as explained before, this is not enough to estimate the energy efficiency of quantum computers. For a given period of time, two devices may be able to perform very different numbers of quantum algorithms. In addition, for the same device, two algorithms can take various computing times. 

\subsection{Number of algorithms performed in a time period}

The number of quantum algorithms $N^{\pi}(t)$ that a platform $\pi$ can perform in a time $t$ is a much more delicate quantity to estimate than the flat energy consumption. It highly depends on the connectivity between qubits, the gate implementation, the sampling overhead, and many other features that can vary from different architectures, that can even belong to the same qubit technology. In addition, one can account for an extra-layer of complexity due to error correction or mitigation, which increases the number of qubits, gates, classical overhead, and the algorithm compilation time. We leave the discussion on error correction and fault tolerant quantum computing to Section~\ref{sec:errorcorrection}.

In general, the number of times a platform $\pi$ is able to perform an algorithm $\mathcal{A}$ in a given time $t$ is given by:

\begin{equation}
\label{eq:numberofalg}
    N^{\pi}_{\mathcal{A}}(t) =   \frac{t}{t_{{\mathcal{A}}}^{\pi}},  
\end{equation}

where $t_{{\mathcal{A}}}^{\pi}$ is the execution time of $\mathcal{A}$. We thus consider partial executions of an algorithm as useful computing time. This means that for example, if $N^{\pi}_{\mathcal{A}}(24h) = 300.75$, the computer $\pi$ can complete the algorithm 300 times in 24h, and also the extra $75\%$ of the next execution. Not considering partial execution can also be done by rounding down the the above quantity. In that case we would have that if $t<t_{{\mathcal{A}}}^{\pi}$, then the computer does not have time to perform $\mathcal{A}$ and $N^{\pi}_{\mathcal{A}}(t)=0$. For readability in this paper, we will omit rounding down this quantity and consider partial execution as useful time.

Combining Equations~\ref{eq:energycons} and~\ref{eq:numberofalg} leads to the following time-independent equation for the energy efficiency of a quantum computer $\pi$ running algorithm $\mathcal{A}$:

\begin{equation}
\label{eq:EEexplicit}
    EE_{\mathcal{A}}^{\pi} = \frac{N^{\pi}_{\mathcal{A}}(t)}{E^{\pi}(t) } = \frac{1}{t_{{\mathcal{A}}}^{\pi} \cdot P^{\pi}}.
\end{equation}

By varying $\mathcal{A}$, one can benchmark the energy consumption of a quantum computer for various use cases. In the remainder of this section, we explain the model we use to estimate $t_{{\mathcal{A}}}^{\pi}$, the time it takes to perform a quantum algorithm, taking into account the native gate set and other specifications of $\pi$. This study focuses on a circuit-based computation model, while we leave the discussion of other paradigms, such as quantum annealing, or measurement based quantum computing to Section~\ref{sec:otherparadigms}.

\subsubsection{Running time of a quantum algorithm}

In this work, we make a distinction between a \textit{quantum algorithm}, which is a high-level description of a task, and a \textit{quantum circuit}, which corresponds to a translation of the task in terms of quantum gates, ready to be run on a specific quantum computing device. The process of adapting the sequence of quantum gates to be run on a device is the so-called compilation, which we discuss in the next section.

Since quantum computers give access to measurement outputs, \textit{i.e.} samples from probability distributions, circuits have to be implemented a certain number of times $N_{\rm{samples}}$ to get sufficient statistics. This number of shots depends on the noise of the hardware used and on the specifics of the algorithm. This leads to the following formula for the time necessary to perform a quantum algorithm $\mathcal{A}$ in a device $\pi$:

\begin{equation}
\label{eq:talg}
    t_{\mathcal{A}}^{\pi}=t_{\rm{circuit}}^{\pi}\cdot N_{\rm{samples}},
\end{equation}

Throughout this work, we examine five quantum computing platforms that can be grouped in three families: \textit{solid-state} devices, comprising superconducting and spin-based qubits, \textit{atom-based} devices, which use neutral-atom and trapped-ion qubits, and finally \textit{photonic} devices. To factor in each platform's specificity, a different model to estimate the time to execute a circuit $t_{\rm{circuit}}^{\pi}$ is derived for each of these families.

In the case of solid-state platforms, we consider that a circuit can be broken down into different operations: reset of the qubits, gate application, and measurement. In general, a quantum circuit has a certain \textit{depth} $D$, which represents the number of time-steps necessary to perform the circuit gates. We consider that synchronous programming languages are used to translate algorithms on quantum hardware. This means that the computer uses an internal clock to define time steps. Each of these time-steps is composed of different basic operations that happen in parallel, namely implementation of single- and two-qubit gates. This leads us to the following formula:
\begin{equation}
    \label{eq:tcircuit_solid_state}
    t_{\rm{circuit}}^{\rm{solid-state}} = t_{\textrm{reset}} + D \cdot t_{\rm clock} + t_{\rm meas},
\end{equation}
where $t_{\textrm{reset}}$ is the time necessary to reset the qubit between every sample of the algorithm and $t_{\rm meas}$ is the time needed to read the final state of the qubits at the end of the circuit. The clock time $t_{\rm clock}$ corresponds to the time the computer requires to execute a single time-step. These three time parameters depend on several hardware features and are discussed in detail in the section pertaining to each platform.

In the case of atom-based platforms, the reasoning is similar, but an additional component needs to be considered in the circuit time estimation: atoms need to be moved to make use of their all-to-all connectivity. This is typically done by moving from an \emph{storage zone} to different \emph{gate zones}, in which gates can be applied. The execution of the circuit is thus similar to solid-state platforms, but an extra term accounts for atom transport between layers that require different sets of qubits. The formula we use is:

    \begin{equation}
    \label{eq:atom-based_t_circuit}
        t_{\rm{circuit}}^{\rm{atom-based}}= t_{\textrm{reset}} + D \cdot t_{\rm clock} + t_{\rm meas}+\beta_{\textrm{trans}}\cdot D\cdot t_{\rm{trans}},
    \end{equation}

where $t_{\rm{trans}}$ is the transport time and $\beta_{\textrm{trans}}$ is the fraction of layers that requires changing the atoms that are held in the gate zones. This quantity highly depends on the structure of the circuits, and the number of gate zones on a given device. In this work, we present results for different values of $\beta_{\textrm{trans}}$. More details regarding atom transport are given in Appendix~\ref{app:coldcomp}.

Lastly, photonic quantum computers operate in a totally different manner. A circuit is not decomposed in layers of gates, but rather seen as a unitary that can be implemented in optical components arranged on a chip. Single photons are produced by a source, sent on the chip and measured using single photon detectors. Since the travel time of the photons from the source to the detector is negligible, the main parameters that quantify the time to perform a successful measurement on a photonic device are the rate of the source $r_{\rm{source}}$ and the end-to-end transmission probability $\eta$ of the chip. We get that:

\begin{equation}
        t_{\rm circuit}^{\rm{photonic}}=\frac{n}{r_{\rm{source}}}\cdot\frac{1}{\eta^{n}},
\end{equation}

where $n$ is the number of photons involved. The end-to-end transmission $\eta$ depends on the optical depth, \textit{i.e.} the number of layers of optical component on the chip. We explain it in more detail in the corresponding section. 

Beyond finding values for the relevant hardware parameters for each platform, the above equations present a challenge that is common to both solid-state and atom-based platforms. Namely: how can we compute the depth of a circuit implementing an algorithm on a specific platform? 

\subsubsection{Compilation of an algorithm into a quantum circuit}
\label{sec:estimating_depth}

Compilation of a quantum algorithm into a hardware-aware quantum circuit is a complex optimization process. In photonic chip-based platforms, compilation consists in a single classical computation step which translates the algorithm into a configuration of linear optics elements. The chip reconfiguration procedure is discussed in section~\ref{sec:photon}. The rest of this section is relevant for compilation of algorithms executed on solid-state and atom-based platforms, where quantum circuits consists of $D$ layers of gates. Gates can be arranged in multiple ways to optimize the overall depth, map logical operations to physical qubits, properly schedule operations etc. In this work, we simplify this compilation process in two steps: (1) \textit{transpilation}, where the algorithm is translated into a circuit using the native gate set of a given platform and (2) \textit{routing}, where this circuit is adapted to the constraints of the platform, as could be, for example, the limited connectivity of superconducting chips or the movement of atoms to interaction zones in atom-based computers.

\begin{figure}[h!]
    \centering
    \includegraphics[width=1\linewidth]{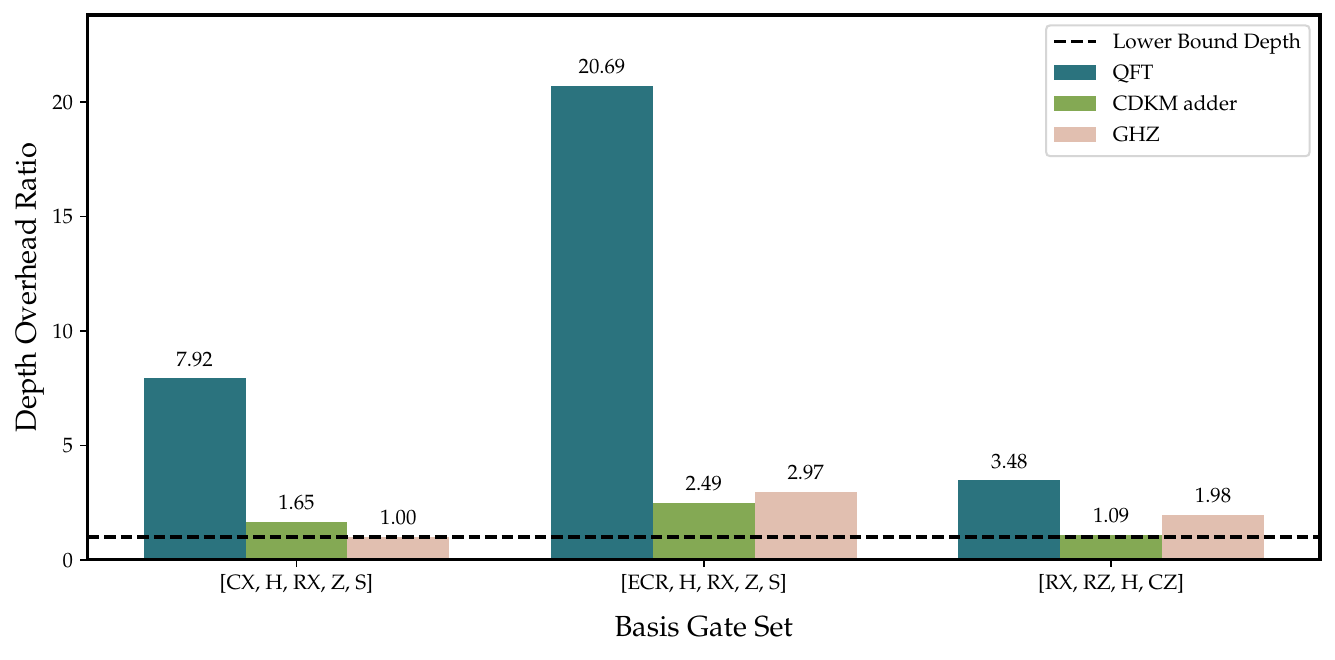}
    \caption{Relative depth overhead ratio of the QFT, CDKM adder and GHZ state preparation circuits for different basis gate sets in a 100 qubit quantum computer, based on the Qiskit transpiler. The labels of the gates in the x-axis correspond to the Pauli Z gate (Z), controlled gates (CX and CZ), the Hadamard gate (H), parametric rotations (RX and RZ), the phase gate (S) and the maximally entangling gate (ECR). The lower bound (dashed line) is computed taking into account all the accessible gates in Qiskit.}
    \label{fig:gate_set_comparison}
\end{figure}

We first examine the effect of the transpilation step. Depending on the native gate set of a platform, less time steps might be necessary to perform an algorithm compared to another, even if they have access to the same number of qubits. Using Qiskit~\cite{Qiskit}, we show in Fig.~\ref{fig:gate_set_comparison} how the set of quantum gates that is accessible to a platform influences the final depth of a circuit for three different examples: the Quantum Fourier Transform (QFT), the CDKM adder~\cite{CDKMadder} and an algorithm for $\ket{\rm{GHZ}}=\frac{\ket{0...0}+\ket{1...1}}{\sqrt{2}}$ state preparation.

Once a gate set is fixed, the depth of a circuit is also influenced by the classical transpilation method used to translate an algorithm into these native gates. Very developed methods can result in a lower circuit depth at the cost of a longer classical computation time. There is thus a trade-off between the energy spent in compilation and the depth of the final circuit to be run on a quantum computer. In this work, we simplify this issue and consider that algorithms are pre-transpiled on classical machines and scheduled to run on the quantum computer. This corresponds to the cloud-based picture that is currently in place, where quantum devices are remotely accessible to end users. In this way, the time where the quantum machine is used is optimized. The transpilation time is assumed to be the same order of magnitude for all platforms, and performed on classical computers in parallel to the quantum computations.

\begin{figure}[!h]
    \centering
    \includegraphics[width=1\linewidth]{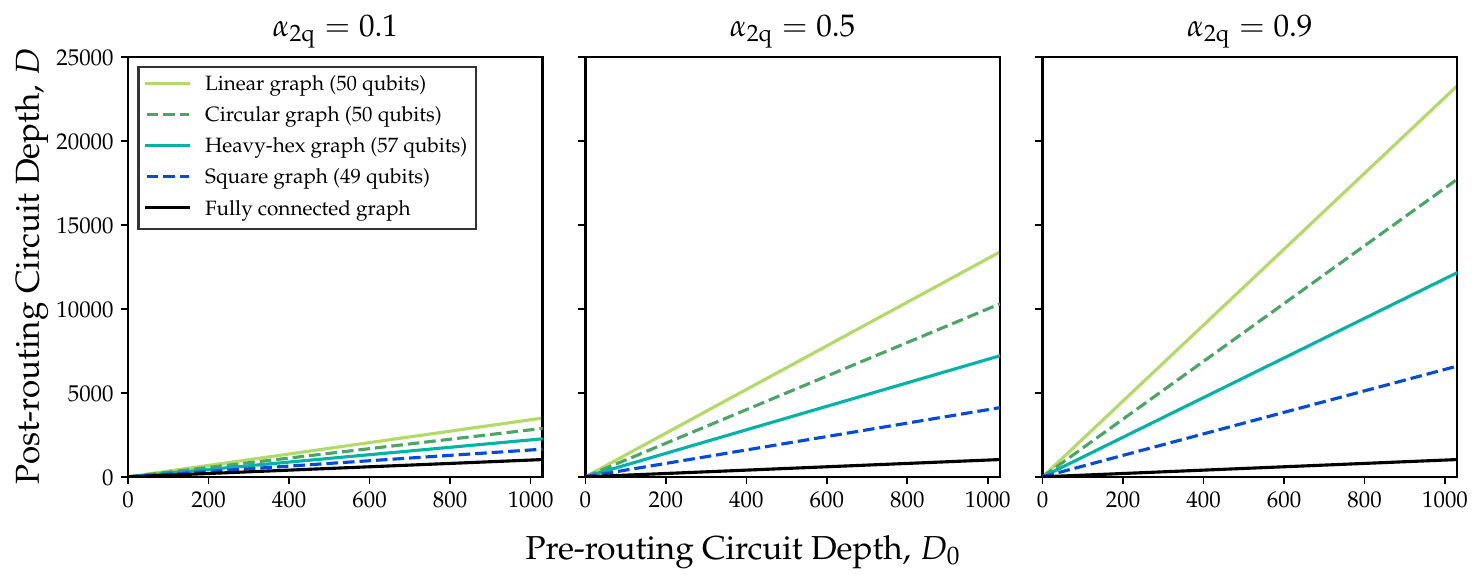}
    \caption{Post-routing circuit depth of a circuit as a function of $D_{0}$. The three panels correspond to different ratios of layers with two-qubit gates $\alpha_{\rm{2q}}$}
    \label{fig:connectivity_graphs}
\end{figure}

The routing step of the compilation consists in adapting the circuit to the specificity of the platform, ensuring that gates can be applied on the appropriate qubits. In the case of solid-state platforms, the main constraint is the connectivity between qubits. Qubits positioning in a superconducting and spin-qubit computers is commonly represented with a graph where the nodes are the qubits and the edges connect pairs of qubits that can interact. In order to perform a gate between non-connected qubits, additional gates are needed to mobe their state into a suitable connected configuration. This can be  done using SWAP gates, that we consider here to be composed of 3 CNOT gates. As a result, each two-qubit gate implies a depth overhead on the post-transpilation depth $D_0$, which depends on the distance between the relevant nodes in the processor. To have a general description of the overhead required for any given circuit, we introduce the parameter $\alpha_{\mathrm{2q}}=\frac{D_{0}^{\mathrm{2q}}}{D_{0}}$ which corresponds to the ratio of layers that include two-qubit gates $D_{0}^{\mathrm{2q}}$ over the total number of layers before routing. We approximate the final depth of a circuit of transpiled depth $D_{0}$ executed in a solid-state qubit based processor as follows:
    \begin{equation}
        D^{\rm solid-state}\approx D_{0}+\alpha_{\mathrm{2q}}\cdot 3\left\lceil\frac{\overline{d(G)}-1}{2}\right\rceil,
    \end{equation} 
    
where $\overline{d(G)}$ is the average distance between nodes of the graph $G$ representing the qubit connectivity in the platform. Fig.~\ref{fig:connectivity_graphs} shows how the graph connectivity and the ratio of two-qubit gates influences the number of times a given circuit can be performed in 24 hours. In the case of an all-to-all connectivity, no overhead would be necessary. In Appendix~\ref{app:compilsolid}, we go in more detail regarding solid-state platforms connectivity.

In contrast, in an atom-based computer, all qubits can interact with each other, but only being at a close distance. Therefore, gates are usually performed in interaction zones (or gate zones), where the qubits are brought together. The amount of interaction zones and the number of qubits they can host depend on the device considered. To address this constraint we consider two general scenarios: (a) the original circuit has less gates per layer than the available number of gate zones, in which case the depth of the circuit remains the same, and (b) on average, the number of gates per layer of the original circuit is higher than the number of gate zones. In this case, the maximum amount of gates performed per layer is limited by the number of zones. The depth of the circuit becomes:
    
\begin{equation}
    D^{\rm atom-based} = \left\lceil\frac{\overline{N_{\rm{g}}^{\rm{L}}}}{N_{\rm{zones}}} \right\rceil\cdot D_{0},
\end{equation}
where $\overline{N_{\rm{g}}^{\rm{L}}}=N_{\rm{g}}/D_{0}$ is the average number of gates per layer of the post-transpilation circuit and $N_{\rm{zones}}$ the number of gate zones of a given device. Here, it is assumed that each gate zone can allocate 1 or 2 qubits, depending on the gate that is being performed. However, some devices can host multiple atoms in a single zone and perform gates on all of them simultaneously. In that case, $N_{\rm{zones}}$ is assumed to be equal to the number of atoms held in the entire zone. We go in more detail, including some special cases, in Appendix~\ref{app:coldcomp}.

Finally, it is worth to mention that there are a plethora of compilation techniques that can be used to compress the circuits further, and are not considered in this work. Research on optimized compilers for quantum algorithms is still ongoing. Some studies~\cite{ShorDepth1,ShorDepth2} estimate that the depth of a circuit factorizing a 2048-bit integer is $1.80 \cdot 10^{12}$. Quantum Fourier Transform on $n$ qubits, which is  a basic component of many quantum algorithms, typically has a depth of $n$, but approximations can be achieved in depth $\log(n)$~\cite{FourierDepth1,FourierDepth2}. In quantum chemistry applications, estimation of the depth of a circuit employed to elucidate reaction mechanisms in complex chemical systems vary in the range of $10^{10}$ and $10^{15}$~\cite{DepthChemistry1,DepthChemistry2}. While these numbers are generally very large, VQE applications seem to require circuit depths of the order of 100~\cite{VQEdepth}. This seems to be more in line with current quantum computer capabilities, whose qubit coherence time limits the depth to a few hundred layers. It is also worth noting that these values do not take into account the platform-dependent constraints that we mentioned in this section. There is a tradeoff to be considered between the resources devoted to the compilation of a circuit, and the resources that will be spent executing such circuit. For the sake of generality, we analyze the energy efficiency of quantum computers for generalized circuits that can be described by their depth, $D$. Therefore, we assume that the same algorithm can result in different circuits with a wide range of depth values, depending on the gate basis, connectivity, or any specific compilation techniques that might be used. 

\section{Superconducting qubits}
\label{sec:superco}
Superconducting qubits are a solid state technology based on superconducting electrical circuits. They are also considered one of the best scaling technologies for quantum computing, thanks to the high adressability of each individual qubit in the chip and the compactness of semiconductor design~\cite{Ezratty2023}. For these reasons, this technology is being adopted and developed by some of the largest companies within the quantum computing industry, such as IBM~\cite{Kim2023}, Google~\cite{Acharya2025} and D-Wave~\cite{King2025}.

\subsection{Technology}
By arranging capacitors, inductors, transmission lines, and other electronic components, one can create circuits with coherent quantized energy levels, where qubits can be encoded~\cite{ReviewSuperconducting}. Any particular circuit can be associated to a Hamiltonian defined by its circuit components. The physical design and manufacturing of its main components results in the so-called characteristic energies. This will determine the anharmonicity and energy spectra of the qubits. In general, these qubits can be coupled capacitively or inductively, which modifies their Hamiltonian, allowing pair-wise interaction. This Hamiltonian development stemming from the design of the circuit is a step in the flourishing field of Circuit Quantum Electrodynamics (circuit QED). 

The central quantum mechanical device that allows circuit properties to be used as qubits are the so-called \textit{Josephson Junctions} (JJ)~\cite{Josephson1974}. They consist of two superconductors separated by a very thin insulator (typically about 1nm thick). When held under an electrical current, the electrons grouped in Cooper pairs can tunnel from one superconductor to the other. JJ anharmonic quantum behavior is shared with the electrical properties of other components in the circuit, making them suitable to encode qubit states. There are three main superconducting-qubit families: charge qubits, where the information is encoded in the number of cooper pairs on one side of the JJ~\cite{Bouchiat1998,Nakamura1999}; flux qubits, where the information is encoded in the direction of the current~\cite{Mooij1999}; and phase qubits, where the information is encoded in the phase difference across a JJ~\cite{Martinis2009}. The most common qubit design is the transmon qubit~\cite{Blais2004, Girvin2009}, which belongs to the charge family. In this work, we consider a flux qubit architecture, the fluxonium~\cite{Fluxonium_1,Fluxonium_2}, adopted by Qilimanjaro.
\begin{figure}[!h]
    \centering
    \begin{subfigure}{0.58\textwidth}
        \includegraphics[width=\textwidth]{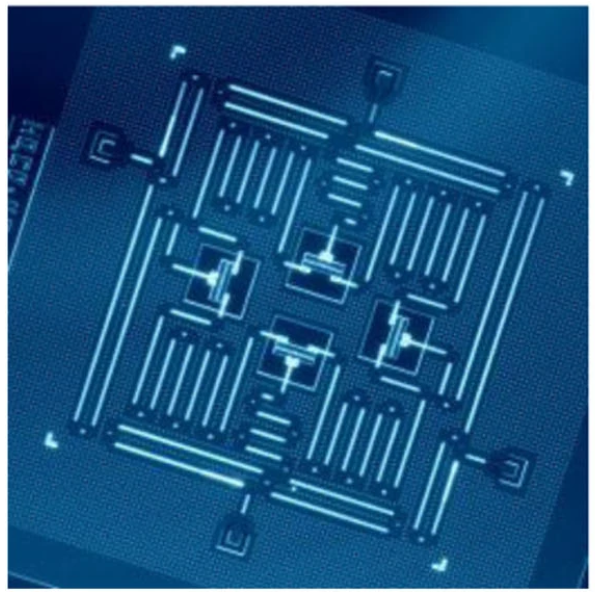}
        \caption{}
    \end{subfigure}
    \begin{subfigure}{0.37\textwidth}
        \includegraphics[width=\textwidth]{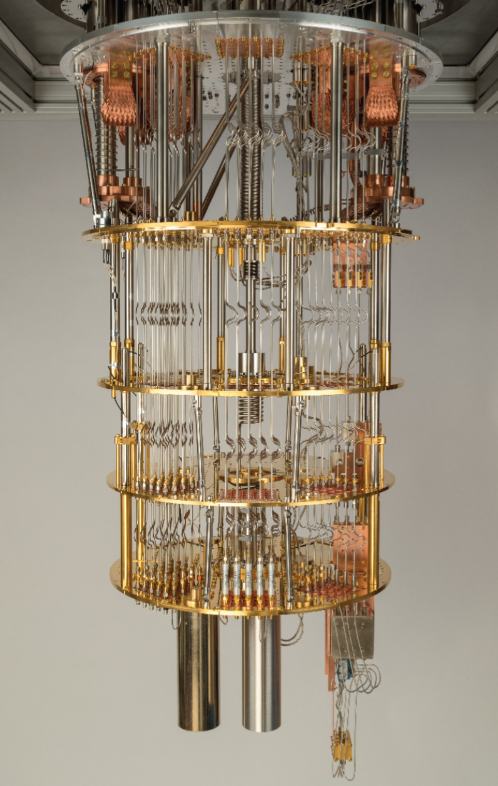}
        \caption{}
    \end{subfigure}
    \caption{Panel a) displays an IBM transmon-qubit processor, consisting of 4 qubits and 4 ressonators. Image taken from Ref.~\cite{Gambetta2017}. Panel b) shows the internal configuration of a dilution refrigerator, hosting a superconducting-qubit quantum computer. This setup, extracted from Ref.~\cite{Krinner2019}, is similar to the required cryogenic, and wire, configuration for spin qubits.}
    \label{fig:superconducting_photos}
\end{figure}

In general, superconducting qubits are controlled with Rabi oscillations produced in the form of microwave oscillations~\cite{Blais2004,Girvin2009, Kjaergaard2020,Huang2020}. In a quantum chip, superconducting qubits would be interconnected within the same electrical circuit. By tuning the frequencies~\cite{Girvin2009} of multiple qubits, they can become resonant and interact~\cite{Kjaergaard2020,Huang2020, Dewes2012}. Typically, the measurement of a superconducting qubit state is done by dispersive readout. This technique consists in measuring the transmission coefficient of a resonator coupled to the qubit. It is important to note however, that the control and readout schemes of superconducting qubits might vary depending on the qubit design.

Superconducting qubits have to be typically kept below $T_{q}=35$ mK, to reduce the thermal noise that affects the fidelity of their quantum state~\cite{Jin2015} and maintain the metal component plates of the Josephson junctions at superconducting state. This condition implies the use of powerful dilution refrigerators capable of reaching millikelvin temperatures~\cite{Zu2022} and evacuating the heat dissipated by the quantum chip and electronics. These refrigerators are known to be the a major factor on the overall energy consumption of superconducting computers.

In summary, superconducting computers are electrical circuits controlled by a series of microwave signals, that have to be held under extremely low temperatures. This setup requires a dilution fridge able to cool down the chip and the wires, which are connected to the control electronics that generate and send the microwave pulses (generally at room temperature).


\subsection{Energy model}
In this section, we break down the hardware components of a superconducting quantum computer and their corresponding consumption. In order to estimate its energy efficiency, we also list the main operations that define the time that this computer takes to execute a circuit. Both the hardware components and operation times are listed in detail in Appendix~\ref{app:superconducting}.

\subsubsection{Power consumption of the hardware}
The components displayed in Table~\ref{tab:superconducting_components} of Appendix~\ref{app:superconducting} correspond to a superconducting-qubit computer that contains around 50 qubits connected in a 2D square lattice. Depending on the qubit architecture, a single feed line can power different numbers of qubits. In this case, we use Qilimanjaro's fluxonium architecture \cite{Fluxonium_1, Fluxonium_2} as a reference. This reference model is intentionally simplified and serves as a baseline architecture for our estimations, rather than a description of a specific deployed system. To complement this reference architecture with a real-world validation,
Appendix~\ref{app:superconducting} also includes a breakdown of the experimentally characterized superconducting quantum
processor QMIO~\cite{QMIO}, deployed in the Galicia Supercomputing Center (CESGA). According to the components shown in Table~\ref{tab:superconducting_components}, the consumption of a superconducting-qubit computer can be estimated to be $P^{\rm{supercond}}\approx20$ kW. This corresponds to $\approx1700$ MJ of energy consumed in 24h.

\begin{figure}[h!]
    \centering
    \includegraphics[width=0.8\linewidth]{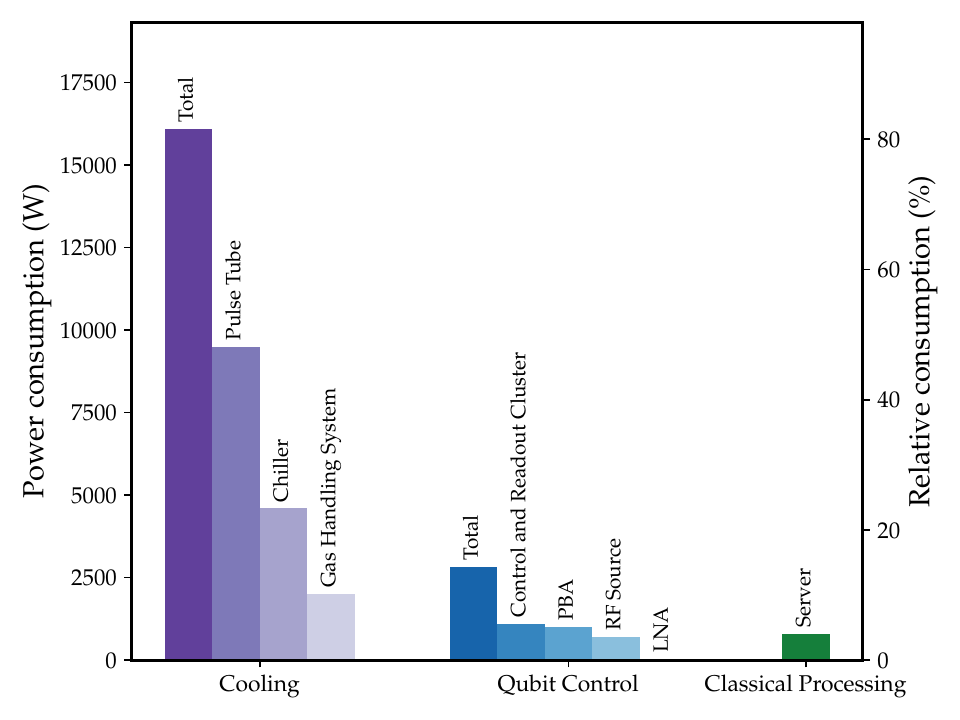}
    \caption{Power breakdown of a superconducting-qubit computer with $\sim50$ qubits, according to the components shown in Table~\ref{tab:superconducting_components} of Appendix~\ref{app:superconducting}. In the x-axis, the components are grouped depending of their purpose in the computer. Each group has a bar that corresponds to the total accumulated power of its components. The left y-axis shows the absolute value of power consumption of the components, and the right y-axis the relative value with respect to the overall consumption of the computer.}
    \label{fig:superconducting_power_breakdown}
\end{figure}

The consumption contribution of the different components in the hardware is shown in Figure~\ref{fig:superconducting_power_breakdown}. In this figure, the hardware components are grouped according in three different areas: cooling, qubit control and classical processing.

As one can see, the highest power-demanding parts of the computer are devoted to the cooling system, with $80\%$ ($16.1$ kW) of the overall consumption. It is unclear how the power of the refrigerator scales with the number of qubits, but we consider this value representative to the superconducting devices of this size ($\sim50$ qubits). As we have seen for almost every platform in this work, there is a common trend on consuming most power to produce the suitable physical environment to host qubits. It is important to point out that we do not account for the power consumption of the vacuum pump in this analysis, since it is only used to initialize the computer, and its power consumption during computations is negligible.

There is a clearer dependency of the electronics needed in the readout lines with respect to the number of qubits. In particular, each readout line need its own radio-frequency (RF) source as well as two low-noise amplifiers (LNA). Considering each line can feed up to 5 superconducting qubits, we can compute the number of control qubits components for a given number of qubits. The control and readout cluster has a max power demand of $1.1~\mathrm{kW}$, and can be modified combining different modules that addresses the different signals needed for qubit control and readout. Therefore, its power consumption also depends on the number of qubits in the computer, and the lines required according to their architecture. For the sake of generality, we include its maximum consumption to our estimations. Finally, this component could be replaced by a combination of FPGAs. This is the case of the QMIO computer shown in Table~\ref{tab:qmio_appendix_style_final}, where each qubit has a dedicated FPGA in charge of producing and sending the corresponding signals. Compared to this particular FPGA design, the cluster turns out to be the most compact and efficient choice.

\subsubsection{Derivation of the energy efficiency}
\label{sec:energysuperc}

To compute the energy efficiency of running an algorithm $\mathcal{A}$ on a superconducting qubit platform, one first needs to compile $\mathcal{A}$ into a quantum circuit of depth $D$. As explained in the introduction and developed in Appendix~\ref{app:compilsolid}, we estimate the gate overhead due to the connectivity constraints of the platform by using the average distance between the qubits in the connectivity graph of the platform. We consider here that $D$ is the final depth of a circuit implementing $\mathcal{A}$, which has been already compiled according to the gate set and connectivity of the device. By taking $D$ as a parameter, the energy efficiency of the computer can by analyzed without having to define a specific task $\mathcal{A}$. The model of the execution time of $\mathcal{A}$ compiled into quantum circuit of depth $D$ on a superconducting-qubit quantum computer is given by:

\begin{equation}
    \label{eq:tcircuit_superc}
    t_{\mathcal{A}}^{\rm{supercond}}= N_{\rm{samples}}\cdot t_{\rm{circuit}}^{\rm{supercond}} = N_{\rm{samples}}\cdot (t_{\textrm{reset}} + D \cdot t_{\rm clock} + t_{\rm meas}),
\end{equation}
where $N_{\rm{samples}}$ is the number of samples, $t_{\textrm{reset}}$ is the time necessary to reset the qubit between every circuit sample, $t_{\rm meas}$ is the time needed to read the final state of the qubits at the end of the circuit and $t_{\rm{clock}}$ is the clock time of the computer. Based on observations of current superconducting qubit devices, we define the following operation times: $t_{\rm{reset}}\approx5~\mathrm{\mu s}$, $t_{\rm{clock}}\approx 50~\mathrm{ns}$ and $t_{\rm{meas}}\approx1.6~\mathrm{\mu s}$, which are further discussed in Appendix~\ref{app:superconducting_baseline}. This reset time corresponds to the active reset technology, which we compare with a passive reset time ($t_{\rm{reset}}^{\rm{pass}}\approx200~\mathrm{\mu s}$) at the end of the section; until then, we assume active reset.

Due to the dependency of $t_{\rm{circuit}}$ with the post-compilation depth $D$, it is important to note that the connectivity constrains of the processor will have a deep effect on the computing time, and therefore, the efficiency of superconducting computers. In this section we do not analyze the efficiency of processors with different graph connectivities, but one can see how the depth overhead scales with the number of qubits for different connectivity architectures in Appendix~\ref{app:compilsolid}. In addition Figure~\ref{fig:connectivity_graphs} shows values of the post-compilation depth $D$ for different densities of two-qubit gates and connectivity graphs, for processors that have $\sim 50$ qubits.

The energy efficiency of a superconducting-qubit quantum computer is thus given by 
\begin{equation}
\label{eq:EEsupercond}
    EE^{\textrm{supercond}} =  \frac{1}{N_{\rm{samples}}\cdot (t_{\textrm{reset}} + D \cdot t_{\rm clock} + t_{\rm meas}) \cdot P^{\textrm{supercond}}},
\end{equation}

where $P^{\textrm{supercond}}$ is the sum of the power consumption of the hardware elements shown in Table~\ref{tab:superconducting_components}.


\subsection{Results and discussions}
\label{sec:superconducting_results}

\begin{figure}[!h]
    \centering
    \includegraphics[width=\linewidth]{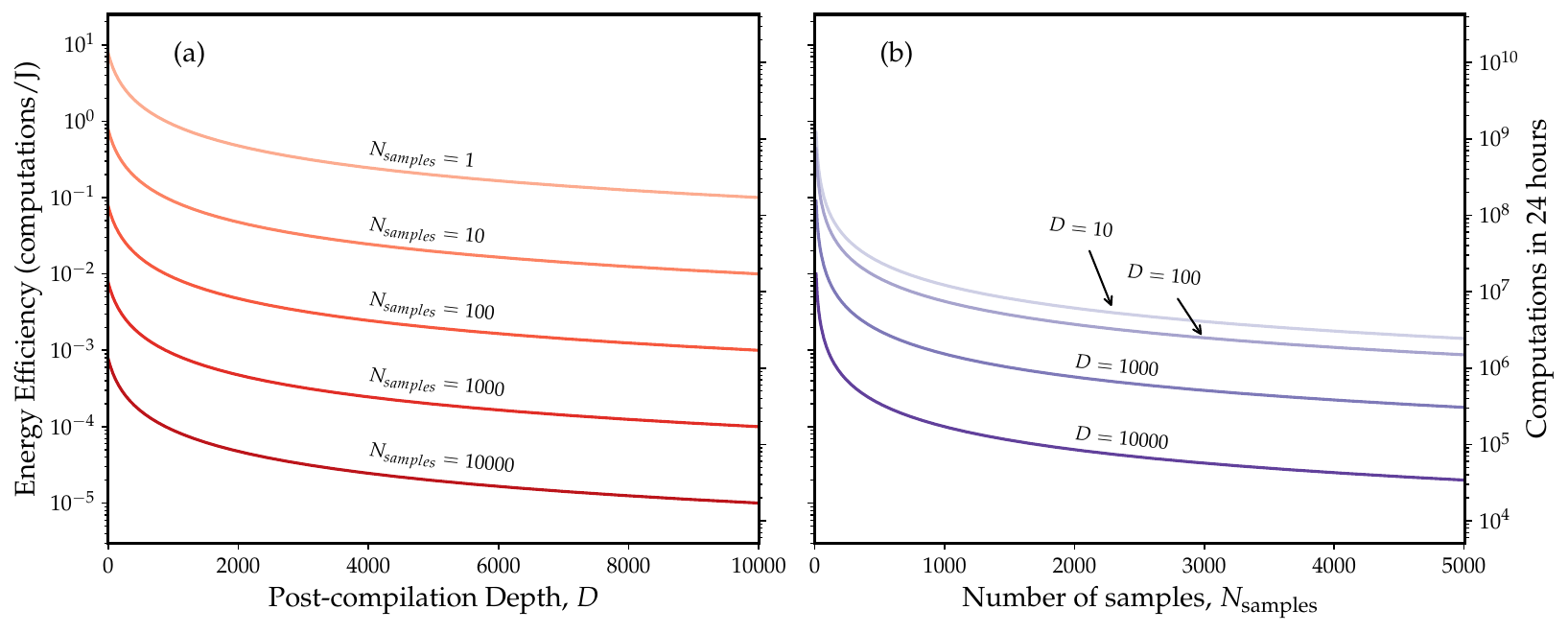}
    \caption{Energy efficiency (left y-axis) and number of computations (right y-axis) in 24 hours of a superconducting-qubit computer depending on the final circuit depth, $D$ and number of samples, $N_{\rm{samples}}$. Panel (a) contains several series with different $N_{\rm{samples}}$ values and the x-axis corresponds to $D$. Series in panel (b) have different $D$ values and its x-axis corresponds to $N_{\rm{samples}}$.}
    \label{fig:superconducting_EE_Npi_D}
\end{figure}

Figure~\ref{fig:superconducting_EE_Npi_D}, shows the energy efficiency and number of computations of a superconducting computer executing an algorithm with given $D$ and $N_{\rm{samples}}$ for 24 hours. The relation between the energy efficiency and the number of samples is inversely proportional, according to Eq.~\ref{eq:EEsupercond}. For this reason, the series shown in panel (a) are equidistant in the logarithmic scale, separated exactly by an order of magnitude.

As one can see, the dependency of $EE$ with the post compilation depth $D$ presents a different behavior. In panel (b), the series corresponding to $D=10$ and $100$ are closer together than the rest. This is due to the effect of the measurement and reset times at lower depths. For lower values of $D$, $t_{\rm{reset}}$ and $t_{\rm{meas}}$ are several orders of magnitude larger than $t_{\rm{clock}}$, and therefore, they take most of the computing time. As an example, a sample of a $D=10$ layers circuit would take $t_{\rm{circuit}}=7.1~\rm{\mu s}$, which does not increase substantially at $D=100$ layers, for which $t_{\rm{circuit}}=11.6~\mathrm{\mu} s$. In both instances, most of the time is spent measuring and resetting the qubits ($6.6~\mathrm{\mu s}$). For $D=1000$ and $10000$ layers, the clock time becomes more relevant; $t_{\rm{circuit}}=56.6$ and $506.6~\mathrm{\mu s}$, respectively. 

\begin{figure}[!h]
    \centering
    \includegraphics[width=0.7\linewidth]{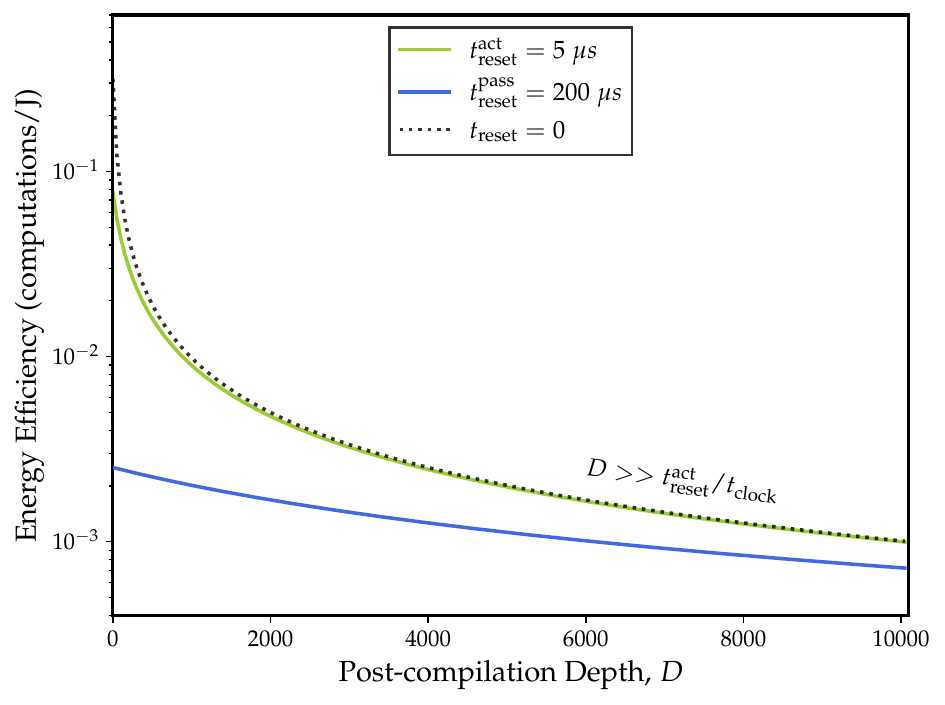}
    \caption{Energy efficiency of a superconducting-qubit computer, depending on the depth of the executed circuit and the reset time. The green line corresponds to a computer with active reset, the blue line to passive reset, and the dotted line to instantaneous reset.}
    \label{fig:superconducting_t_init}
\end{figure}

To analyze further the relation between $t_{\rm{clock}}$, $t_{\rm{reset}}$ and $t_{\rm{meas}}$, we compare the efficiency between computers with active and passive qubit reset. Figure~\ref{fig:superconducting_t_init} displays the energy efficiency of the computer with active reset ($t_{\rm{reset}}^{\rm{act}}=5$ $\mu$s), passive reset ($t_{\rm{reset}}^{\rm{pass}}=200$ $\mu$s), and instantaneous reset of the qubits ($t_{\rm{reset}}=0~\mathrm{s}$). As one can see, the efficiencies of the computer with passive reset display a lower and flatter profile than the active one. This is due to the high impact of the passive reset time during computation, which forces short circuits to take a very similar computing time. On the contrary, the active reset favors the execution of lighter circuits. In particular, for short circuits of depth $D\approx100$ layers, the device with passive reset has an efficiency of $EE=2.5\cdot10^{-3}$ computations/J, $20$ times lower than its counterpart.

In accordance with Eq.~\ref{eq:EEsupercond}, Figure~\ref{fig:superconducting_t_init} shows how the efficiency becomes inversely proportional to $D$ as the depth of the circuits increase, independently of $t_{\rm{reset}}$. This happens when the circuit is deep enough ($D>>\frac{t_{\rm{reset}}+t_{\rm{meas}}}{t_{\rm{clock}}}$) for the time taken to measure and initialize the qubits to be negligible in comparison with the operation time $D\cdot t_{\rm{clock}}$. In that regime, the energy efficiency curve converges to the black dotted curve that represents an ideal computer with no reset time. In the case of the computer with active reset, this can be already seen at $D\sim10^{3}$. However, for the computer with passive reset this regime is shifted to higher circuit depths ($\sim 10^{5}$ layers).

Superconducting qubits come in a wide variety of designs. The choice of qubit architecture and chip topology, together with the type of coupler used (whether tunable or fixed), shapes the effective Hamiltonian of the system and directly influences its anharmonicity, decoherence rates, and susceptibility to noise. Beyond the physics, these design choices also determine the number of drive and readout lines required, which in turn sets the scale of the energy needed for control and readout operations. As a general rule, each loop in the superconducting circuit requires its own drive line. A standard transmon, for instance, contains a single loop and therefore needs only one drive line~\cite{Roth2023}. However, architectures that incorporate a Superconducting Quantum Interference Device (SQUID) introduce an additional loop, meaning that a single qubit in such a design would require two drive lines rather than one~\cite{Huang2020}.

This inherent diversity in superconducting qubit design makes it remarkably difficult to produce a unified estimate of energy consumption and efficiency across the full landscape of architectures. Parameters such as native gate sets, number of loops, coupler type, and resulting qubit connectivity all vary significantly from one design to another, and each of these factors influences how efficiently a given algorithm can be executed. While the individual components making up these architectures are generally expected to behave homogeneously, the specifics of circuit depth required for a particular algorithm implementation, the number of drive lines, and how both of these scale with increasing qubit count are highly architecture-dependent and far from straightforward to generalize.

\section{Silicon spin qubits}
\label{sec:spinqubit}

Loss and DiVincenzo proposed a spin-based approach to quantum computation by using the spin of trapped charges, such as electrons confined in quantum dots, to encode quantum information\cite{Loss1998}. This approach exploits the spin-$\frac{1}{2}$ nature of electrons, which inherently provides a two-level quantum system. Leveraging complementary metal-oxide-semiconductor (CMOS) technologies, spin qubits are a strong candidate for scaling up future quantum computers. Recent work has demonstrated key building blocks for quantum error correction with silicon spin qubits, such as weight-four parity checks, though these implementations rely on more complex architectures involving electron shuttling~\cite{Undseth2026}. Spin qubits are being investigated by many companies (Intel) and startups (Quantum Motion, Diraq, Quobly, Equal1 and many more).
    
\subsection{Technology}

Several physical platforms have been developed to realize spin qubits, each relying on a different way to confine and manipulate individual spins. The first proposal, introduced by Loss and DiVincenzo in 1998~\cite{Loss1998}, encodes quantum information in the spin state of a single charge carrier—either an electron or a hole—trapped inside a gate-defined quantum dot in a semiconductor. In this configuration, the qubit logical states correspond to the spin-up and spin-down projections of that confined particle.

Alternative implementations exist. One approach uses a donor atom embedded in a semiconductor lattice, where the electron remains bound to the impurity potential~\cite{kane1998}. Other designs exploit the collective spin states of two or more electrons confined in coupled quantum dots: for instance, the singlet–triplet qubit in a double quantum dot~\cite{levy2002}, or the exchange-only qubit realized in a triple quantum dot~\cite{divincenzo2000}. These multi-spin encodings can provide some resilience to charge or magnetic noise, at the cost of more complex control schemes. This study restrains itself exclusively to single-spin qubits confined in gate-defined quantum dots, as this platform currently offers the most mature technology for semiconductor-based quantum computing. Spin-qubit arrays with up to twelve~\cite{george2025} and ten~\cite{john2025} qubits in silicon and germanium quantum wells, respectively, were demonstrated. In parallel, single- and two-qubit gate fidelities have overcome the fault-tolerance threshold for quantum error correction \cite{yoneda2018, veldhorst2014, xue2022, noiri2022,mills2022}.
\begin{figure}
    \centering
    \begin{subfigure}{0.35\textwidth}
        \includegraphics[width=\textwidth]{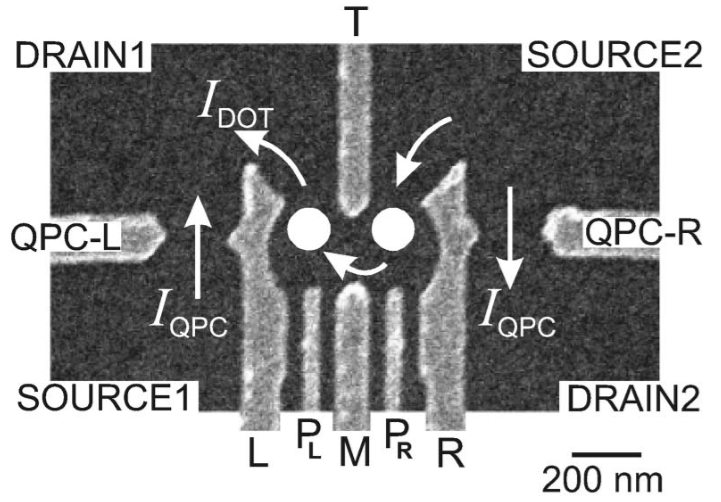}
        \caption{}
    \end{subfigure}
    \begin{subfigure}{0.45\textwidth}
        \includegraphics[width=\textwidth]{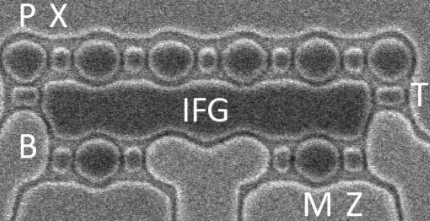}
        \caption{}
    \end{subfigure}
    \begin{subfigure}{0.38\textwidth}
        \includegraphics[width=\textwidth]{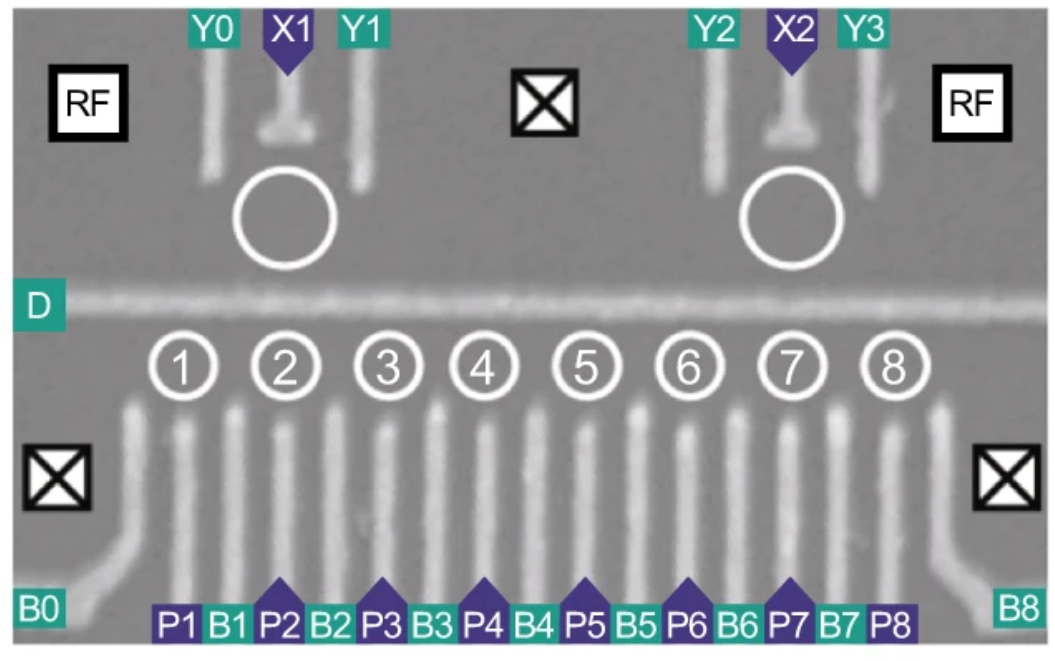}
        \caption{}
    \end{subfigure}
    \begin{subfigure}{0.42\textwidth}
        \includegraphics[width=\textwidth]{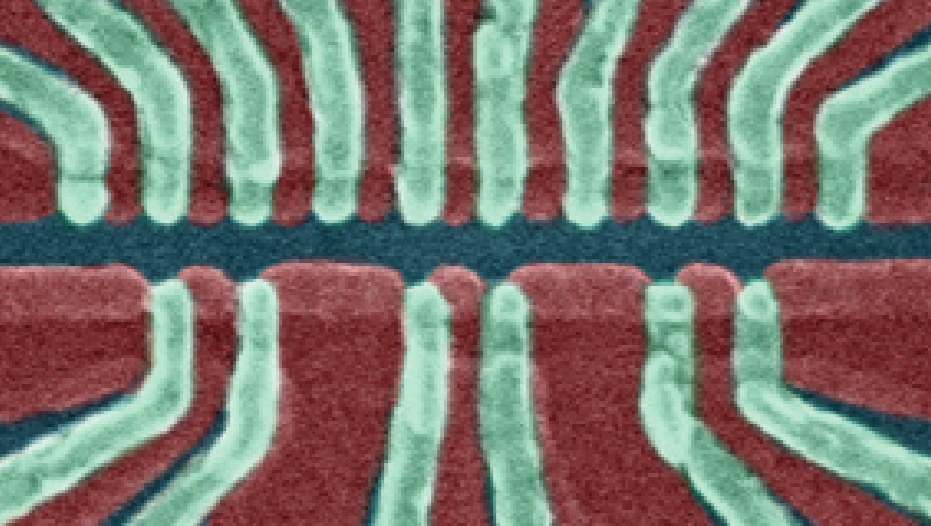}
        \caption{}
    \end{subfigure}
    
    \caption{Electron microscope images of spin-qubit processors, with a) two qubits, b) six qubits, c) eight qubits, and d) nine qubits. Pictures taken from~\cite{Elzerman2003, Volk2019,Ha2022,Zajac2016}, respectively. The semiconductor technology used in each design might differ. Images not at scale.}
    \label{fig:placeholder}
\end{figure}

A magnetic field lifts the degeneracy between the spin-up and spin-down states through the Zeeman effect, defining the qubit energy splitting. Spin rotations (i.e. single-qubit gates) can be driven electrically by taking advantage of spin–orbit coupling. This coupling can be intrinsic, as is the case for holes in silicon or germanium, where the spin and orbital motion are naturally entangled~\cite{maurand2016}, or engineered by introducing a magnetic field gradient using on-chip micromagnets, enabling electrical control of electron spins in otherwise weakly spin–orbit-coupled materials~\cite{pioro-ladriere2008}.

Two-qubit gates are typically realized through the exchange interaction, which arises when the wavefunctions of neighboring electrons overlap. Because of the Pauli exclusion principle, electrons with parallel spins experience a higher energy cost than those with opposite spins, effectively producing a spin–spin coupling that can be tuned by gate voltages~\cite{veldhorst2015}. Beyond nearest-neighbor coupling, long-range two-qubit interactions have recently been demonstrated using photon-mediated coupling through superconducting microwave cavities~\cite{dijkema2025}, opening the door to modular spin-based quantum processors.

The state of a spin qubit is read out indirectly by converting spin information into charge information. One of the most common methods relies on Pauli spin blockade: two electrons with the same spin cannot occupy the same quantum dot orbital, even if it would minimize the electrostatic energy~\cite{koppens2006}. This spin-dependent charge configuration can then be detected by a nearby charge sensor, such as a quantum point contact or a single-electron transistor. An alternative approach is RF gate-based sensing, where the charge configuration is inferred from the change in impedance of an RF resonant circuit, which can be either classical~\cite{hamonic2025} or superconducting~\cite{petersson2012}.

Spin qubit devices are typically operated at millikelvin temperatures in a dilution refrigerator, comparable to the conditions required for superconducting qubits. Operating at such low temperatures minimizes thermal excitations that could otherwise randomize the spin state. Remarkably, recent experiments have shown that coherent spin-qubit operation remains feasible up to 4 K~\cite{camenzind2022}, which would enable simpler and more compact cryogenic setups without the need for $^3$He-based refrigeration. Similarly to superconducting quantum processors, spin-based chips are mounted on the coldest plate of the dilution refrigerator. Electrical control and readout signals are delivered through DC and coaxial lines connecting the cryogenic environment to the electronic control hardware at room temperature.

\subsection{Energy model}
This section proceeds to discuss the hardware elements present in a spin qubit quantum computer, and the operation times required to execute circuits. Due to the current state of spin qubit development, the present devices fall short in terms of number of qubits in comparison with the other platforms. For that reason, the estimations shown in this section are calculated by extrapolating the number of hardware components that the current devices require to a larger amount of qubits, to provide estimations that correspond to a computer of similar size to those shown in other sections. In addition, we assume that the cryogenic setup for a spin qubis computer is equal to the one corresponding to a superconducting computer of the same size, presented in the previous section. Appendix~\ref{app:spin_qubits} contains the detailed list of components and operation times of the considered spin qubits processor design.

\subsubsection{Power consumption of the hardware}
The hardware components of the baseline spin-qubit computer are shown in Tables~\ref{tab:spin_components} and~\ref{tab:spin_components_2} of Appendix~\ref{app:spin_qubits}. In this case, the number of components correspond to a $7\times 7$ 2D square lattice architecture, with $N_{\rm{q}}=49$ qubits. We consider a full one-to-one control scheme where each qubit is addressed with an individual driving line. Resonators are placed on half of the qubit sites to perform parity readout. The number of resonators in this architecture would therefore be $N_{R}=\lceil N_{\rm{q}}/2\rceil=25$. The end of this section discusses the energy efficiency of different architectures with more complex qubit connectivity based on multiplexing.

\begin{figure}[!h]
    \centering
    \includegraphics[width=0.8\linewidth]{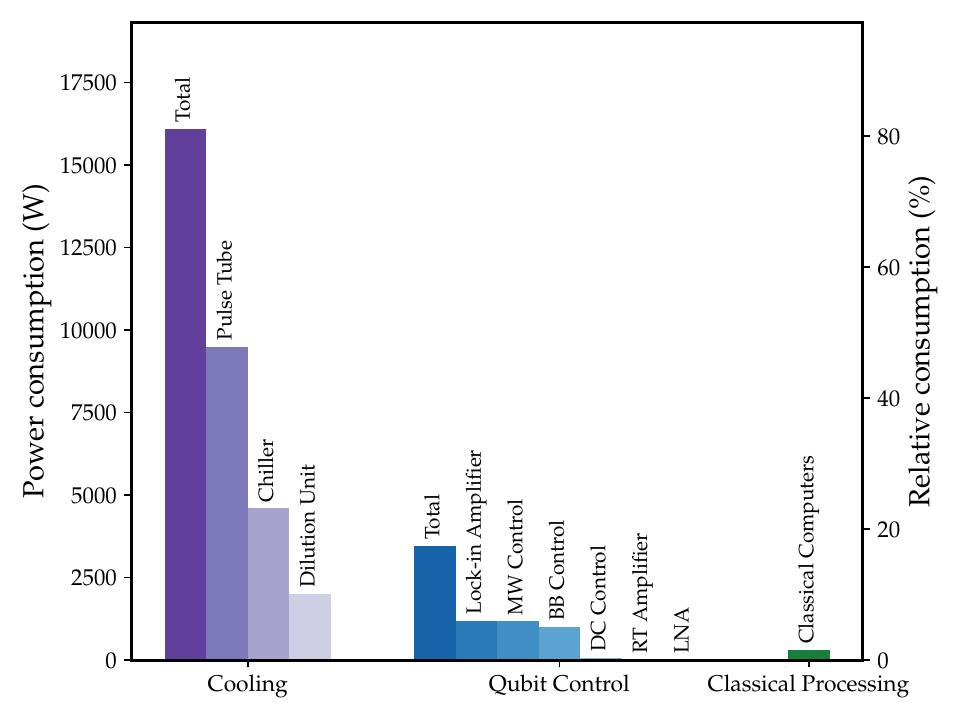}
    \caption{Power breakdown of a spin-qubit computer with 49 qubits, according to the components shown in Table~\ref{tab:superconducting_components} of Appendix~\ref{app:spin_qubits}. In the x-axis, the components are grouped depending of their purpose in the computer. Each group has a bar that corresponds to the total accumulated power of its components. The left y-axis shows the absolute value of power consumption of the components, and the right y-axis the relative value with respect to the overall consumption of the computer.}
    \label{fig:spin_power_breakdown}
\end{figure}

Taking the components shown in Table~\ref{tab:spin_components}, a spin qubits quantum computer with $N_{\rm{q}}=49$ qubits, would consume $P^{\rm{spin}}\approx20~\mathrm{kW}$ of power. Therefore, having this computer turned on for a day would cost $\approx1700$ MJ of energy. Figure~\ref{fig:spin_power_breakdown} shows the power demand of all the components in the computer, split in three groups: cooling, qubit control and classical processing. As one can see, the power consumption is distributed similarly to the superconducting computer examined in Sec.~\ref{sec:superconducting_results}. Likewise, the cooling hardware of a spin-qubit computer consumes $16.1~\mathrm{kW}$ of power, which corresponds to the $81\%$ of the total consumption of the device.

As we mention above, we consider the signals that control the qubits to be addressed individually by dedicated electronics. Then, the number of lines does not only depend on the number of qubits, but also the connectivity of the processor. In the following section we explore the energy efficiency of different architectures, based on their connectivity graph and their multiplexing capabilities. Although the increase in the power consumption of qubit control electronics can be clearly analyzed, it is unclear how increasing the number of lines and qubits will affect the power consumption of the cryogenic setup.


\subsubsection{Derivation of the energy efficiency}
\label{sec:spin_time}
The derivation of the energy efficiency of running an algorithm $\mathcal{A}$ on a spin qubit platform is identical to the one for the superconducting-qubit platform presented in Section~\ref{sec:energysuperc}. Given the final depth $D$ of a circuit implementing $\mathcal{A}$ on a spin qubit quantum computer, the execution time of $\mathcal{A}$ is given by: 

\begin{equation}
    \label{eq:tcircuit_spin}
    t_{\mathcal{A}}^{\textrm{spin-qubit}}= N_{\rm{samples}}\cdot t_{\rm{circuit}}^{\textrm{spin-qubit}} = N_{\rm{samples}}\cdot (t_{\textrm{reset}} + D \cdot t_{\rm clock} + t_{\rm meas}),
\end{equation}

where $N_{\rm{samples}}$ is the number of samples, $t_{\textrm{reset}}$ is the time necessary to reset the qubit between every circuit execution, $t_{\rm meas}$ is the time needed to read the final state of the qubits at the end of the circuit and $t_{\rm{clock}}$ is the clock time of the computer. In order to provide estimations, we base these values on the updated review of spin-qubit computers by Stano and Loss~\cite{stano2022}. According to their survey, these operation times can vary several orders of magnitude, even for spin qubits of the same species. In this work, we consider the following average values: $t_{\rm{reset}}\approx 0.1~\mathrm{ms}$, $t_{\rm{meas}}\approx10~\mathrm{\mu s}$ and $t_{\rm{clock}}=t_{\textrm{2q}}\approx 10~\mathrm{ns}$. Comparing these operation times, we expect to see the computations in spin-qubit computers to be mainly determined by the reset and measurement time of the qubits, which are several orders of magnitude larger than the clock time.
The energy efficiency of a spin-based quantum computer is thus given by 

\begin{equation}
\label{eq:EEspin}
    EE^{\rm{spin}} =  \frac{1}{N_{\rm{samples}}\cdot (t_{\textrm{reset}} + D \cdot t_{\rm clock} + t_{\rm meas}) \cdot P^{\rm{spin}}},
\end{equation}

where $P^{\textrm{spin}}$ is the sum of the power consumption of the hardware elements of Table~\ref{tab:spin_components}.

\subsection{Results and discussion}
\label{sec:spin_results}

\begin{figure}[!h]
    \centering
    \includegraphics[width=\linewidth]{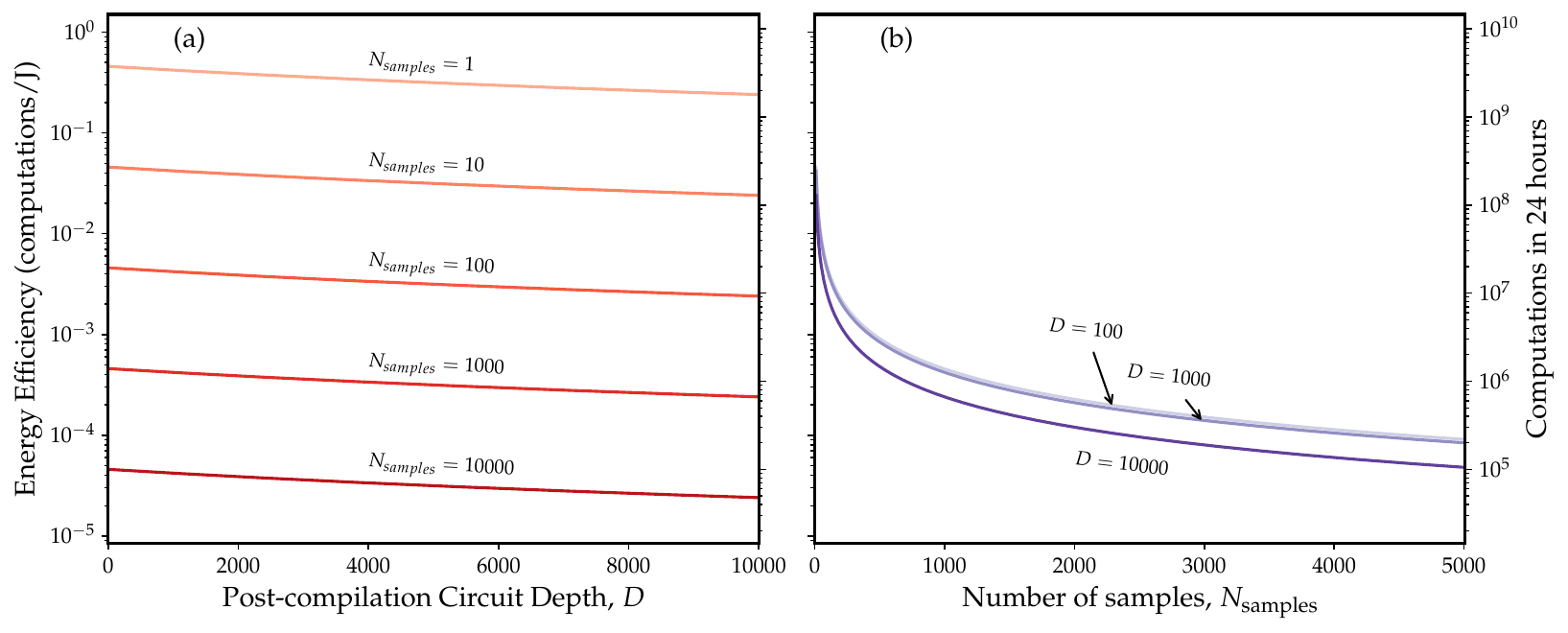}
    \caption{Energy efficiency (left y-axis) and number of computations (right y-axis) in 24 hours of a spin-qubit computer depending on final circuit depth, $D$ and number of samples, $N_{\rm{samples}}$. Panel (a) shows several series with different $N_{\rm{samples}}$ values and the x-axis corresponds to $D$. Series in panel (b) have different $D$ values and its x-axis corresponds to $N_{\rm{samples}}$.}
    \label{fig:spin_EE_vs_Nsamples}
\end{figure}

In Figure~\ref{fig:spin_EE_vs_Nsamples}, we show the energy efficiency (left y-axis) and number of computations (right y-axis) of a spin-qubit computer executing a circuit with given $D$ and $N_{\rm{samples}}$ for 24 hours. Since $EE^{\rm{spin}}$ and $N^{\rm{spin}}$ are inversely proportional to the number of samples, $N_{\rm{samples}}$, the different series in panel (a) are equidistant on a logarithmic scale, shifted by an order of magnitude.

Interestingly, the $EE$ curve profiles appear almost flat, in contrast with the results discussed in the superconducting qubits section~\ref{sec:superconducting_results}. This behavior is directly related to the relatively high initialization time of spin qubits. As mentioned above, the clock time does not play any major role until the number of layers is large enough. Therefore, one can see how the energy efficiency of the device barely decreases as the depth of the circuit grows. As an example, running a single execution of a circuit with $D=100$ layers would take $t_{\rm{circuit}}=111~\mathrm{\mu s}$, where reset and measurement operations took $99\%$ of the time. However, if we increase the depth of the circuit enough ($D\approx t_{\rm{reset}}/t_{\rm{clock}}=10000$), the clock time starts taking over the majority of computing time. To put this into perspective, a single sample of an extremely long circuit with $D=10000$ layers would take $t_{\rm{sample}}=210~\mathrm{\mu s}$. In this case, the spin-qubit device would spend almost half of the computing time initializing the qubits.

With that in mind, it is clear that near term machines that will run short circuits can improve their energy efficiency by optimizing the initialization time of the qubits. However, this improvement would only be perceived for circuits with a relatively short depth ($D<t_{\rm{reset}}/t_{\rm{clock}}$). A representation of this behavior can be seen in Figure~\ref{fig:superconducting_t_init}, where decreasing the initialization time of superconducting qubits leads to pronounced gradients of the $EE$ with respect to $D$, for relatively short circuits.

The panel (b) of Figure~\ref{fig:spin_EE_vs_Nsamples} displays the values of $EE$ and $N^{\pi}$ with respect to the number of samples, $N_{\rm{samples}}$, taken for three different circuits with depths $D=100$, $1000$ and $10000$. Here, the gap between lines is also determined by the initialization time of the qubits, for the same reason we explained above: the clock time is not the main contribution of the computing time, until the depth of the circuit is the order of $D\approx10000$ layers. Therefore, the $D=100$ and $D=1000$ series in panel (b) have almost the same values, since these circuits take almost the same time to be computed.

Additionally, we analyze two extra features  that have a major impact on the computation time and energy expenditure. Until now, spin qubits processors with a a 2D qubit connectivity have been considered. However, due to the small number of qubits in actual processors, current devices are often displayed with linear graphs~\cite{stano2022}. We proceed to demonstrate how the energy efficiency of spin-qubit computers would scale if they were developed according to different qubit connectivities, rather than a two-dimensional one. Section~\ref{sec:estimating_depth} already displays the poor scaling of linear graphs, in comparison with higher-dimensional ones. The same comparison can be seen here, applied to the specific case of spin-qubit components and operation times.

\begin{figure}[!h]
    \centering
    \includegraphics[width=0.8\linewidth]{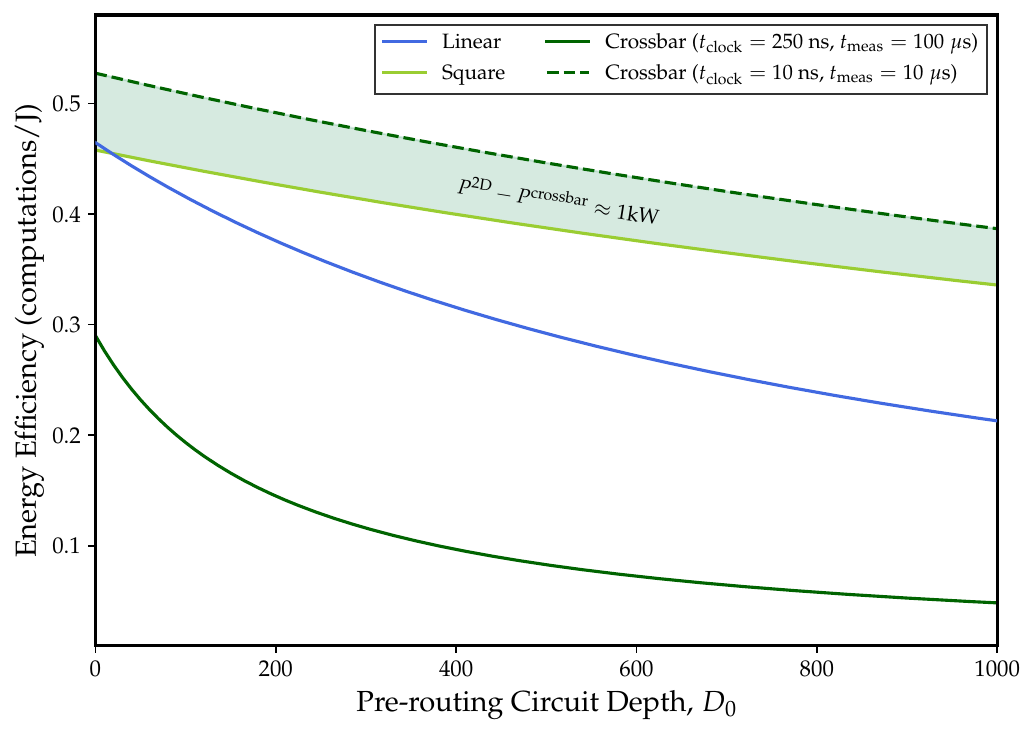}
    \caption{Energy efficiency of a spin-qubit computer executing 1 sample of a circuit, depending on its pre-routing depth $D_{0}$. The blue curve corresponds to a computer with a linear connectivity graph, the solid light green to a 2D square-lattice connectivity graph, the dark green to a crossbar device with computational times taken from Ref.~\cite{li2018}, and the dashed dark green line to a crossbar device with the computational times detailed in Sec.~\ref{sec:spin_time}. The shaded area marks the difference in $EE$ between the 2D processor and a Crossbar architecture that share the same operation times, thanks to the reduced number of lines. To compute the final circuit depth of the compiled circuits, we assume a ratio of layers including two-qubit gates $\alpha_{\rm{2q}} = 0.5$.}
    \label{fig:spin_planar_vs_linear}
\end{figure}

Figure~\ref{fig:spin_planar_vs_linear} shows the energy efficiency as a function of the pre-routing depth of the circuit, $D_{0}$, for a ratio of layers including two-qubit gates $\alpha_{\rm{2q}}=0.5$. This value corresponds to a circuit that has at least one two-qubit gate on half of their layers, after transpilation. The figure includes values corresponding to linear (solid light green) and 2D (solid dark green) connectivities. In addition, we include the energy efficiencies of a processor with a "Crossbar" architecture, that is discussed at the end of the section.

For this comparison, we consider that the computers with linear and 2D connectivities share the majority of components, besides the number of DC control lines and base-band (BB) control lines. Since the spins in the linear processor have fewer nearest neighbors than the 2D processor, they also have fewer coupling signals, thus having reduced numbers of DC and BB lines. In Table~\ref{tab:spin_components} and~\ref{tab:spin_components_2}, we break down all the number of lines for each computer. 

Due to this line reduction, the power consumption of the linear processor saves $200~\mathrm{W}$ of power. Despite this, the linear processor quickly becomes less efficient than the 2D one as we increase the pre-routing depth of the executed circuit. Such efficiency loss is due to the extra-computing time that the processor needs to execute a larger amount of SWAP gates. It should be noted that, for a given $D$, both processors would take the same time and consequently, the linear processor would be slightly more efficient, due to the $200~\mathrm{W}$ saved in qubit control. In order to interpret Figure~\ref{fig:spin_planar_vs_linear} correctly, one has to consider that for the same value of pre-routing depth, $D_{0}$, the linear processor has a larger SWAP gate overhead than the 2D one, and thus executes the same circuit with a larger final depth $D$. 

With that in mind, it is clear how the large SWAP overhead due to the limited linear connectivity results in a substantial energy efficiency drop. To put things in perspective, for a precompiled depth of $D_{0}=500$ and $\alpha_{\rm{2q}}=0.5$ the final depth of the circuit for the 2D-connected processor is $D^{\rm{2D}}=2000$. Meanwhile, the final depth of the same circuit for a linearly-connected processor is $D^{\rm{linear}}=6500$. The computation time of these circuits are $t^{\rm{2D}}_{\rm{circuit}}=0.13~\mathrm{ms}$ and $t^{\rm{linear}}_{\rm{circuit}}=0.17~\mathrm{ms}$. This is another example on how the initialization time is still relevant even for circuits that are far apart in terms of depth. We estimate the energy efficiency of these devices for such circuits to be $EE^{\rm{2D}}(D_{0}=500,\alpha_{\rm{2q}}=0.5)=0.387~\mathrm{computations/J}$ and $EE^{\rm{linear}}(D_{0}=500, \alpha_{\rm{2q}}=0.5)=0.292~\mathrm{computations/J}$. 

Finally, we extend the previous comparison to an additional architecture proposed by Li \emph{et al.} in Ref~\cite{li2018}. In this architecture, a 2D lattice of quantum dots is controlled using superconducting strips that carry multiple control signals among rows, columns and diagonals of qubits. All the signals controlled by each bar can be multiplexed into a single wire, reducing the total number of lines and their corresponding active electronic components. With these multiplexed signals, we estimate that a 49 spins processor could save up to $\approx 1~\mathrm{kW}$ of power, thanks to the reduction of lines and control electronics. This estimation, however, only takes into account the power directly delivered to the hardware that we removed from the previous non-multiplexed architecture, but does not consider the potential reduction in terms of cryogenic demands thanks to fewer sources of heat dissipation inside the cryostat. In Appendix~\ref{app:spin_qubits} we include more details regarding the number of lines, electronics, and operation times of the crossbar architecture.

Figure~\ref{fig:spin_planar_vs_linear} shows the energy efficiency as a function of the pre-routing circuit depth $D_{0}$, assuming it has a 2D connectivity graph. First, we estimate the $EE$ of a crossbar processor based on the clock time $t_{\rm{clock}}^{\rm{crossbar}} =t_{\rm{gate}} \approx 250~\mathrm{ns}$ given in the original article. Additionally, we assume that the measurement time of this device would scale with $\sqrt{2N_{\rm{q}}}$, since the corresponding signals of entire rows of qubits should be time multiplexed into individual wires, hence, $t_{\rm{meas}}^{\rm{crossbar}}\approx\sqrt{2N_{\rm{q}}}\times 500~\mathrm{\mu s}$. With these numbers in mind, one can see how the energy efficiency of such device (dark green dashed line) would drop drastically as the pre-routing depth of the circuits increase, in comparison with the previous architectures with a clock $25$ times faster. Interestingly, we see how the linearly-connected processor outspeeds the crossbar one, even with its constrained connectivity, which leads to higher efficiencies. This result stresses the importance of computational speed for energy efficiency, as given by Eq~\ref{eq:EE}.

Additionally, Figure~\ref{fig:spin_planar_vs_linear} shows the same profile for a crossbar processor that has the same operation times as the Linear and 2D processors (dashed dark green line). It is unclear whether such times would be achievable with this architecture, but this result shows how, even with a substantial power reduction of $\approx1~\mathrm{kW}$ of saved power, computational times can have a larger effect on the overall efficiency of the computer.

\section{Trapped Ions}
\label{sec:trappedions}
Trapped ions~\cite{Bruzewicz2019, Bernardini2023, Cirac1995}, are an example of a different kind of quantum computer with respect to the previous sections, based on entangling the electronic state of different atoms through controlled collisions.  That is to say, the electronic states of different atoms are correlated by controlling the centre-of-mass degrees of freedom to modulate the interaction between the internal states~\cite{Turchette1998,Leibfried2003}. In addition, the internal electronic states of individual ions can be separately controlled and measured. Atom-based platforms like trapped ions, or neutral atoms, proved to be a feasible alternative to build quantum computers at room temperature. Some of the main adopters of trapped ion quantum computers are IonQ~\cite{Ye2025} and Quantinuum~\cite{Liu2025}.

\subsection{Technology}
Trapped ion computers consist of a chain of ionized atoms that are trapped using time-dependent electromagnetic fields. State preparation and gate drive is performed with lasers and/or microwave fields. Each ion in the chain behaves as a physical qubit, and their internal energy states are used to encode information. Typically, those are internal electronic states, corresponding to those of an uncoupled, valence electron (known as optical levels), those of the electron coupled with the atom nucleus (known as hyperfine levels)~\cite{Bernardini2023}, or the electronic energy states on a magnetic field (Zeeman states). High fidelity gates have been demonstrated for different choices of states~\cite{Brown2011, Ballance2016, Hughes2025}. Laser cooling is used to reduce the kinetic energy of the ions so that motional dephasing is suppressed. On top of that, a ultra-high vacuum environment is required, as background collisions are the main driver of atom loss. 

Single- and two-qubit gates can be implemented by irradiating the ions with electromagnetic field pulses, either using lasers or microwaves. For single qubit gates, the driving field only changes the internal electronic states while for two qubit gates the internal electronic states and centre-of-mass motional states are coupled, requiring 'phonons' to mediate the interaction as well~\cite{Cirac1995,Bernardini2023}. Movement of the ions can be achieved by changing the electromagnetic potential that confines them, a process known as shuttling. Furthermore, a modular architecture based on moving the ions around different regions of the trap was proposed, with the aim to enable scalability. This modular structure enables parallel gate operations and scalable quantum charge-coupled device (QCCD) architectures~\cite{Kielpinski2002ArchitectureFA}. Shuttling the ions is a fundamental process that allows all-to-all connectivity in ion trapping devices. However, motional heating can occur during ion shuttling and this needs to be mitigated by using extra cooling steps. Therefore, both short and long range transport, and  laser cooling have significant impact in the duty cycle and therefore the power consumption.

The state of the qubits is measured by exciting the ions into an auxiliary excited state that rapidly emits light (a cycling transition), which can be captured by a camera or a photomultiplier tube~\cite{Todaro2021}. Ion traps can be implemented at room temperature; the ions are isolated from the external noise using ultra high vacuum (UHV) and cooled down using laser cooling. However, ion trap chips~\cite{Allcock2013, Weber2024} are often operated at cryogenic temperatures in order to reach a better vacuum conditions and suppress electric noise from the trap.

Our baseline analysis considers a generic, linear, QCCD architecture, operating at room temperature. We refrain from choosing laser or microwave gates, and instead use an average gate drive which would encompass both. The time duration of the gates are based on laser gates, but are of the same order of magnitude of the fastest microwave gates. A single ionic species is assumed to be used, and no sympathetic cooling is performed, only Doppler and sideband cooling~\cite{Itano1995, Metcalf1999}. 

\subsection{Energy model}

This section discusses the energy efficiency of a trapped-ion device constructed from the hardware elements displayed in Appendix~\ref{app:trapped_ions}. As a baseline, we consider a trapped-ion computer that has a single gate zone, which means it can only perform one single- or two-qubit gate at a time. In the next section, the analysis is extended to architectures with multiple gate zones where parallelization is possible, in order to study the potential increase in energy efficiency they enable by decreasing computation time.
\begin{figure}[!h]
    \centering
    \includegraphics[width=0.9\linewidth]{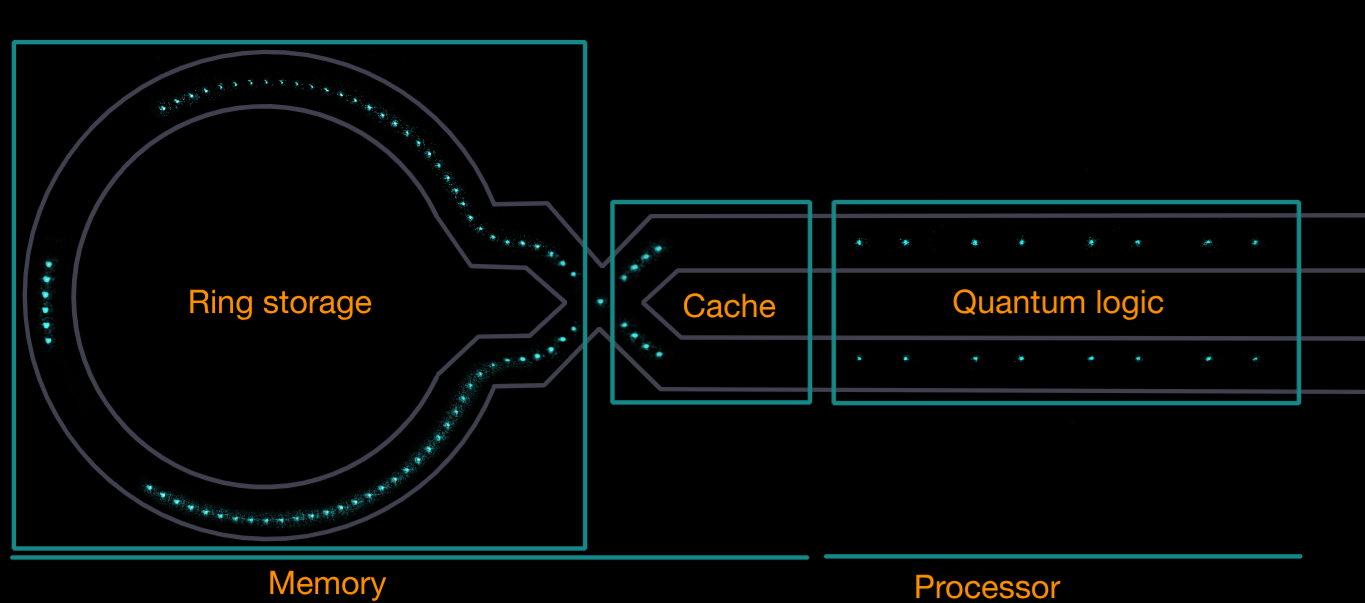}
    \caption{Image of Quantinuum's Helios computer, with 98 qubits. This particular device is composed of a ring-shaped storage, connected to the interaction zone (labeled as "Quantum logic"), with a cache in-between. Image taken from~\cite{Ransford2025}.}
    \label{fig:photo_helios}
\end{figure}

\subsubsection{Power consumption of the hardware}
The detailed list of components of a trapped-ion device with a single gate zone is shown in Table~\ref{tab:trapped_ions_components}. The majority of data in this table has been measured in the NQCC trapped ions laboratory. The power consumption related to the Heating, Ventilation and Air-Conditioning (HVAC) system, which stabilizes the temperature and humidity required by the laser systems, corresponds to an average estimation based on both our trapped ions and neutral atoms data, assuming they require similar conditions. However, HVAC power consumption strongly depends on the facilities hosting the computer. In this work, a value was chosen in accordance with observations in real facilities, but this value might vary depending on the efficiency of the entire infrastructure, the size of the room, the climate conditions of its location, etc. Moreover, the following estimations do not include the cryogenic setup that is used in the NQCC laboratory, based on the fact that most atom-based devices do not operate at cryogenic temperatures. 

\begin{figure}[h!]
    \centering
    \includegraphics[width=0.85\linewidth]{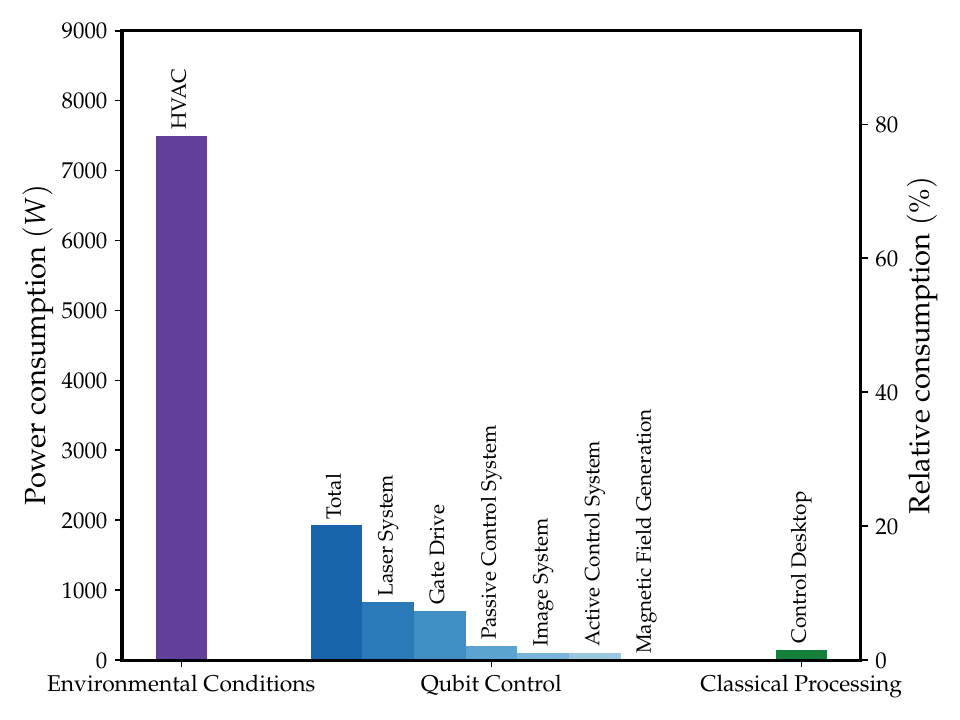}
    \caption{Power breakdown of a trapped ions computer, according to the components shown in Table~\ref{tab:trapped_ions_components} of Appendix~\ref{app:trapped_ions}. In the x-axis, the components are grouped depending of their purpose in the computer. Each group has a bar that corresponds to the total accumulated power of its components. The left y-axis shows the absolute value of power consumption of the components, and the right y-axis the relative value with respect to the overall consumption of the computer.}
    \label{fig:trapped_power_breakdown}
\end{figure}

Figure~\ref{fig:trapped_power_breakdown} shows the consumption of the hardware listed in Table~\ref{tab:trapped_ions_components} in three groups: Environmental Conditions, Qubit Control and Classical Processing. This particular device would be able to host the order of $\sim 10$ qubits, but we expect the same order of magnitude in power demand for a $\sim100$ qubits computer.

According to our estimations, the overall consumption of this particular device is $P_{\rm{trapped\;ions}}\approx10~\mathrm{kW}$, which corresponds to $\approx900~\mathrm{MJ}$ of energy spent in $24$ hours. As one can see, there is a large gap between the power needed to ensure the suited environmental conditions and the rest of the components. This is similar to the high demand of cryogenic fridges we see in solid-state platforms. In particular, the consumption of the HVAC system represents $78\%$ of the total consumption of the device. This value, however, can be highly variable for the reasons we comment above. In addition, this computer can be expected to function under a less demanding HVAC system, since its laser setup is lighter than the one included for the neutral-atom computer in the next section.

The aggregated consumption of all the qubit control components is $\approx1.9~\mathrm{kW}$. Similarly to previous platforms, there is a clearer relation between the size and power consumption of the qubit control components and size of the quantum processor. In this case, the components are mainly determined by the number of gate zones in the device and the ability to perform independent gates on those gate zones. This is further discussed in Appendix~\ref{app:trapped_ions}.


\subsubsection{Derivation of the energy efficiency}

This section explains the model used to estimate the running time of an algorithm $\mathcal{A}$ composed of $N_{\rm{samples}}$ shots of a quantum circuit with depth $D$. As explained in the introduction, the running time $t_{\mathcal{A}}^{\textrm{trapped}}$ of an algorithm is given by the running time $t_{\rm{circuit}}^{\textrm{trapped}}$ of a quantum circuit applied on a trapped ion device:

\begin{equation}
    \label{eq:tcircuit_spin}
    t_{\mathcal{A}}^{\textrm{trapped}}= N_{\rm{samples}}\cdot t_{\rm{circuit}}^{\textrm{trapped}} = N_{\rm{samples}}\cdot (t_{\textrm{reset}} + D \cdot t_{\rm clock} + t_{\rm meas} + \beta_{\textrm{trans}}\cdot D\cdot t_{\rm{trans}}),
\end{equation}

where $N_{\rm{samples}}$ is the number of circuit executions, $t_{\textrm{reset}}$ is the time necessary to reset the qubit between every sample of the algorithm, $t_{\rm meas}$ is the time needed to read the final state of the qubits at the end of the circuit and $t_{\rm{clock}}$ is the clock time of the computer. For the most part, the model is equivalent to the solid-state one, with the addition of the time $t_{\rm{trans}}$ needed to transport the atoms from the storage to the interaction zone. We introduce the parameter $\beta_{\textrm{trans}}$, the ratio of the number of layers requiring long-range transports over the total depth. This parameter characterizes how many times qubit need to be moved in and out of the gate zones and highly depends on the circuit. Appendix~\ref{app:coldcomp} provides more detail regarding atom shuttling. 

In order to provide a state-of-the-art estimation of the computing time, the following values are based on Refs.~\cite{Pino2021a, Ransford2025}, which demonstrate experimentally the viability of QCCD computers and provide values of their operational times. However, we adjust their numbers in order to have a general representation of a trapped ions device that have a storage zone connected to a single, or multiple, gate zones distributed linearly. First, we consider $t_{\rm{reset}}\approx 150~\mathrm{ms}$, which corresponds to the time needed to move the qubits from the gate zones to the storage zone ($50~\mathrm{ms}$) and the time to perform Doppler and sideband cooling to reset the qubits into their $\ket{0}$ state ($100~\mathrm{ms}$). 

There are different shuttling operations that need to be performed to operate with the ions in the different gate zones. First, we consider the transportation time $t_{\rm{trans}}\approx50~\mathrm{ms}$, which is the result of shuttling the ions from the storage zone and the gate zone, and any additional cooling necessary before performing the gates. Next, there is a shuttling operation needed to move the ions within the gate zones, if the device has multiple of them. This time is typically shorter than the transport time, and depends on the number of gate zones in the computer, as well as the velocity at which the ions can be moved and the distance between zones. For multi-zones devices, this operation dominates the clock time, and can be estimated by $t_{\rm{clock}}^{\rm{multi-zone}}\approx250~\mathrm{\mu s}\times (N_{\rm{zones}}^{\rm{gate}}-1)$. When it comes to the single-gate-zone device, this shuttling is not needed, thus we consider its clock time to be the time needed to perform two-qubit gates: $t_{\rm{clock}}^{\rm{single-zone}}\approx 170~\mathrm{\mu s}$. In this case, this value accounts for a two-qubit entangling gate. Finally, the state of the qubits is measured using a digital camera or a photomultiplier tube (PMT), that take $t_{\rm{meas}}\approx 0.5~\mathrm{ms}$.

In the case of a single-gate-zone device, $t_{\rm{trans}}$ is three orders of magnitude larger than $t_{\rm{clock}}$. Therefore, the execution time of a circuit in this device depends highly on the ratio of transports $\beta_{\textrm{trans}}$. For instance, a sample of a circuit of depth $D=100$ with $\beta_{\textrm{trans}}=1$ would take $\approx5.1~\mathrm{s}$. This circuit corresponds to a series of gates that are always applied on different qubits, \emph{i.e.} the device has to bring a new qubit (or pair of qubits) from the storage to the gate zone every time step. In contrast, the time to execute a circuit with $\beta_{\textrm{trans}}=0.5$, is $\approx2.6~\mathrm{s}$. In this case, the circuit performs, on average, two consecutive gates to the same qubit (or pair of qubits) before reaching for a new one from the storage zone. The computing time of a circuit with no transport ($\beta_{\textrm{trans}}=0$) would be $\approx0.1~\mathrm{s}$. This circuit would be a series of $100$ gates applied on the same qubit, but serves as an example of how the structure of the circuit can affect the final execution with such a limited number of gate zones.

Taking all of this into account, the equation for the energy efficiency of trapped-ion devices is :

\begin{equation}
\label{eq:EEtrapped}
    EE^{\textrm{trapped}} =  \frac{1}{N_{\rm{samples}}\cdot (t_{\textrm{reset}} + D \cdot t_{\rm clock} + t_{\rm meas}+ \beta_{\textrm{trans}}\cdot D\cdot t_{\rm{trans}}) \cdot P^{\textrm{trapped}}},
\end{equation}

where $P^{\textrm{trapped}}$ is the sum of the power consumption of the hardware elements of Table~\ref{tab:trapped_ions_components}.

\subsection{Results and discussion}
\begin{figure}[!h]
    \centering
    \includegraphics[width=\linewidth]{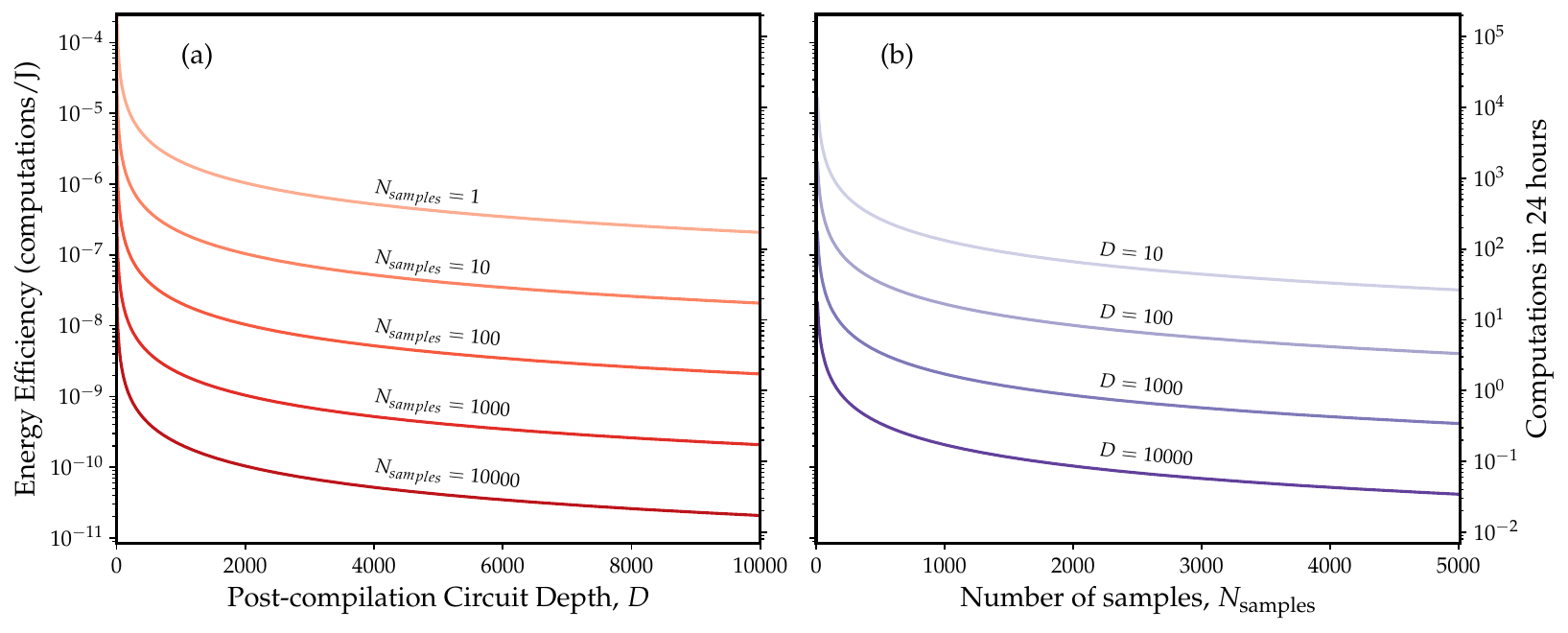}
    \caption{Energy efficiency (left y-axis) and number of computations (right y-axis) in 24 hours of a single-gate-zone trapped-ion computer depending on the final circuit depth, $D$ and number of samples, $N_{\rm{samples}}$. Panel (a) shows several series with different $N_{\rm{samples}}$ values and the x-axis corresponds to $D$. The series in panel (b) have different $D$ values and its x-axis corresponds to $N_{\rm{samples}}$. The circuits displayed in the panels have a ratio of transport per layer $\beta_{\textrm{trans}}=1$.}
    \label{fig:trapped_ions_D_Nsamples}
\end{figure}

Figure~\ref{fig:trapped_ions_D_Nsamples} shows the energy efficiencies and number of computations in 24h of a single-gate-zone device, depending on the depth of the circuit and number of samples.

For the reasoning explained above, we assume that the circuits in Figure~\ref{fig:trapped_ions_D_Nsamples} have a ratio of transport-per-layer $\beta_{\textrm{trans}}=1$, \emph{i.e.} the device performs long-range transportation every layer. Effectively, this would correspond to $t_{\rm{clock}}\approx50~\mathrm{ms}$. This causes the profile of the curves in panel (a) to be inversely proportional to $D$, since the time to execute a single time-step of the circuit is close to the reset and measurement time of the qubits. Therefore, even for short circuits, the transport time takes over the total computation time.

Now, we want to explore the potential energetic advantage of adding additional gate zones in order to parallelize the gates of the circuits. In principle, this would highly decrease the depth of the circuits, thus optimizing the energy efficiency of the device. In this case, we do not expect an increase of hardware components due to the addition of gate zones. In fact, it would only require a negligible increase of power supply to the laser system. However, in order to take full advantage of all the gate zones in the device, we include one more Active Control System, which enables the device to perform single- and two-qubit gates at the same time. Otherwise the device would be limited to perform a single gate type per layer, and the depth of the circuits would increase. In addition, we also estimate the clock time to be longer than the single-gate-zone device. As we explain above, $t_{\rm{clock}}^{\rm{multi-zone}}\approx250~\mathrm{\mu s}\times (N_{\rm{zones}}^{\rm{gate}}-1)$. For the 5-zone device, the clock time would be $t_{\rm{clock}}^{\rm{5-zone}}\approx 1000~\mathrm{\mu s}$, whereas for the 10-zone device it would be $t_{\rm{clock}}^{\rm{5-zone}}\approx 2250~\mathrm{\mu s}$. Both of these times are still an order of magnitude below the transport time, and we can expect the transport to be the most time-consuming operation for circuits with $\beta_{\rm{trans}}$ close to 1.

\begin{figure}[!h]
    \centering
    \includegraphics[width=0.9\linewidth]{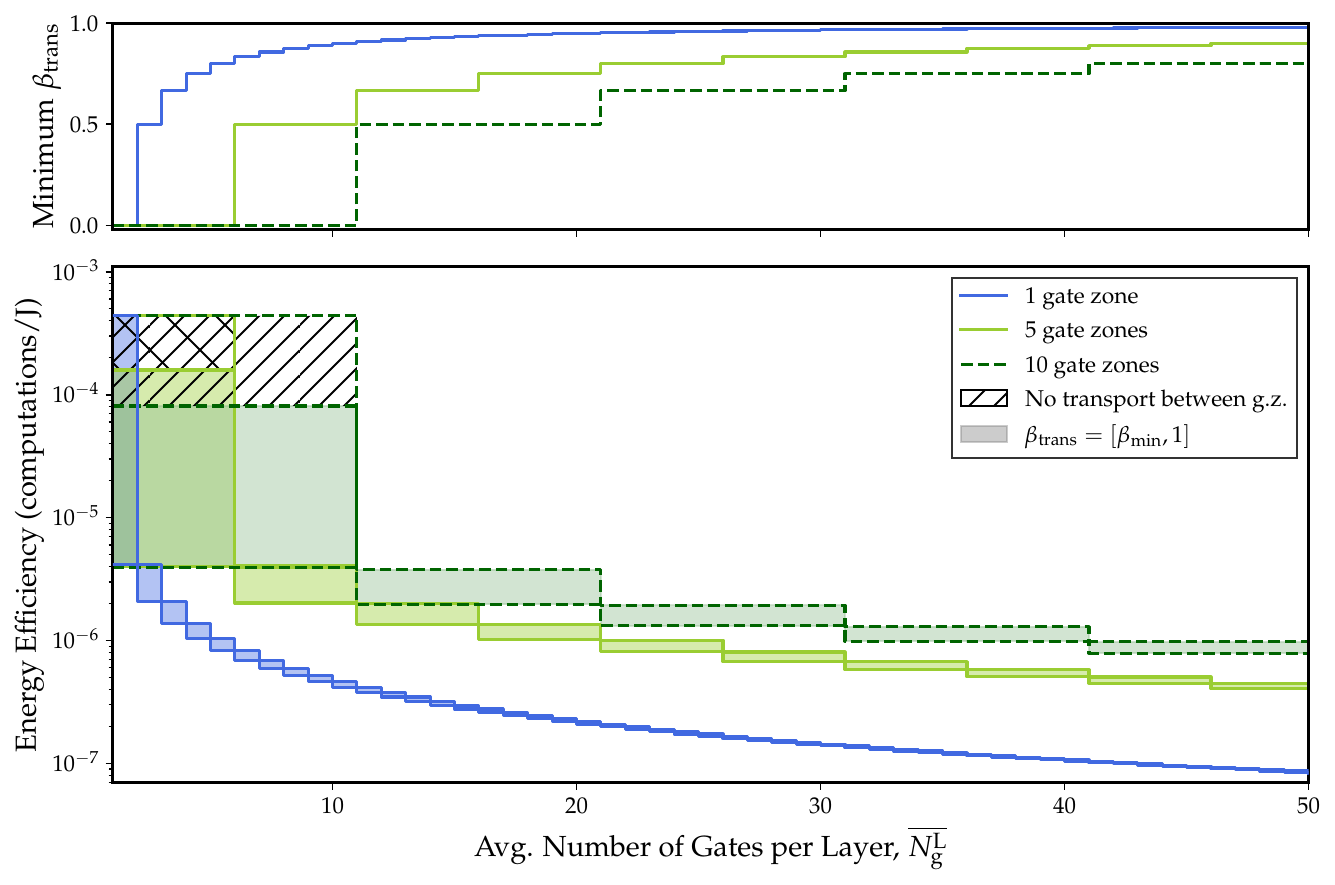}
    \caption{The upper panel shows the minimum value of the transport per layer ratio $\beta_{\textrm{trans}}$ for the 1,5 and 10 gate-zone devices, according to the average number of gates per layer $\overline{N_{\rm{g}}^{\rm{L}}}$. The lower panel shows the energy efficiencies of such devices when running a single sample of a $D_{0}=500$ circuit. The shaded areas display the possible efficiencies between the worst- and best-case scenarios $\beta_{\textrm{trans}}=[\beta_{\rm{min}},1]$. The black crossed area represents the improvement of the multi-gate-zone devices for circuits that transport betwee gate zones, for which $t_{\rm{clock}}^{\rm{multi-zone}}=t_{\rm{clock}}^{\rm{single-zone}}=170~\mathrm{\mu s}$.}
    \label{fig:trapped_ions_1vs5vs10_gatezones}
\end{figure}

In Figure~\ref{fig:trapped_ions_1vs5vs10_gatezones} we display the energy efficiencies of devices with 1, 5 and 10 gate zones, executing one sample of circuits with pre-routing depth of $D_{0}=500$ layers. The results depend on the average of gates per layer $\overline{N_{\rm{g}}^{\rm{L}}}$ and the ratio of transports $\beta_{\textrm{trans}}$.
The upper panel of Figure~\ref{fig:trapped_ions_1vs5vs10_gatezones} shows the minimum value of the ratio $\beta_{\textrm{trans}}$ of the three devices, depending on $\overline{N_{\rm{g}}^{\rm{L}}}$. Here, we assume that, if the average number of gates per layer is higher than the number of gate zones, the device is forced to perform a certain amount of transports. For instance, in order to execute a circuit with $\overline{N_{\rm{g}}^{\rm{L}}}=20$ gates per layer on the $N_{\rm{zones}}=10$ gate-zone device, the original layers of the circuit will have to be split in two. Since the 20 gates of the original layers use necessarily different qubits, the device will be forced to change the qubits between the two new layers that result from splitting the original ones. In Appendix~\ref{app:coldcomp}, we detail how we compute this minimum value of $\beta_{\textrm{trans}}$. On the contrary, the maximum value of this ratio can always be $\beta_{\rm{max}}=1$.

In the lower panel of Figure~\ref{fig:trapped_ions_1vs5vs10_gatezones}, we show the envelopes between the worst- and best-case scenarios of transport. The upper bounds of these envelopes correspond to the minimum value $\beta_{\rm{min}}$, while the lower bounds correspond to $\beta_{\textrm{trans}}=1$, \emph{i.e.} all layers of the circuit require transport. There is an extreme case where the single-gate-zone device would outperform the others, when $\overline{N_{\rm{g}}^{\rm{L}}}=1$ and $\beta_{\textrm{trans}}=0$. This corresponds to a circuit that only performs gates on a single qubit (or pair of qubits). In this particular case, the assumption that we make above, where we consider the clock time of the multiple-gate-zone computers to be equal to the transport between gate zones does not hold, since this operation is not needed. To make the comparison between devices fair, Figure~\ref{fig:trapped_ions_1vs5vs10_gatezones} also includes the energy efficiencies of the multi-gate-zone devices with $t_{\rm{clock}}\approx 170~\mathrm{\mu s}$, where its improvement is highlighted with a black pattern. Then, the energy efficiency of the single-gate-zone computer is marginally better, due to the $100~\mathrm{W}$ saved with respect to its counterparts. That being said, there are virtually no algorithms that can be executed faster in this device; even for extremely well-suited circuits, a device with multiple gate zones can always perform the same operations at least with the same speed.

As the density of the circuit grows, the non-existing parallelization capacity of the single-gate-zone device results in a necessary depth overhead. Moreover, the device has to constantly move qubits from the storage to the single gate zone, which raises the amount of transports that it has to perform. These circumstances make the efficiency of the single-gate-zone device to drop drastically for $\overline{N_{\rm{g}}^{\rm{{L}}}}>1$ gates per layer.

When it comes to the computers with 5 and 10 gate zones, one can see how they are slightly apart for circuits with $\overline{N_{\rm{g}}^{\rm{L}}}<5$ and $\beta_{\rm{trans}}=0$, due to the fastest clock time of the 5-zone computer. However, since this clock time corresponds to the time to move the ions between gate zones, we assume that, with a good compilation strategy, the same clock times could be achieved for the 10-zone device, by restricting the amount of gate zones used to 5, similarly to what we discuss above for $\overline{N_{\rm{g}}^{\rm{L}}}=1$. The efficiencies of these devices do not drop until $\overline{N_{\rm{g}}^{\rm{L}}}>N_{\rm{zones}}$. Before that, there is no overhead due to their limited number of gate zones, assuming an even distribution of gates per layer through the circuit. Although there is an overlapping area of efficiencies between the two devices, a given circuit might have different values of $\beta_{\textrm{trans}}$ for each of them. For instance, the well-known GHZ state preparation circuit~\cite{Nielsen2010}, which has a staircase structure with one CNOT gate per layer, will require twice as many transports for the 5-gate-zone device than the 10-gate-zone one.

As the average number of gates per layer increases, the minimum value of $\beta_{\textrm{trans}}$ grows, and the area between the worst- and best-case scenarios shrinks. Interestingly, the efficiency gap between the multi-zone devices is never as large as the gap between $\beta_{\rm{min}}$ and $\beta_{\textrm{trans}}=1$ whenever $\overline{N_{\rm{g}}^{\rm{L}}}<N_{\rm{zones}}$. This result leads to the conclusion that the most relevant advantage can be provided by devices with large number of gate zones, capable to execute the majority of layers of the circuit without adding extra layers and transport. Moreover, the use of advanced compilation techniques is key to optimize the ratio of transport $\beta_{\textrm{trans}}$ and reduce the execution time of the circuit. Since the power demands of an HPC-based compiler could lead to a significant increase in consumption, there is a trade-off to consider between the energy spent in classical optimization and the one used by the quantum device itself. 

\section{Neutral atoms}
\label{sec:neutralatoms}
Neutral-atom quantum processors, much like trapped-ion systems, encode quantum information in individual atoms. Unlike ions, however, neutral atoms carry no net charge and therefore are not trapped via electric fields. Ordered lattices and tweezer arrays of neutral atoms have developed into a leading platform for large scale analog simulation, where they provide highly controllable realizations of many-body systems  \cite{Georgescu2014}.  More recently, advances have been made to perform digital quantum computation with neutral atoms trapped in optical tweezers via discrete Rydberg entangling gates \cite{Jaksch2000, levine_parallel_2019, evered_high-fidelity_2023}. These experiments have emerged as a promising platform for quantum computation and have demonstrated error-correction \cite{bluvstein_logical_2024, reichardt_fault-tolerant_2024} and digital simulation over hundreds of qubits \cite{evered_probing_2025, chinnarasu_variational_2025}.
Commercial adoption of neutral atoms has become increasingly widespread with the establishment of companies such as Pasqal, QuEra, Atom Computing, Infleqtion and many others. 

\subsection{Technology}

In neutral atom devices, individual atoms, most commonly Alkali \cite{bluvstein_logical_2024} and Alkali-earth atoms \cite{shaw_benchmarking_2024, jenkins_ytterbium_2022, ma_universal_2022}, are trapped and laser cooled in the tweezer potentials of tightly focused beams of light. Each choice of atomic species, provides energy levels that can be used to store and manipulate quantum information. Although a variety of approaches exists and most of the following discussion is generally applicable, this work restricts itself to rubidium-87 atoms for digital quantum computation. 

Tweezers are initially loaded from a cloud of laser-cooled atoms in a magneto-optical trap (MOT), prepared inside and ultra-high vacuum chamber. In digital neutral atom processors, qubits are encoded within the ground state hyperfine manifold of the atom. These states offer long coherence times ($T_2 > 2s$) during which quantum circuits can be executed. Quantum gates are implemented by irradiating atoms with specific laser pulses. Arbitrary single-qubit rotations are performed within the hyperfine manifold using qubit-specific, parallel Raman excitations. Entangling operations are performed by exciting atoms into highly-excited atomic Rydberg states. Then, the entangled qubits can be controlled by either only illuminating certain atoms within a dense 2D lattice with local Rydberg beams \cite{chinnarasu_variational_2025}, or by rearranging atoms with movable beams between each gate layer \cite{bluvstein_quantum_2022}. Only atoms placed within the Rydberg blockade radius will be entangled. In the latter mode, atoms are trapped in two types of tweezers: static tweezers generated via phase manipulation with a spatial light modulator (SLM) or movable tweezers manipulated on the time-scale of circuit operations in a highly parallel way via acoustic-optic deflectors (AODs). After the computation, qubit states are readout on a camera via local or global florescence imaging through an objective. 
\begin{figure}[!h]
    \centering
    \begin{subfigure}{0.4\textwidth}
        \includegraphics[width=1\textwidth]{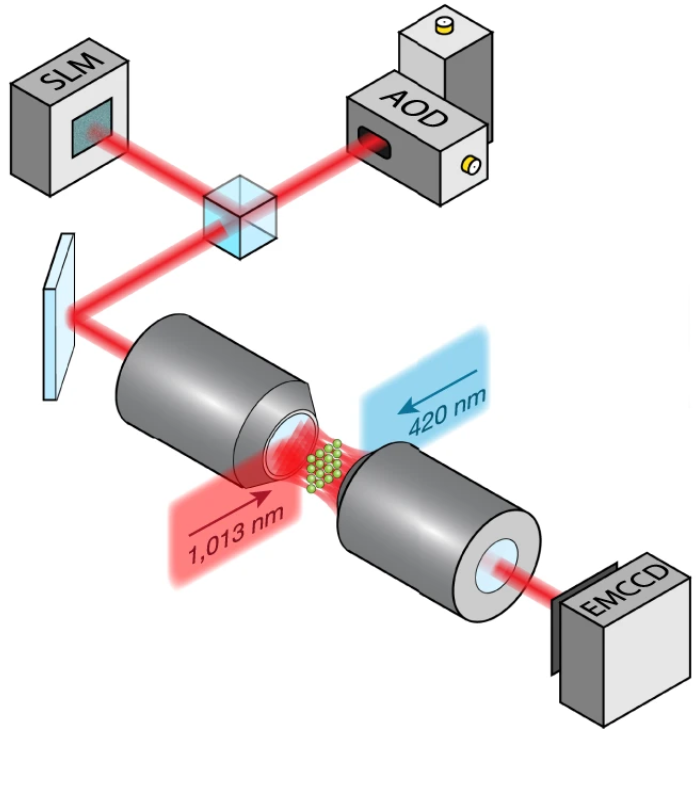}
        \caption{}
    \end{subfigure}
    \begin{subfigure}{0.53\textwidth}
        \includegraphics[width=1\textwidth]{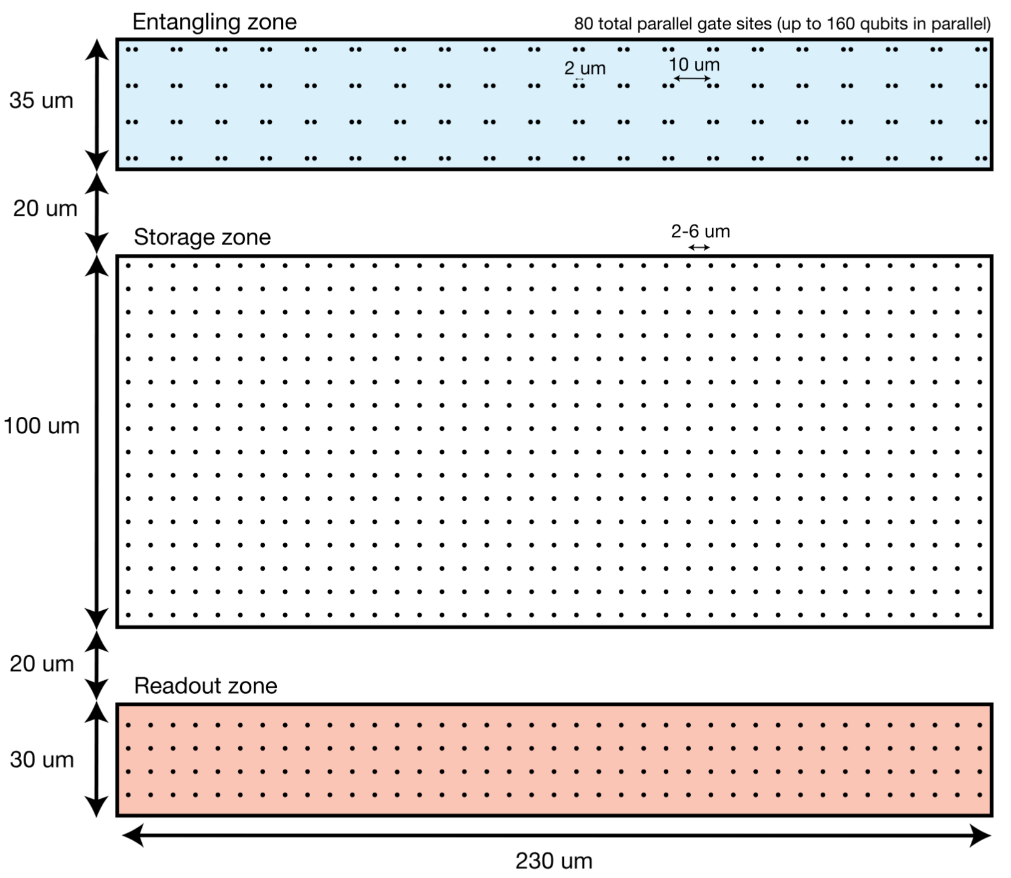}
        \caption{}
    \end{subfigure}
    \caption{Diagrams of the neutral-atom setup considered in this work. Panel a) depicts the basic laser configuration for atom control, according to Ref.~\cite{Ebadi2021}. Panel b) shows the array configuration of atoms inside the processor, arranged in three main zones: a large storage zone in the center, an entangling zone, and a readout zone. The figure is taken from~\cite{Bluvstein2024}.}
    \label{fig:diagram_neutral_atoms}
\end{figure}

In summary, qubits are encoded in the energy levels of laser-cooled neutral atoms. Universal quantum computation is achieved by performing single and multi-qubit gates by exciting atoms via specific laser pulses. Furthermore, by shuttling atoms within the computational region, a highly-programmable connectivity between qubits is accessible, enabling crucial tools of error-correction and algorithms such as non-local computation and transversal gates.

\subsection{Energy model}

This section discusses the hardware elements of a neutral atom quantum computer and the expected power consumption of each component. To ensure comparability with other platforms, the following table has been estimated for a system of  $N_{\rm{q}} = 400$ qubits, although several experiments have demonstrated using several thousands of qubits \cite{manetsch_tweezer_2025, chiu_continuous_2025}. We also specifically consider the power requirements for a neutral atom processor based on rubidium-87 with entangling operations performed through a two-photon Rydberg transition where atoms are rearranged between gates, although these considerations should be broadly applicable.

\subsubsection{Power consumption of the hardware}

Current neutral atom quantum computers setups are dominated by free-space optics with often considerable laser power requirements. These laser systems are either entirely free-space or delivered to the atoms via fibers. Its important to note that as error-rates improve, it becomes increasingly clear that maintaining the stability of these various macroscopic optical components is crucial to achieving algorithmically relevant error rates \cite{evered_high-fidelity_2026}. In addition to the laser and vacuum components, we therefore also highlight the HVAC requirements to ensure stability in room temperature and humidity, necessary for maintaining alignment of many precise beams paths. As it is discussed in the trapped ions section, the power consumption of the HVAC system might vary heavily depending on the facilities hosting the quantum computer. In this work, we chose a mean value between the data measured for the NQCC trapped-ion device and the Harvard neutral-atom device. The HVAC system will likely remain a key, and potentially increasing, form of power consumption until technological stability advances, such as on-chip photonic circuits, are more readily available at the required optical powers.  

\begin{figure}[!h]
    \centering
    \includegraphics[width=0.85\linewidth]{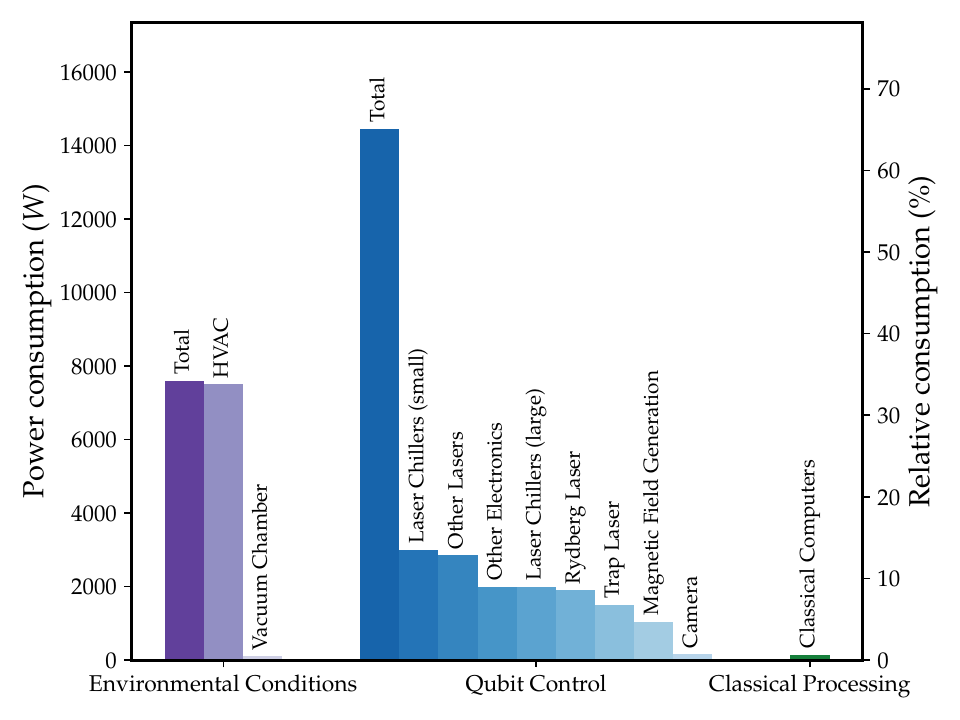}
    \caption{Power breakdown of a neutral-atoms computer, according to the components shown in Table~\ref{tab:neutral_atoms_components} of Appendix~\ref{app:neutral_atoms}. In the x-axis, the components are grouped depending of their purpose in the computer. Each group has a bar that corresponds to the total accumulated power of its components. The left y-axis shows the absolute value of power consumption of the components, and the right y-axis the relative value with respect to the overall consumption of the computer.}
    \label{fig:neutral_atoms_power_breakdown}
\end{figure}

The overall power consumption of the considered device is broken down in Figure~\ref{fig:neutral_atoms_power_breakdown}. This figure shows the power consumption of all the hardware components displayed in Table~\ref{tab:neutral_atoms_components} in three different groups: Environmental Conditions, Qubit Control and Classical Processing.

With these components in mind, the overall power consumption of the computer is $P^{\rm{neutral}}\approx22~\mathrm{kW}$, which corresponds to $1900~\mathrm{MJ}$ of energy consumed in 24 hours. In contrast with all the other platforms included in this work, the energy expenditure of this neutral-atoms computer is not majorly devoted to produce the suited environmental conditions for the qubits. In particular, components consume up to $7.6~\mathrm{kW}$ of power, which corresponds to the $35~\%$ of the overall consumption. In contrast with solid-state and photonic computers, neutral-atoms computers do not require a cryogenic setup to keep the chip enclosed in cryogenic temperatures. However, atoms that carry quantum information are still effectively "cooled" using lasers present in the Qubit Control group.

In contrast, the power consumed by Qubit Control components is $14.5~\mathrm{kW}$, which corresponds to $65~\%$ of the overall consumption. In this group, we include different chillers, that evacuate the heat dissipated by the lasers, but are not devoted to cool down the register itself.

\subsubsection{Derivation of the energy efficiency}

In order to estimate the energy efficiency of a device based on neutral atoms, we now explain our model to estimate the time it takes to perform an algorithm $\mathcal{A}$ composed of $N_{\rm{samples}}$ of a quantum circuit with depth $D$. Although the technology is similar to the one of  trapped-ion circuits, neutral atoms architectures impose different constraints on the running time of algorithm. In particular, the device needs to periodically refill atoms that are typically lost during the computation process. It has been shown that it is possible to load several thousands of atoms with a single set of tweezers directly from the MOT \cite{manetsch_tweezer_2025}. There, qubits can be addressed and gates can be performed. However after some time, new qubits need to be reloaded into the optical traps, adding a time $t_{\rm{reload}}$~\cite{evered_probing_2025, bluvstein_architectural_2025}.  It was observed that, on average, 40 to 80 samples of a circuit can be run before requiring to refill the traps. Moreover, the current state-of-the-art allows to run circuits of depth $D\sim 30$. As a broad estimate, we set to $30\times80=2400$ the number of layers after which a time $t_{\rm reload}$ is added to the circuit running time, which reads:
\begin{equation}
\label{eq:alg_time_neutral}
t_{\mathcal{A}}^{\rm{neutral}}=N_{\rm{samples}}\cdot t_{\rm{circuit}}^{\rm{neutral}} = N_{\rm{samples}}\cdot (t_{\textrm{reset}} + D \cdot t_{\rm clock} + t_{\rm meas} + \beta_{\textrm{trans}}\cdot D\cdot t_{\rm{trans}}+t_{\rm{reload}}\cdot \frac{D}{2400}) ,
\end{equation}
where $t_{\textrm{reset}}$ is the time necessary to reset the qubit between every shot of the algorithm, $t_{\rm meas}$ is the time needed to read the final state of the qubits at the end of the circuit, $t_{\rm{clock}}$ is the clock time of the computer, $\beta_{\textrm{trans}}$ the ratio of layers of the circuit requiring moving the atoms to and from interaction zones and $t_{\rm{trans}}$ the transport time. For more details on $\beta_{\textrm{trans}}$, we refer the reader to the Appendix~\ref{app:coldcomp}.

Equation~\ref{eq:alg_time_neutral} assigns the corresponding fraction of the reloading time that takes place every 2400 layers to the individual samples of the circuit. For instance, during the execution of circuits with 1600 layers, the register would be reloaded every 1.5 circuit samples. Therefore, each circuit execution would carry $1600/2400=2/3$ of the total reloading time. Here, we assume that one can reload the register during a circuit execution. This would require the quantum information to be copied on the new set of atoms (e.g. using SWAP gates), which we assume to take a negligible amount of time compared to the $500~\mathrm{ms}$ spent on the reloading itself. In Ref.~\cite{Bluvstein2024}, the authors describe multiple experiments being run on a neutral atom device, with different operation times. For our estimations, we consider $t_{\rm{reset}}\approx10~\mathrm{ms}$ and $t_{\rm{meas}}\approx10~\mathrm{ms}$. We assume that most of the time taken during every layer is devoted to configure the atom array inside the interaction zone according to the gates to be performed, which leaves us with $t_{\rm{clock}}\approx 500~\mathrm{\mu s}$. Finally, we consider the time needed to transport atoms from the storage to the interaction zone is $t_{\rm{trans}}\approx500~\mathrm{\mu s}$.

Taking into account that the reload process happens every 2400 layers, the reloading time per layer of the considered computer is $t_{\rm{reload}}/2400 \approx 210~\mathrm{\mu s}$. Therefore, the operations that depend on the depth of the circuits are two orders of magnitude below the measurement and the reset times. With this in mind, one can expect the latter to dominate the circuit time until $D\sim100$. From that point on, the other operations start taking over the computing time, until $t_{\rm{circuit}}$ becomes proportional to the depth. To put this into perspective, a single sample of a $D=10$ layers circuit would take $t_{\rm{circuit}}=27~\mathrm{ms}$, mostly dominated by $t_{\rm{reset}}+t_{\rm{meas}}=20 ~\mathrm{ms}$. For larger circuits with $D=100$ and $1000$ layers, the circuit time grows significantly to $t_{\rm{circuit}}=91$ and $720~\mathrm{ms}$, respectively.

The energy efficiency of a neutral atom quantum computer can be written as:
\begin{equation}
\label{eq:EEneutral}
    EE_{\mathcal{A}}^{\rm{neutral}} =\\  \frac{1}{N_{\rm{samples}}\cdot \left(t_{\textrm{reset}} + D \cdot t_{\rm clock} + t_{\rm meas}+ \beta_{\textrm{trans}}\cdot D\cdot t_{\rm{trans}}+t_{\rm{reload}}\cdot \frac{D}{2400}\right) \cdot P^{\rm{neutral}}},
\end{equation}

where $P^{\rm{neutral}}$ is the sum of the power consumption of the elements in Table ~\ref{tab:neutral_atoms_components}.

\subsection{Results and discussion}
\begin{figure}[!h]
    \centering
    \includegraphics[width=\linewidth]{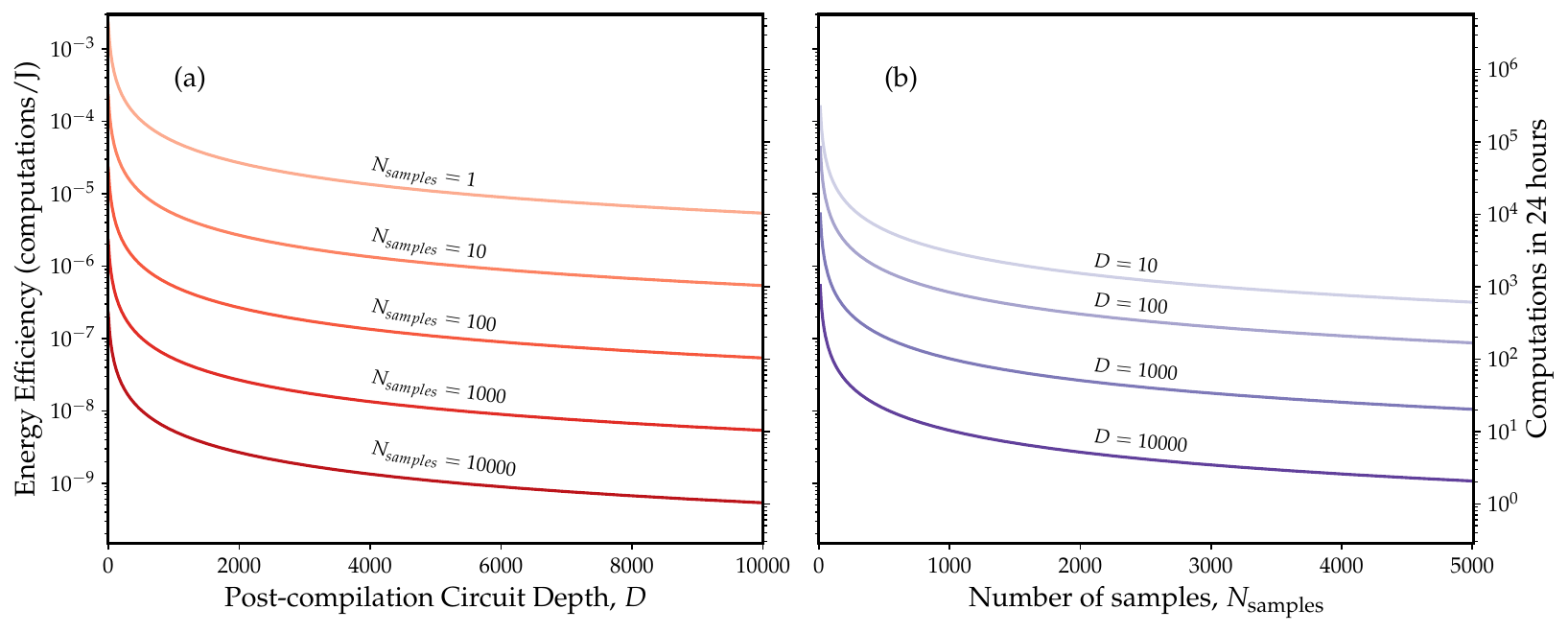}
    \caption{Energy efficiency (left y-axis) and number of computations (right y-axis) in 24 hours of a neutral-atoms computer depending on the post-compilation circuit depth, $D$ and number of samples, $N_{\rm{samples}}$. Panel (a) shows several series with different $N_{\rm{samples}}$ values and the x-axis corresponds to $D$. The series in panel (b) have different $D$ values and its x-axis corresponds to $N_{\rm{samples}}$. The circuits displayed in the panels have a ratio of transport per layer $\beta_{\textrm{trans}}=0.25$ transports per layer.}
    \label{fig:neutral_atoms_EE_D_Nsamples}
\end{figure}

Figure~\ref{fig:neutral_atoms_EE_D_Nsamples} shows the energy efficiencies and number of computations in 24h of a the neutral atom device described above, depending on the depth of the circuit and number of samples.

In accordance with Eq.~\ref{eq:EEneutral}, the left panel of Figure~\ref{fig:neutral_atoms_EE_D_Nsamples} displays equidistant lines for the different samples values, indicating that $EE$ is inversely proportional to $N_{\rm{samples}}$. On the contrary, we see the $D=10$ and $100$ lines are closer to each other, with respect to the rest of them, in the right panel of the figure. As it discussed above, for small circuits, the reset and measurement times are dominant in the overall circuit time. This makes circuits around $D\sim 10$ to take similar $t_{\rm{circuit}}$. As the depth grows past 100, the $EE$ of the neutral-atom computer becomes inversely proportional to $D$. Note that a value of $\beta_{\textrm{trans}}=0.25$ transports per layer have been chosen, corresponding to circuits where atoms are moved to the interaction zone every four layers on average. This value can be optimized and is highly dependent on the algorithm, but its not as impactful as in the trapped-ions device discussed above, due to the lower transport time in neutral atom devices.

To scale neutral atom system to large, practical algorithms it has become increasingly clear that the atom array must be loaded and used continuously. In a few recent experiments this has been executed by laser cooling atoms in one chamber and then transporting them via, for instance, an optical lattice to be loaded into an array of prepared qubits \cite{chiu_continuous_2025, gyger_continuous_2024, li_fast_2025}. This modality allows for fast circuit cycle rates and can be scaled to longer computations. We therefore consider it as an extension of the energy model. Essentially, the terms related to the reloading and resetting of atoms in Eq.~\ref{eq:alg_time_neutral} vanish and, in return, additional components, for instance an additional transport lattice and more complex vacuum systems, are considered in the total power consumption of the device. These additional hardware components, shown in Table~\ref{tab:neutral_atoms_components} of Appendix~\ref{app:neutral_atoms}, increase the power consumption of the studied device by $2550~\mathrm{W}$.

\begin{figure}[!h]
    \centering
    \includegraphics[width=0.85\linewidth]{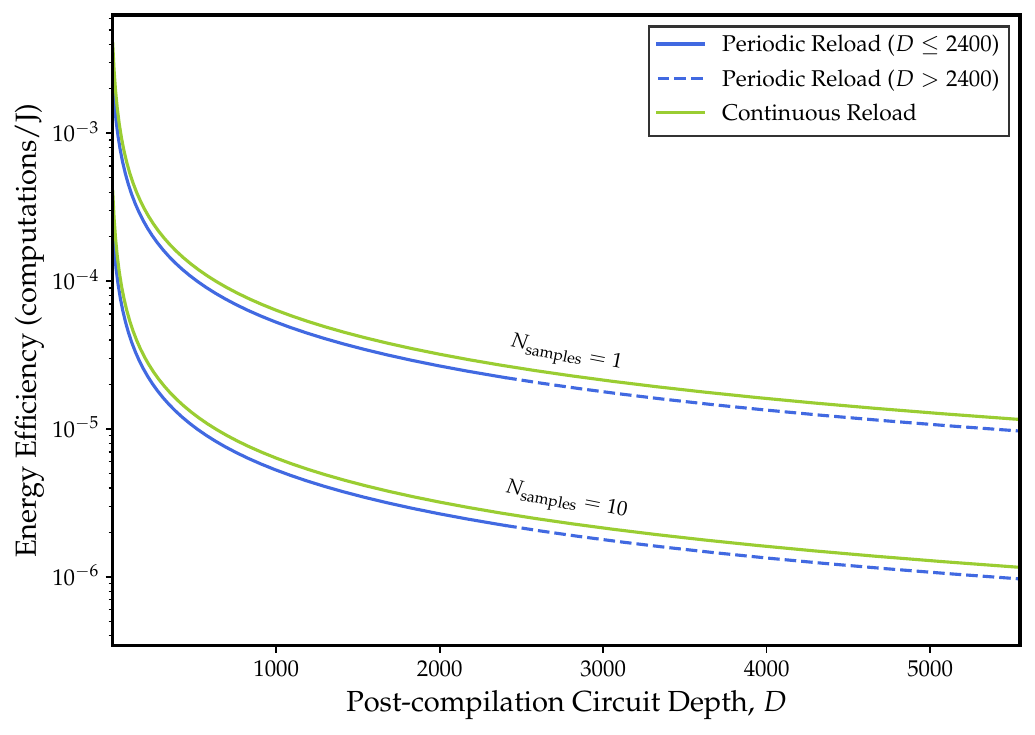}
    \caption{Energy efficiency of a neutral-atoms computer executing an algorithm, depending on its post-compilation depth. The blue curve corresponds to a computer that reloads its atom periodically, and the green line to a device that can reload them continuously, without interruption. The dashed blue line corresponds to those circuits that surpass the reloading frequency $D=2400$, and thus would need to perform the reloading operation during its execution. All the circuits displayed have a ratio of $\beta_{\textrm{trans}}=0.25$ transports per layer.}
    \label{fig:neutral_periodic_vs_continuous}
\end{figure}

Figure~\ref{fig:neutral_periodic_vs_continuous} shows the energy efficiency comparison between a neutral-atoms computer that loads its register periodically and a similar computer that has the ability to load the register continuously.

As one can see, the computer that reloads the register continuously displays higher energy efficiencies for all $D$ values, regardless of its extra power consumption. For short circuits, the reset and measurement operations dominate computing times. In that regime, the continuous reload computer needs $10~\mathrm{ms}$ less time for each sample, since it is always refilling the register with atoms prepared at state $\ket{0}$. However, the higher speed of this computer overcompensates the $2550~\mathrm{W}$ power increase with respect to its counterpart. As the depth of the circuits grow, the energy efficiency gap increases, until it surpasses the $D\sim500$ layers. This regime corresponds to the moment where $EE$ starts becoming inversely proportional to $D$. From that point on, the improvement ratio between the computers tends asymptotically to $EE^{\rm{cont}}/EE^{\rm{periodic}}\rightarrow4/3$ (for $\beta_{\rm{trans}}=0.25$). The ratio of improvement $EE^{\rm{cont}}/EE^{\rm{periodic}}$ does not depend on the number of samples executed. 

Figure~\ref{fig:neutral_periodic_vs_continuous} includes circuits with depths past the reloading frequency $D=2400$, which, in order to be executed in the Periodic Reload computer, would need to perform the reloading operation in the middle of a sample. As mentioned above, we assume that this could be done by swapping the information to the newly set of atoms entering the register, but this process might impose new physical challenges and constrains to these computations, which would probably affect the computing time and fidelity of the algorithm. Our results show an efficiency improvement due to the adoption of the continuous reload technique in neutral-atoms computers. This could be even more significant under a fault-tolerant computation scheme, where sets of qubits are measured and reset after each syndrome detection round, creating windows for constantly substituting the used atoms with freshly loaded ones.

\section{Photonic computing}
\label{sec:photon}
Photonic quantum computing architectures represent a paradigm change compared to the other platforms studied in this article. Unlike matter-based platforms, photonic quantum computing uses light to perform computation. This approach has the advantage of avoiding the relatively short coherence time of matter-based qubit compared to photonic ones. In addition, the manipulation of light using linear optics components can be done at room temperature, which reduces the cost of cryogenic power. However, these relative advantages are counterbalanced by the difficulty of storing light for long periods of time. 

There are two approaches to photonic quantum computing: continuous variable (CV) and discrete variable (DV). In the former, information is encoded in the properties of the electromagnetic field, that can take any value in a continuous interval (e.g. position and momentum). In the latter, information is encoded in discrete modal properties, such as polarization or the spatial mode of single photons. For both schemes, operations require using of linear optics elements, such as beam splitters or phase shifters, but in order to perform universal computing, some additional elements are necessary. The CV scheme requires non-Gaussian states of light, while in DV computing, a key ingredient is the generation of entangled single-photon states. This entanglement is introduced either through entangled cluster states, also called resource states, on which measurements are performed following the so-called measurement-based quantum computing paradigm~\cite{MBQC}, or through entangling gates using projective measurements~\cite{klm2001klm}. 

The implementation of these schemes on physical hardware is currently being developed by several companies, such as Xanadu, implementing CV computing with Gaussian states of light~\cite{xanadu2019source, xanadu2019archi}, PsiQuantum, implementing a novel, fault-tolerant, measurement based, DV computing scheme called fusion-based quantum computing~\cite{FBQC}, or Quandela, implementing linear optics DV quantum computing~\cite{Qandela}.

In this work, we focus on DV photonic quantum computing, and its implementation on quantum computers using quantum dot based single photon sources. For an overview of the field of photonic quantum computing, we refer to \cite{milburn2025photonic-qc}.

\begin{figure}
    \centering
    \includegraphics[width=0.8\linewidth]{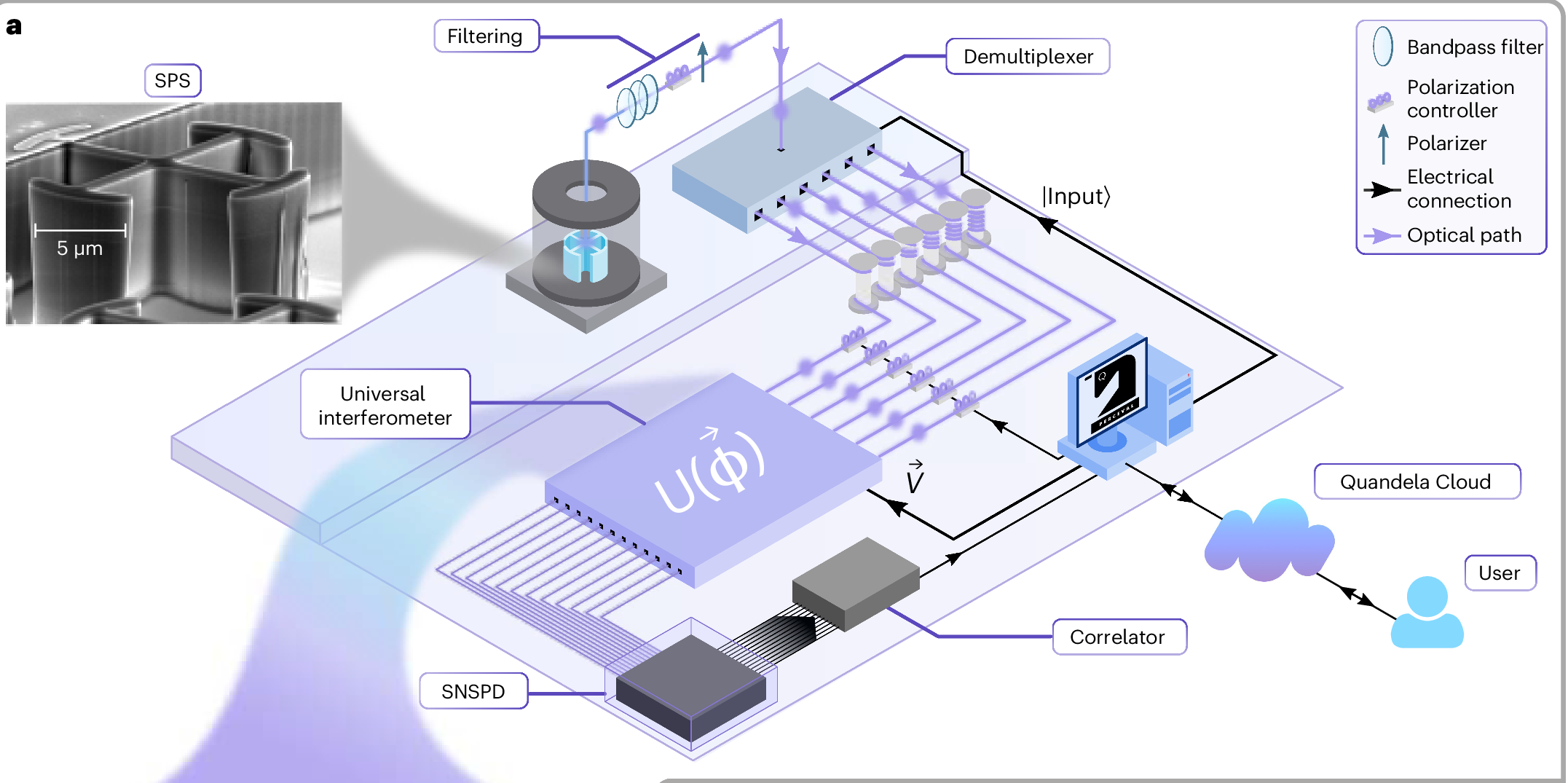}
    \caption{Example of a photonic computer, taken from~\cite{Qandela}. A quantum-dot based single photon source (SPS) sends photons to a demultiplexer, encoding information in different optical modes. They go through a cloud-controlled universal interferometer and are then measured using SNSPDs.}
    \label{fig:Ascella}
\end{figure}

\subsection{Technology}
In DV photonic quantum computers, information is encoded in discrete degrees of freedom of single-photon states such as their path. 

The first essential building blocks of these computers are single-photon sources. These sources need to meet high efficiency requirements and deliver photons of near-unity purity and indistinguishability. The development of such sources is an active field of research and two main schemes are under development. On the one hand, probabilistic sources based on nonlinear materials, exploiting either spontaneous parametric down-conversion~\cite{ramelow2013, weston2016, kaneda2016} or four wave mixing~\cite{spring2017, francisjones2016} are being studied. In order to overcome the probabilistic nature of these mechanisms, heralding and multiplexing strategies need to be implemented which results in a high overhead in the resources required. On the other hand, deterministic sources using semiconductor based quantum dots can produce high quality photons on demand \cite{senellart2021bright}. In this work, we focus on the latter, which are used in Quandela's architecture~\cite{Qandela} shown in Fig.~\ref{fig:Ascella}.

In this setup, the source is operated at a temperature around $5$ K and emits a temporal train of photons. The photons are then sent through an active demultiplexer followed by fibered delay lines which convert the train of photons into photons arriving simultaneously on an $m$ mode photonic chip. The chip contains a universal interferometer, capable of implementing any $m\times m$ unitary matrix. Such an interferometer can be realized according to three different schemes which implement all-to-all connections between the modes using linear optics components: beamsplitters, which couple the spatial modes, and phase shifters, which adjust their interferences. The first, due to Reck et al. is comprised of a triangular mesh \cite{Reck1994}. The second, due to Clements et al. is comprised of a square mesh and, by a consequence of its symmetry, offers improved spatial footprint, stability and balanced losses \cite{clements2017}. The third, due to P\'{e}rez et al. utilizes a Hexagonal mesh which, while using more active elements, offers additional functionalities due to its ability to create recirculating structures \cite{Perez2016}. Tunable beamsplitters are generally implemented via Mach-Zehnder interferometers with a phase shifter in one or both arms. The chip is then completely controlled by adjusting phase shifters and allows to implement any unitary matrix by following algorithms defined for each type of mesh. 

Note that these unitaries act on the optical modes, not on the photons directly. Finally, photons are detected using superconducting nanowire single-photon detectors (SNSPDs), whose coincidence counts are connected to a correlator and a classical software. Depending on these outputs, a new electronic control signal can be applied to change the unitary applied to the next train of photons. 

The time required to configure the chip depends on various parameters, for instance on the phase shifter technology. As an example, thermo-optical phase shifters require about $1~\mu$s while electro-optical phase shifters are much faster, around $10$ ps (see Appendix \ref{app:hardwarephoton}). In practice, configuration times are longer due to crosstalk between phase shifters.

\subsection{Energy model}

\subsubsection{Power consumption of the hardware}

In Table~\ref{tab:photonic_component} of Appendix~\ref{app:hardwarephoton} one can see the hardware components of a photonic quantum computer. In the architecture that we consider as a baseline, the number of most components does not increase straightforwardly with the number of optical modes. For example, the number of sources needed to reach a large number of optical modes is related to the size of the demultiplexers. We consider here quantum dot sources and a demultiplexer able to send at most 12 photons on the chip per source, in line with current state of the art devices. Several sources can be operated simultaneously using a single laser and beam splitters. Note that the performance of the source, usually characterized by its brightness, and by the purity and indistinguishability of the photons produced, is independent of how much energy is used to operate it. Specifically, making a high quality source requires leveraging complex physical effects in cavity quantum electrodynamics and condensed matter, which make their design technologically challenging. 

A demultiplexer consists of several delay lines which allow to synchronize the photons emitted sequentially by the source, so they enter the chip simultaneously. For an $m$ mode chip, a demultiplexer uses $m-1$ Pockels cells, each consuming 0.5 W. Controlling these cells requires using an amplifier which uses about 4 W.

Superconducting nanowire single photon detectors (SNSPDs), on the other hand, scale with the number of modes. Although both sources and detectors operate at cryogenic temperatures, they can be placed in the same cryostat. We take as an example current small cryostats that can host up to one quantum dot source and $\sim25$ SNSPDs per cryostat. For larger systems, larger cryostats would be more appropriate. We note that bigger cryostats have a larger power consumption, but are also more efficient: small air-cooled cryocoolers, such as those considered here, typically have around 1\% Carnot efficiency, whereas larger cryostats can reach 30-50\% Carnot efficiency. In any case, if the number of modes $m$ increases above the maximum number of SNSPDs that a cryostat can hold, we will need either more than one cryostat or a larger cryostat. Consequently, the cryogenic power does not increase linearly with the system size.  

The chip on which the computation is performed requires strict control of the phase shifters and other active elements. For our example architecture, controlling the chip efficiently requires monitoring its temperature and keeping it stable in order to avoid cross talk effects. Cooling the chip at room temperature requires a Peltier thermo-cooler; the power spent is related to the stabilization time of the phase shifters: for $m$ modes, one needs $O(m^2)$ optical components. The control systems for the phase shifters are generally provided by field programmable gate arrays (FPGAs) due to their high speed and compact form factor. However, commercially available high-speed FPGAs are limited to 16 channels. As the number of active components grows so does the required number of FPGAs which require $\sim$ 15 $W$ each and can impose a significant power cost. The exact power cost of an FPGA is highly specific to the application but we can conservatively estimate the scaling of this power contribution as the number of electrical channels required to control $N_q$ photonic qubits divided by the number of channels in one FPGA. This results in a factor of ${{(2N_q)}^2 \over {16}}$. Details of these classical control electronics relevant to the energy efficiency calculations may be found in Table~\ref{tab:photonic_component}.

\begin{figure}[!h]
    \centering
    \includegraphics[width=0.85\linewidth]{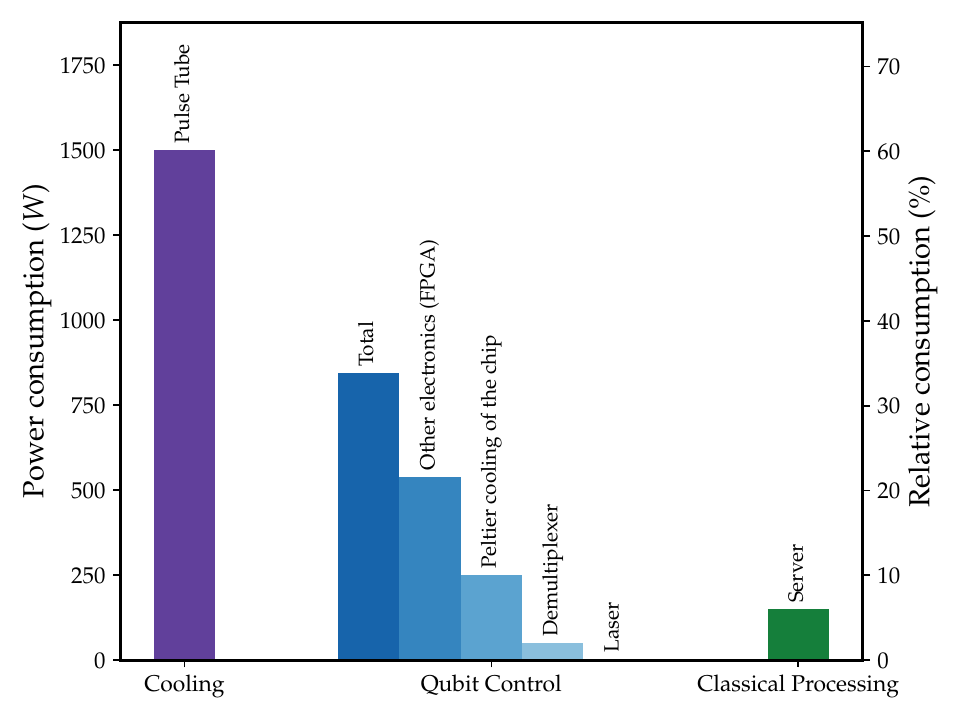}
    \caption{Power breakdown of a photonic quantum computer of 12 qubits (24 modes) according to the components shown in Table~\ref{tab:photonic_component} of Appendix~\ref{app:hardwarephoton}. In the x-axis, the components are grouped depending of their purpose in the computer. Each group has a bar that corresponds to the total accumulated power of its components. The left y-axis shows the absolute value of power consumption of the components, and the right y-axis the relative value with respect to the overall consumption of the computer.}
    \label{fig:photonic_power_breakdown}
\end{figure}

\subsubsection{Derivation of the energy efficiency}

The framework that we have used in the previous sections to estimate the running time of algorithms does not hold for photonic computing architectures, since the computation happens very fast once the photon is created. After compiling an algorithm $\mathcal{A}$ into a circuit, the running time is still given by 

\begin{equation}
    t_{\mathcal{A}}^{\textrm{photonic}}= N_{\rm{samples}}\cdot t_{\rm{circuit}}^{\textrm{photonic}}.
\end{equation}

However, a photonic circuit is not decomposed in a series of gates but rather seen as a single unitary that can be implemented in optical components arranged on a chip. The circuit running time is given by the time it takes for photons to go through the chip and be detected. Changing the depth of a circuit does change the length of this path, but the effect on the circuit time is negligible. However, as the depth grows, the probability that a photon is not lost while traveling through the chip decreases exponentially with the number of optical components. More precisely, the end-to-end transmission probability from the source to the detector is given by \cite{soret2026quantumenergeticadvantagecomputational}:

\begin{equation}
    \eta(m) = \eta_\text{of}\eta_\text{d}\eta_\text{dmx} 10^{-2c_\text{coup}/10} 10^{-D_{\textrm{optical}}(m)c_\text{MZI}/10} \, ,
    \label{eq:eta-photonic}
\end{equation}

where $\eta_\text{of}$ is the efficiency of the source at the fiber output, $\eta_\text{d}$ the efficiency of the SNSPDs, $\eta_\text{dmx}$ the efficiency of the demultiplexer, $c_\text{coup}$ the attenuation at the fiber to waveguide interface (there are two interfaces, hence the factor 2), $c_\text{MZI}$ the attenuation per optical component, and $D_{\textrm{optical}}(m)$ the depth of the chip as a function of the number of modes $m$. The connection between \eqref{eq:eta-photonic} and photon loss is the following: given $N$ input photons in a circuit of $m$ modes, the transmitted photons follow a binomial distribution, of average $\eta(m) N$ and standard deviation $\sqrt{N\eta(m)[1-\eta(m)]}$. Moreover, the time necessary to detect $N$ photons is proportional to $\eta(m)^{-N}$ (see \eqref{eq:t-photonic} below).

In this work, we assume that the chip represents a universal interferometer, designed following the Clements architecture~\cite{clements2017}. In this case, $D_{\textrm{optical}}(m)=m$, and the total number of components scales like $m^2$. The values for the other components can be found in Table~\ref{tab:photonparam1} and~\ref{tab:photonparam2}.

The average time necessary to successfully create one sample of a circuit on $N_q$ qubits is given by 

\begin{equation}
    t_{\text{circuit}}^{\text{photonic}} = \frac{N_q}{N_{\textrm{source}}\cdot r_{\rm{source}}} \cdot\frac{1}{\eta(2N_q)^{N_q}},
    \label{eq:t-photonic}
\end{equation}
where $N_{\textrm{source}}$ is the number of single-photon sources and $r_{\rm{source}}$ their rates. Here, two modes of light are used to encode a single qubit. The rate of the sources can be limited by either the lasers themselves or the detectors dead time, but not by the optical depth. Here, choose the state-of-the-art value $r_\text{source} = 1$ GHz and assume that $N_{\textrm{source}}=1$. The energy efficiency of photonic quantum computers of $N_q$ qubits is thus given by 

\begin{equation}
    EE^{\textrm{photonic}}= \frac{1}{N_{\text{samples}}\cdot\frac{N_q}{N_{\textrm{source}}\cdot r_{\rm{source}}} \cdot\frac{1}{{\eta(2N_q)^{N_q}}} \cdot P^{\textrm{photonic}}},
\end{equation}

where $P^{\textrm{photonic}}$ is the total power consumption of the hardware presented in Table~\ref{tab:photonic_component}.

\subsection{Results and discussion}

\begin{figure}[!h]
    \centering
    \includegraphics[width=0.85\linewidth]{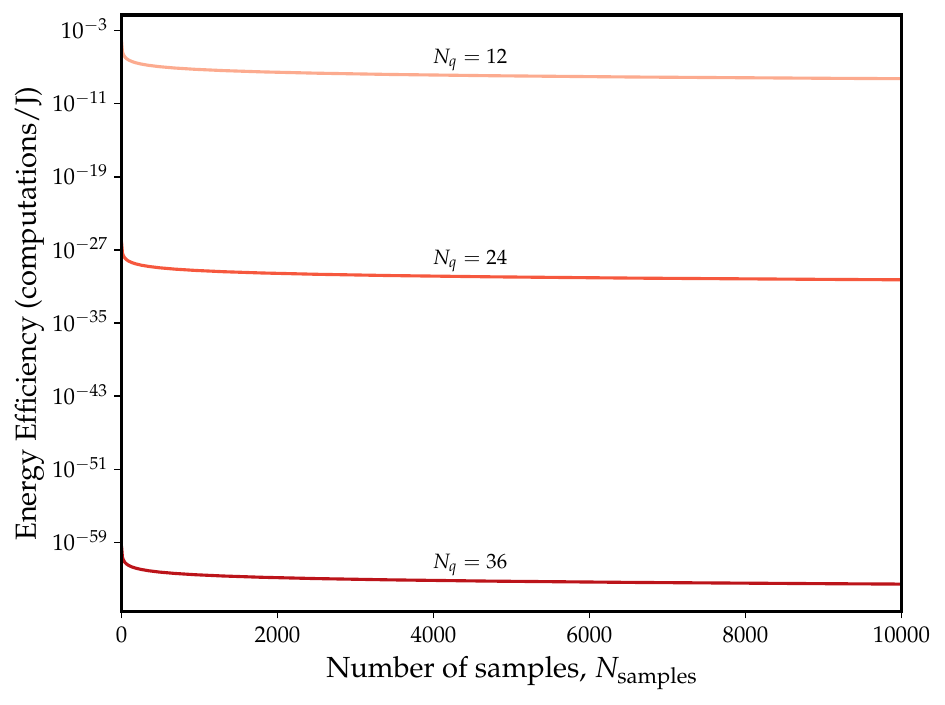}
    \caption{Energy efficiency of a photonic quantum computer as a function of the number of samples, for different number of qubits $N_{\rm{q}}$.}
    \label{fig:photonic_EE_vs_sample}
\end{figure}

In Fig.~\ref{fig:photonic_EE_vs_sample}, we show the energy efficiency of photonic quantum computers as a function of the number of samples $N_{\rm{samples}}$, for different number of qubits. 

From this plot, it can be immediately seen that current photonic architectures do not scale well with the number of qubits. This is mainly due to the quadratic scaling of the number of optical components on the chip as the number of modes grows. Not only does this lower the end-to-end transmission probability of the chip, but it also requires commensurate scaling of electronic qubit control components (mainly FPGA) to the list of hardware components. On top of these components, appropriate numbers of cryostats and demultiplexers are also added to match the required number of detectors. However, the current Quandela architecture has proven to be competitive in terms of energy efficiency.

\begin{figure}[!h]
    \centering
    \includegraphics[width=0.85\linewidth]{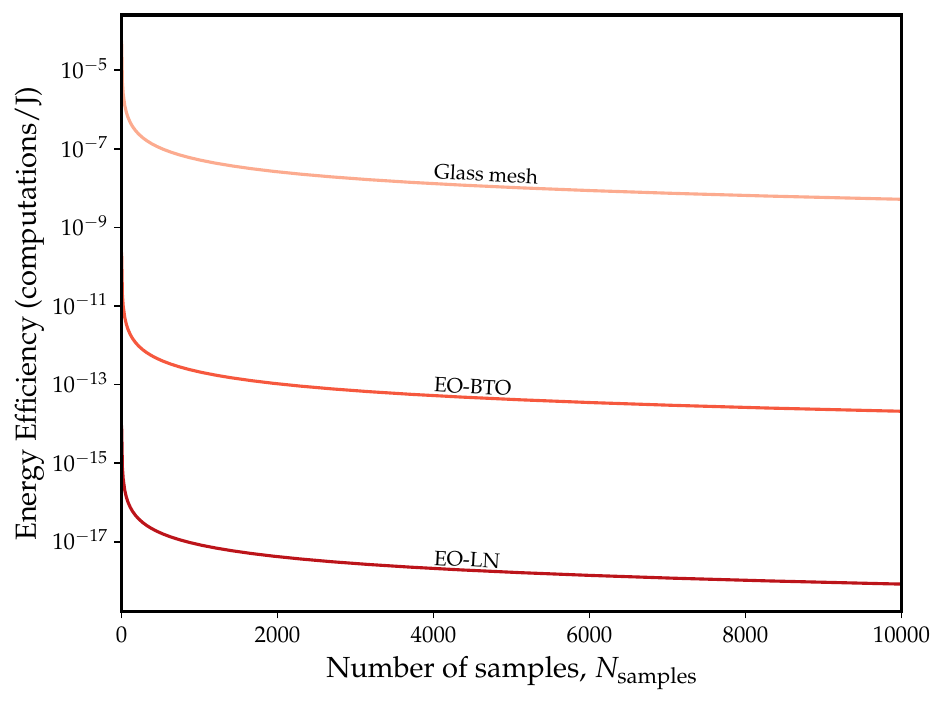}
    \caption{Comparison between state-of-the-art chip technologies assuming a 12-qubit photonic quantum computer in a Clements topology. Glass mesh refers to amorphous Silicon waveguides utilizing TO phase shifters, EO-BTO refers to Silicon Nitride waveguides evanescently coupled to Barium Titinate EO phase shifters while EO-LN refers to monolithic Lithium Niobate on insulator waveguides which provide native EO switching capabilities.  }
    \label{fig:photonic_EE_vs_chip}
\end{figure}

Although today’s photonic quantum computers are energy-dominated by cryogenics, the programmable interferometer mesh that implements the unitaries can also draw substantial power. As systems scale, this cost grows with the number of programmable elements, and different material platforms result in very different energy consumption. In Figure \ref{fig:photonic_EE_vs_chip} we compare the state of the art in thermo-optic (TO) and electro-optic (EO) platforms and project the energy efficiency of each for a Clements interferometer mesh.

The optimal choice of phase shifter technology for photonic QC depends on regimes determined by the ratio of sampling and reconfiguration time as these correspond to dominance of static or dynamic components of power. The caveat is that the losses of the platform must not increase sampling time so much that additional cryogenic energy outweighs mesh savings. This is the regime of current technologies depicted in Figure \ref{fig:photonic_EE_vs_chip}, where the higher losses of EO platforms compared to TO platforms result in a longer sampling time. Consequently, the EO platforms require more cryogenic energy, making negligible the benefits of their high switching speed and low static power consumption. Detailed specifications of the different mesh platforms can be found in Table~\ref{tab:photonparam2}. Reducing coupling and component losses of EO platforms could reduce sampling time such that they operate in a regime where energy efficiency is differentiated by static power costs of the chips. Within this regime, where sampling time still dominates reconfiguration time, minimizing mesh static power strongly favors lumped element (LE) EO phase shifters over TO and traveling-wave (TW) EO. As photon sources, chip losses and detectors improve further and sampling time falls further, dynamic power and switching rate will matter more. If the sampling times fall to the point where reconfiguration rates on the scales of 10-100 GHz become essential, TW approaches may become practical and necessary as they can minimize dynamic energy and maximize throughput. The discussion in Appendix~\ref{app:chiptech}, includes more technical details regarding the specifications and tradeoffs of these technologies.

We highlight that we took a conservative approach in the computation of the sampling time: we imposed that all of the input photons should be detected, yielding the exponential factor $\eta(2N_q)^{N_q}$ in \eqref{eq:t-photonic}, which implies a poor scaling of the energy efficiency with the system size. However, the energy efficiency could be significantly improved using loss-mitigation strategies \cite{mills2026mitigatingphoton}, which are in fact routinely used in photonic quantum computing.

Moreover, although linear optical quantum computing constitutes one of the first schemes for universal quantum computing, the scalability issue has motivated the development of alternative architectures to realize fault-tolerant photonic quantum computing, such as fusion-based quantum computing \cite{FBQC} and the hybrid spin-optical architecture \cite{deGliniasty2024spinopticalquantum}. In addition, in the current era of NISQ photonic quantum computing, photon native schemes \cite{salavrakos2025photonnative} are being developed which are not necessarily gate based, in particular in quantum machine learning \cite{gyurik2024exponentialseparationsclassicalquantum}, photonic quantum reservoir computing \cite{garcia2023photonicqrc, rambach2025photonicquantumacceleratedmachinelearning,Nerenberg2025}, photonic quantum neural networks \cite{Monbroussou2025photonicqcnn, wang2026qcircnn}. An example of native task for NISQ photonic quantum computers is boson sampling \cite{aaronson2011computational}, which has been extensively used as a benchmark of quantum advantage \cite{clifford2018classical, go2025bs, oszmaniec2025complexitytheoreticfoundationsbosonsamplinglinear, wu2018benchmark, wang2019bs-20-photons}, and can be exploited to solve certain problems, such as graph problems \cite{mezher2023solvinggraphproblemssinglephotons}, classification tasks \cite{sakurai2025qorc,Nerenberg2025}, or in hybrid classical-quantum algorithms \cite{anguita2025experimentaldemonstrationbosonsampling}. The energy efficiency of these alternative schemes is still poorly known, although it is generating an increasing interest, and has motivated recent work on an in depth investigation of the energy efficiency of boson sampling \cite{soret2026quantumenergeticadvantagecomputational}.

\section{Future Directions}
\label{sec:future}
\subsection{Error correction and fault tolerant quantum computing}
\label{sec:errorcorrection}

Noise in quantum computing systems prevents the successful implementation of large-scale algorithms with application to real-world problems. Fully addressing this limitation requires the use of quantum error correcting codes (QEC) and, more generally, fault-tolerant quantum computing. However, moving to a fault-tolerant regime leads to a substantial increase in resource requirements, both in terms of hardware and energy consumption.

Replacing physical qubits by logical qubits implies a huge increase in the amount of resources, inversely proportional to the code encoding rate \textit{i.e.} the ratio between the number of logical qubits per physical qubits. As an example, for a distance $d=30$ surface code, the amount of physical qubits increases by a factor of roughly 1000. This overhead arises not only from the large number of physical qubits needed to encode logical qubits, but also from the implementation of fault-tolerant logical gates.

Moreover, the set of transversal gates is code dependent and cannot be universal \cite{eastin2009}. Other techniques such as lattice surgery~\cite{horsman2012surface} or magic state distillation~\cite{bravyi2005universal}, which achieve a universal set of gates, require significant additional resources. Some logical operations such as lattice surgery also require a time which scales with the code distance $d$, often involving $O(d)$ physical operations for a single logical gate.

In addition, decoding, the process of identifying and correcting errors from measured syndromes, represents another major source of computational and energetic cost. Optimal decoding is usually an NP-hard problem, and could thus only be approximated. This entails a trade-off between the accuracy and the speed (or algorithmic complexity) of practical decoders.
Therefore, efficient decoding demands advanced classical algorithms, which may themselves consume considerable computational power and prolong computation times. A realistic assessment of the total energy cost of fault-tolerant quantum computing must therefore include this classical contribution, which will certainly prove to be non-negligible.

Recent studies highlight the scale of these challenges. For instance, factoring a 2048-bit RSA integer using a surface-code-based fault-tolerant quantum computer has been estimated to require about one million physical qubits and five days of computation, assuming standard circuit-level noise and two-dimensional qubit connectivity~\cite{gidney2025factor}. This setup is compatible with superconducting-qubit architectures, but the cooling power hosting this number of physical qubits is significant and there are existing trade-offs between noise and cooling temperature that were already investigated in this setting~\cite{Fellous-Asiani2023}. Other hardware platforms with different connectivity patterns may lead to different resource trade-offs~\cite{cain2026shorsalgorithmpossible10000}. Systems with higher qubit connectivity could, in principle, support quantum LDPC codes with higher encoding rates than surface codes, reducing the number of required qubits~\cite{breuckmann2021quantum}. However, in these systems, implementing fault-tolerant logical gates remains difficult.

As of today, there is a challenge in listing the hardware components involved in fault-tolerant quantum computers, as only small scale examples exist. Simply scaling up the number of components of current computing architectures to millions of qubits leads to huge energy expenditure, which makes clear the necessary engineering and fundamental work that still needs to be done. Moreover, as explained above, syndrome decoding in QEC implies additional classical servers whose power consumption is also evolving.

We believe nonetheless that our work can be adapted to the case of fault-tolerant quantum computing. Once a hardware list of a fault tolerant quantum computer is established, QEC can be treated as any other circuit applied to physical qubits, increasing the time where the hardware is consuming energy. QEC increases the number of gates that are applied to the qubits and thus increases the depth of quantum circuits implementing algorithms. As an example on a superconducting quantum computer, Equation~\ref{eq:tcircuit_superc} can be adapted to the case of circuits running on error corrected qubits. Defining a clock time is even more coherent for QEC codes, since each gate requires a set of operations that is consistent, even for different logical gates. On the other hand, QEC codes can also reduce the number of shots necessary to remove uncertainty due to noise on the qubits, which might reduce the total execution time of algorithms. They then introduce a trade-off on the execution time between depth increase and sample reduction. This does not apply, however, to the number of samples that are necessary because of the algorithm statistics. 

Taking all these aspects into account is essential for understanding the overall energy requirements of large-scale quantum computing. A complete analysis should thus consider the energy consumption associated with both quantum hardware operations and classical decoding processes. Since the field of QEC is going through very rich developments, both theoretical and experimental, we leave for future investigation the computation of the depth of those circuits. 


\subsection{Other computing paradigms and high performance computing}
\label{sec:otherparadigms}

Our study is primarily focused on circuit-based computing, where algorithms are compiled into quantum circuits that are encoded in quantum hardware. We believe it can be adapted to new hardware by studying a baseline hardware architecture and a compilation model taking into account the specifics of the new platform. There also exist other computing paradigms that were not covered in this study, which could be adapted fit in our model. These are left as future research directions, with some initial thoughts on how to benchmark their energy consumption.

As an example, our methodology could be extended to measurement-based quantum computing (MBQC)~\cite{MBQC}, where algorithms are encoded in a series of adaptive measurements applied on universal graph states (also called cluster states or resource states). This can be especially adapted to the case of blind and private delegated computing~\cite{delegated1,LeoSecurity,Broadbent_2009,BroadbentDelegated}. Benchmarking the energy consumption of MBQC in the context of delegated computation over a quantum communication network could give great insights on the cost of quantum cloud computing. In order to do so, one would need to analyze the hardware involved in maintaining highly entangled clusters of qubits, the compilation procedure behind MBQC, as well as the cost involved in back-and-forth communication over a classical or quantum channel.

An interesting photonic-based approach towards fault-tolerant computation is fusion-based quantum computing~\cite{FBQC}, where algorithms are also encoded in measurements of photonic cluster states. This approach being primarily pursued in the private sector, we were not able to get enough insights on the current devices to properly benchmark their energy consumption. The inherently probabilistic nature of quantum optics using state-of-the-art components suggests however a very consequent energy consumption devoted to cooling down the single-photon sources and detectors.

Another paradigm of interest is adiabatic quantum computing~\cite{Albash_2018}, based on quantum annealing~\cite{annealing1,annealing2}, where a computation is decomposed into a slow continuous transformation of an initial Hamiltonian into a final Hamiltonian, whose ground states contain the solution. It is of particular interest in the context of solving complex combinatorial optimization problems by finding the ground state of a system. While it may be highly specialized in certain tasks, adiabatic computing could turn out to be more energy efficient than universal quantum computers in some cases, as less qubit control components are required. A comparison of the energy efficiency of adiabatic and circuit-based approaches solving the same task would be of great interest to understand which is the optimal design towards useful and sustainable quantum devices.

Another research direction is to benchmark the energy consumption of quantum machine learning techniques. Besides quantum neural networks~\cite{Altaisky2001, Schuld2014}, this field also includes other approaches in development like quantum reservoir computing~\cite{dudas2023quantumreservoirneuralnetwork, garcia2023photonicqrc, carles2026experimentalquantumreservoircomputing} or neuromorphic quantum computing~\cite{Markovic_2020}. To fit in the spirit of our model, benchmarking the energy efficiency of quantum machine learning requires defining a performance metric for a learning procedure (\textit{e.g} accuracy of the output) and understanding the hardware involved in the training and use of a quantum neural network.

Finally, a relevant research direction is to study how quantum computing would combine with high-performance computing (HPC). It is believed that quantum processors can be embedded into classical servers to enhance their ability to solve certain problems. A strong synergy between highly-advanced classical devices and emerging quantum platforms could prove to be the fastest way towards quantum-enhanced computing capabilities for research and society. A dedicated study of the energy efficiency of hybrid quantum-HPC machines would thus be of great interest.

\section{Concluding remarks}
\label{sec:conclusion}
In this work, we have developed a method for benchmarking the energy efficiency of quantum computers and provided examples of five of the leading hardware platforms currently envisioned. It quickly appears that one precise energy model adapted to all platform is not obvious and that it is safer to adapt a general metric, such as the energy efficiency presented in Eq.~\ref{eq:EE}, to the specifics of each quantum computer architecture. We presented nonetheless different methods to compute a value of the energy efficiency for each platform, taking into account  platform specification such as qubit connectivity, native gate, number of interaction zone or probabilistic operations.

With the insights of a diverse author list, ranging from academia to industry, we were able to investigate several state-of-the-art hardware architectures and to derive the power consumption of their different components (see Appendix~\ref{app:hardware}). This allowed estimations of the order of magnitude of the current energy consumption of quantum computers. While these numbers are related to specific technologies, it appears that most laboratories and companies actually use very similar state-of-the-art hardware from a few different vendors that have similar power consumption. As expected in the current state of quantum devices, maintaining the qubits in a controlled environment, either through cooling or maintaining the temperature of the room, is the most energy costly operation in most quantum computer architectures.

The core of this work resides in the compilation model (\textit{i.e.} the translation of an algorithm in a hardware-aware circuit) that was established for each platform in order to estimate the running time of quantum algorithms. Analyzing the functioning of devices is essential to properly benchmark their energy consumption taking into account their specificity. We compiled key time-related parameters on each platform, such as the initialization time, the measurement time and the clock time of current state-of-the-art hardware. These parameters are crucial to understand how fast quantum computers can complete a given task and thus compute their energy efficiency.  We believe that, while these compilation models can be further detailed and improved, they allow to already provide a good estimate of the computing speed of NISQ quantum computers.

Scaling up the results to quantum computers with thousands or millions of qubits is a tricky endeavor, as there is a consensus that radical changes and innovation in the engineering of the hardware are required for a sustainable scale-up. It is challenging to draw a strict line between the number of qubits and the number of hardware elements. Clever ways of sustaining and manipulating more qubits with less hardware are yet to be discovered through fundamental research. It can be however expected that the cost of the qubit control electronics will quickly outgrow the cooling cost as the number of qubit grows. Many more qubits could theoretically fit in current environment control systems without significantly altering their energy cost. The current road taken by the quantum computing industry, where technologies are evolving faster than fundamental research, is to fit as many qubits as possible without any major technology improvement, no matter the cost. Moreover, the implementation of QEC codes will imply a significant power dedicated to classical decoding algorithms. The cost of control electronics and classical computing might thus rapidly grow in a significant way.

We would like to emphasize that the results of this study alone cannot be used directly to make a fair comparison between platforms, since they are at different stages of development. While it seems that superconducting-qubit computers have a higher energy efficiency, other platforms might be more efficient at resolving specific tasks. As we showed, the native gate set of a platform can impose very large overhead during the compilation of an algorithm into a quantum circuit. Moreover, the specifics of a device constrain the execution time of certain operations and favor some families of algorithms. Consequently, the time required for a specific task on different platforms varies greatly. On top of that, as mentioned in the introduction, other aspects, such as the manufacturing cost of the hardware elements present in the quantum computer, might outgrow by several orders of magnitude the running cost of the device.

In addition, our work highlights the importance of understanding how the structure of quantum circuits interact with the quantum computing hardware and its constraints. Due to this interaction, a given circuit, decomposed using the same gate set, can led to different computing times depending the quantum processor runs it. This observation points out the importance of co-designing the hardware and software aspects of quantum computing, or at least, to design quantum processors that are optimized with certain circuit structures in mind. This can be particularly interesting with specific computational schemes, like variational algorithms, that tend to be more user-oriented, and there is a high degree of adaptability when it comes to circuit design.

Our work can be used to guide engineering and policy-making decisions in the near future of quantum technologies. Our framework can readily be used to benchmark different materials, \textit{e.g.} different chip materials for photonic computing, architectures, \textit{e.g.} different qubit connectivity in solid-state computers, or techniques, \textit{e.g.} continuous reloading of atoms in neutral-atom computers. In general, it appears that significant improvement in energy efficiency can originate from increasing computing speed without reducing the power demand of particular elements in the hardware. With clever experimental advances, the computing time can already be reduced by huge factors, while reducing the power demand of the majority of the hardware requires skills that are often out of reach for researchers in quantum technologies. This also emphasizes the necessity of building strong synergies between hardware constructors and researchers. We hope that our results will inspire more researchers to take on the task of benchmarking the energy of their device and guide our field towards a more sustainable future.

\section*{Acknowledgments}
The authors deeply thank Thierry Lahaye for his insights, Alexia Auffèves for her inspiration, Antonio Acin for his support and Sergio Lucio de Bonis for sharing his expertise on cryogenic coolers. 

The authors acknowledge financial support from the European Union (ERC, ASC-Q, 101040624) and (EQC, 101149233), European Union’s Horizon Europe research and innovation program under the grant agreement No 101114043 (QSNP) and QUCATS, together with Vinnova and Wallenberg Centre for Quantum Technology through the NQCIS project (1011113375), the EC through HORIZON-EIC-2022-PATHFINDEROPEN-01-101099697 (QUADRATURE) the ERC (AdG CERQUTE, 834266), the PEPR integrated project QCommTestbed ANR-22-PETQ-0011, as well as support from the Government of Spain (Severo Ochoa CEX2019-000910-S and FUNQIP, and QUANTUM ENIA project call - QuantumSpain), Fundació Cellex, Fundació Mir-Puig, Generalitat de Catalunya (CERCA program), Programa Fundamentos de la Fundación BBVA 2022-Electronic quantum simulator, Plan de Recuperación, Transformación y Resiliencia-Financiado por la Unión Europea- NextGenerationEU, and by the OECQ i-démo project as part of France 2030. 

Views and opinions expressed are however those of the author(s) only and do not necessarily reflect those of the European Union or the European Research Council. Neither the European Union nor the granting authority can be held responsible for them.

\appendix

\section*{Appendix}
\section{Hardware parameters}
\label{app:hardware}
In this appendix, we give the detailed list of hardware components, and operation times, used to estimate the energy efficiencies of all the quantum computing platforms and their architectural variations present in this work.

We report the number of components in the quantum computer architecture, their power consumption, and the  references used to make our estimation. In general, hardware components can scale differently with the number of qubits $N_{\rm{q}}$. Some components, such as the amplifiers for readout lines in a solid-state computer, scale linearly with the number of qubits. If possible, we explicit the scaling with $N_{\rm{q}}$, should the overall design of the architecture remain the same as the number of qubit grow. However, there is no fundamental scaling law for the hardware components in quantum computers. A combination of fundamental and engineering work should improve these scaling laws as technology progresses.

On the other hand, there are hardware components that do not directly scale with the number of qubits. This is the case for example of cryostats or lasers that produce optical traps. Current cooling systems can host thousands of qubits and, should the system size grow beyond that point, one would have to replace these components with bigger, more powerful ones. The scaling of the power consumption of these device is not a direct function of the number of qubits, as many companies are working on more efficient cooling systems. In this case, we specify the number of components reported in our observations on current experimental devices. When possible, the power consumption of each element was measured in real life implementations.

To account for scheduling and classical processing that may happen during the execution of quantum algorithms, classical servers are included in the component lists. In some of the current implementations, one or two computers have enough computational capabilities to handle the required operations. When this is the case, classical computers with a power consumption of 150W (which includes a screen) are added to the component list. In the future however, classical processing happening in parallel of quantum computation needs to be carefully studied, especially when considering syndrome decoding in quantum error correction codes.

In addition, we report notes on the energy consumption of cryogenic systems, since they are the most energy consuming hardware elements and they can be present in many quantum computer architectures.

\subsection{Superconducting qubits}
\label{app:superconducting}
\subsubsection{Baseline architecture}
\label{app:superconducting_baseline}
Table~\ref{tab:superconducting_components} describes a simplified reference architecture depending on the number of readout lines $N_{R}$, which can host approximately five qubits each, and considering a cryogenic setup and a qubit control cluster capable to host $~50$ qubits. Based on Qilimanjaro's superconducting architecture, we survey a Qblox Cluster with 20 module slots~\cite{qblox2025activereset}; how these slots are populated determines the available control and readout capacity. For example, up to 80 baseband control outputs with QCMs, 40 RF control outputs with qubit control RF modules (QCM-RF) IIs, or 20 RF  readout lines with QRM-RFs. RF signal generation is integrated within the cluster rather than supplied by a separate RF source. The rated maximum power of the cluster is $1.1~\mathrm{kW}$, though experimental measurements have shown a  steady-state consumption of approximately $600~\mathrm{W}$, with spikes of around $10~\mathrm{W}$ during circuit execution; nevertheless, we use the stated max power consumption. Each readout line requires two room-temperature low-noise amplifiers (LNAs) for warm-side signal amplification, and one cryogenic HEMT amplifier at the $4~\mathrm{K}$ stage for the first stage of qubit signal amplification~\cite{cha2020hemt}. The qubit control and readout electronics are integrated into a modular cluster.  In our architecture, a single readout line serves approximately five qubits. The cryogenic system has been chosen according to CESGA's QMIO computer, discussed in the following section. The compressor drives the pulse tube pre-cooler of the dilution refrigerator, while the Gas  Handling System (GHS) manages helium circulation and vacuum pumping~\cite{bluefors_ghs2}. Some installations deploy a second GHS for redundancy and improved operational  stability. 

\begin{table}[!h]
\centering
\begin{tabular}{|l|c|c|} 
\hline
\textbf{Component}    & \textbf{Number} & \textbf{Power consumption (W)} \\ \hline
Compressor & 1 & 9500 \\ \hline
Gas Handling System & 1 & 4600 \\ \hline
Chiller & 1 & 2000 \\ \hline
Qubit control and readout cluster & 1 & 1100 \\ \hline
RF Source & $N_{\rm{R}}$ & 70 \\ \hline
Room Temperature LNA & $2\cdot N_{\rm{R}}$ & 1.4 \\ \hline
HEMT PBA & $N_{\rm{R}}$ & 100 \\ \hline
Server & 1 & 800 \\ \hline
\end{tabular}
\caption{Main components found in a superconducting qubit based quantum computer with a single feed line and $N_{\rm{q}}$ qubits. As a reference, the average feed line can host 5 superconducting qubits. These values are based on components from both CESGA's and Qilimanjaro's quantum computers.}
\label{tab:superconducting_components}
\end{table}

Table~\ref{tab:superconducting_optime} reports average operation times observed in current superconducting quantum computers. Based on Qilimanjaro's superconducting devices, we define the clock time as the duration of a native two-qubit gate operation. Depending on the architecture and native gate being implemented, two-qubit entangling gates such as the CZ or iSWAP typically take between $30$ and $200 ~\mathrm{ns}$~\cite{postquantum_sc_qubits}. For concreteness, we adopt $t_{\rm{clock}} \approx 50~\mathrm{ns}$ as a representative value. We observed that measurements take approximately $t_{\rm meas}\approx1600~\mathrm{ns}$, consistent with the range of hundreds of nanoseconds to a few microseconds reported in the literature~\cite{transmon_qubit}. Between each circuit sample, qubits can be reinitialized either passively, by thermalization with the cold environment, or actively, using circuit elements to extract the excited-state population. Passive reset requires waiting at least five to seven times the qubit relaxation time $T_{1}$, adding a minimum of $50-500 ~\mathrm{\mu s}$ per cycle for typical transmon devices~\cite{qblox2025activereset}. We observed $t_{\rm reset}^{\rm pass} \approx 200~\mathrm{\mu s}$, while active reset reduces this to $t_{\rm reset}^{\rm act} \approx 5 \mathrm{\mu s}$, a roughly two-order-of-magnitude improvement.

\begin{table}[!h]
\centering
\begin{tabular}{|l|l|l|}
\hline
\textbf{Parameter} & \textbf{Symbol} & \textbf{Value} \\ \hline
Clock time & $t_{\textrm{clock}}$        & $50\cdot 10^{-9}$ s  \\ \hline
Measurement time & $t_{\textrm{meas}}$         & $1.6\cdot 10^{-6}$ s \\ \hline
\begin{tabular}[c]{@{}l@{}}Active reset time \\ between shots\end{tabular} & $t_{\textrm{reset}}^{act}$  & $200\cdot 10^{-3}$ s \\ \hline
\begin{tabular}[c]{@{}l@{}}Passive reset time\\ between shots\end{tabular} & $t_{\textrm{reset}}^{pass}$ & $5\cdot 10^{-6}$ s   \\ \hline
\end{tabular}
\caption{Operation times for a superconducting quantum computer.}
\label{tab:superconducting_optime}
\end{table}

\subsubsection{Example of CESGA's QMIO computer}
QMIO is a 32-qubit superconducting quantum processor based on coaxmon technology developed by Oxford Quantum Circuits and operated at the Galicia Supercomputing Center (CESGA). Unlike the reference model described above, QMIO represents a fully deployed system with measured steady-state power consumption at the facility level, including cryogenic infrastructure, control electronics and classical processing.

\begin{figure}[!h]
    \centering
    \includegraphics[width=0.8\linewidth]{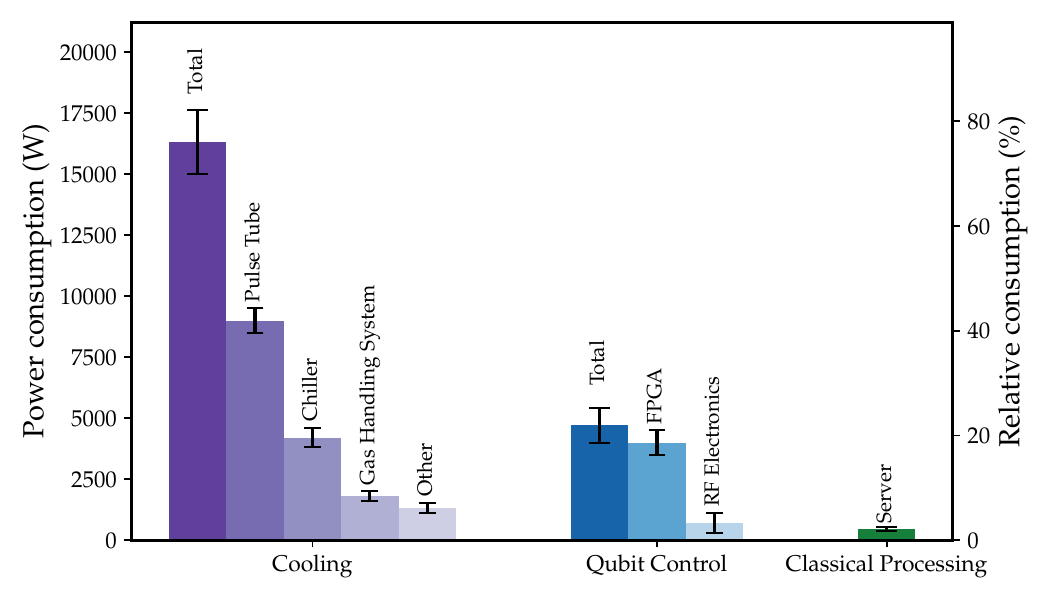}
    \caption{Energy breakdown of a QMIO superconducting-qubits computer (32 qubits), according to the components shown in Table~\ref{tab:qmio_appendix_style_final}. In the x axis, the components are grouped according to their purpose in the computer. Each group has a bar that corresponds to the total accumulated power of its components. The left y axis shows the absolute value of power consumption of the components, and the right y axis the relative value with respect to the overall consumption of the entire computer. Error bars correspond to variations in the power consumption over time.}
    \label{fig:qmio_superconducting_power_breakdown}
\end{figure}

The comparison between the literature based superconducting model described above and the experimentally characterized QMIO 32-qubit processor reveals that, within the $~50$ qubits system size, the steady-state power consumption of these superconducting platforms remains confined to a relatively narrow interval dominated by cryogenic overhead. Specifically, the architecture discussed above, reaches 20 kW of power consumption, and the QMIO processor operates at $21.4 \pm 2.4$ kW steady-state (Table~\ref{tab:qmio_appendix_style_final}). The component-level power breakdown of QMIO is summarized in Table~\ref{tab:qmio_appendix_style_final} and the measured power breakdown for QMIO is shown in Figure~\ref{fig:qmio_superconducting_power_breakdown}. It confirms that cooling alone accounts for roughly 76\% of the total system power. 

\begin{table}[!h]
\centering
\small
\begin{tabularx}{\textwidth}{|l|c|c|X|}
\hline
\textbf{Component} & \textbf{Number} & \textbf{Power (W)} & \textbf{Notes} \\
\hline
Compressor (pulse tube) & 1 & $9000 \pm 500$ & Derived from cryogenic increment (baseline subtraction method). \\
\hline
Chiller (water cooling) & 1 & $4200 \pm 400$ & Derived from cryogenic increment. \\
\hline
Gas Handling System (GHS) & 1 & $1800 \pm 200$ & Derived from cryogenic increment. \\
\hline
Cryostat auxiliaries (pumps, controllers) & 1 & $1320 \pm 200$ & Residual to close cryogenic increment budget. \\
\hline

Per line FPGA control/readout modules & 32 & $125 \pm 19$ & Allocated from 

measured control electronics (FPGA-per-qubit architecture). \\
\hline
RF/IF up- and down-conversion electronics & 1 & $600 \pm 100$ & Included in measured control electronics. \\
\hline
Clock distribution system (2 GHz differential) & 1 & Included & Described in QMIO architecture; included in control electronics power. \\
\hline
Control server and orchestration & 1 & $440 \pm 80$ & Included in measured control electronics. \\
\hline
\end{tabularx}
\caption{Main components found in CESGA's QMIO superconducting quantum computer ($N_q=32$) and their operational power consumption (steady-state). The power consumption of each element was directly measured on the existing setup. The error range corresponds to variations in the energy consumption over time.}
\label{tab:qmio_appendix_style_final}
\end{table}
 \newpage

\subsection{Spin qubits}
\label{app:spin_qubits}

This appendix presents the architectures used in the main text to compute estimations of the energy efficiency of spin-qubit quantum computers. Table~\ref{tab:spin_components} describes a simplified reference component list, constructed having in mind a certain architecture requiring $N_{DC}$ number of DC connections, $N_{BB}$ number of baseband control for fast charge manipulation, $N_{MW}$ number of microwave control line for qubit operations and necessitates $N_{R}$ number of readout lines. Table~\ref{tab:spin_components_2} presents how these numbers relate to the number of qubits in the computer for three different architectures as well as their reset time between shots, clock time, and measurement time.

\begin{table}[!h]
\centering
\begin{tabular}{|l|c|c|c|} 
\hline
\textbf{Component}   & \textbf{Number}  & \textbf{Power consumption (W)} & \textbf{Ref.}\\ \hline
Pulse tube      & 1  & 9500     & \\ \hline
Chiller  & 1 & 4600  & \\ \hline
GHS & 1       & 2000    & \\ \hline
Classical computer  & 2 & 150  &\\ \hline
DC control   & $N_{DC}$  & $\sim 0.5$ per unit   & \cite{QbloxCurrentControler}\\ \hline
Baseband control   & $N_{BB}$  & $\sim 7.5$ per unit   & \cite{QbloxBasebandControler}\\ \hline
MW control   & $N_{MW}$   & $\sim 24$ per unit    & \cite{QbloxMWControler}\\ \hline
Readout lockin      & $N_{R}$ & $\sim 48$ per unit    & \cite{QbloxReadOut}\\ \hline
Low-noise amplifiers & $N_{R}$ & $\sim 0.1$ per unit   & 
\cite{LNFlownoise} \\ \hline
Room-temperature amplifiers  & $N_{R}$ & $\sim 0.5$ per unit   & \cite{RoomTempAmpl}\\ \hline
Magnet power supply    & 1  & neg. in steady state  &\\ \hline
\end{tabular}
\caption{Main components found in a spin-based quantum computer}
\label{tab:spin_components}
\end{table}

\begin{table}[!h]
\centering
\begin{tabular}{|l|c|c|c|} 
\hline
          & \textbf{2D ($\mathbf{N_q = n\times n}$) } & \textbf{Linear ($\mathbf{N_q = n}$)}& \textbf{2D crossbar ($2\mathbf{N_q = n\times n}$)}\\ \hline\hline
$N_{DC}$  & $3n^2-2n=3N_q-2\sqrt{N_q}$& $2n-1=2N_q-1$   &  $4n+1= 4 \sqrt{2N_q}+1$\\
$N_{BB}$  & $3n^2-2n=3N_q-2\sqrt{N_q}$& $2n-1=2N_q-1$   &  $4n+1= 4 \sqrt{2N_q}+1$\\
$N_{MW}$  & $n^2=N_q$                 & $n^2=N_q$       & 1\\
$N_{R}$   & $n^2/2=N_q/2$             & $n/2=N_q/2$     &  $n =  \sqrt{2N_q}$\\\hline\hline
$t_{\rm{reset}}$ (ms) & 0.1 & 0.1 &0.1 \\
$t_{\rm{clock}}$ (ns) & 10 & 10 & 250 \\
$t_{\rm{meas}}$ (ms) & 0.01 & 0.01 & $0.01\times \sqrt{N_q/2}$ \\\hline
\end{tabular}
\caption{Number of hardware components and operation times in a spin-qubit based quantum computer for the 3 different connectivity considered.}
\label{tab:spin_components_2}
\end{table}

The prototypical example of a quantum register can be envisioned as a two-dimensional array of quantum dots arranged in a square lattice of size $n\times n$. The power consumption strongly depends on how this quantum register is controlled. In the worst-case scenario, each qubit and coupling gate is controlled individually (a full one-to-one control scheme). This would require one DC line per quantum dot and one per coupling gate, yielding approximately $N_{DC}\sim 2n^2$. The same reasoning applies to baseband control lines, giving $N_{BB}\sim 2n^2$. Microwave control is only required for qubit manipulation, leading to $N_{MW}\sim n^2$. 

Regarding the readout, several options exist. Charge sensors typically complicate the geometry of the quantum-dot array and still need to be coupled to external resonators to achieve high bandwidth. We therefore restrict ourselves to gate-based dispersive readout, as this choice does not substantially affect the total power consumption. In the worst-case scenario, one would require a resonator per pair of qubits for parity-based readout, and one readout line per resonator, resulting in $N_R \sim n^2 / 2$.

A second strategy consists of reducing the number of control lines. For example, Ref.~\cite{li2018} proposed a crossbar approach, in which control lines are shared among rows, columns, and diagonals of the array. In this case, instead of having a number of control lines that scales quadratically with the number of qubits, the scaling becomes linear in $n$, i.e. proportional to the square root of the total qubit number

\subsection{Trapped ions}
\label{app:trapped_ions}

This appendix presents the architectures used in the main text to compute estimations on the energy efficiency of trapped ions quantum computers. Table~\ref{tab:trapped_ions_components} describes a simplified reference component list, based on the trapped ions device found in the NQCC laboratory. Here we consider a room temperature device, albeit the energy cost of operating at cryogenic temperature is also shown. This considers a processor running with a single ion species, and we split the control system into active and passive, where active means it's expected to respond in real time to perform gates. In this case, the active control system would be the only component expected to vary for the 5 and 10 traps computer. The considered computers can perform single- and two-qubit gates in parallel (in order to take full advantage of their multiple gate zones), thus we need these two active control systems that can target different sets of atoms and perform independent rotations on their state at the same time. However, there are two instances for which we expect the number of active control system to scale further. First, if one wants to perform more species of gates at the same time (such as different kinds of single-qubit gates), $N_{\rm{control}}$ would be equal to the number of independent gates to be executed in a single time step (with $N_{\rm{zones}}^{\rm{gate}}$ being the maximum). Then, if the number of gate zones is large enough, we expect that a single active control system would struggle to execute the same gate on all of them, as an electronic signal can only be split a given number of times before it becomes too small to be detected by the hardware. In that case, we expect $N_{\rm{control}}$ to scale with $N_{\rm{zones}}^{\rm{gate}}$.

\begin{table}[!h]
\centering
\begin{tabular}{|l|c|c|c|} 
\hline
\textbf{Component} & \textbf{Number} & \textbf{Power consumption (W)} \\ \hline
HVAC& 1 & 7500  \\ \hline
Gate drive &    1   & 700   \\ \hline
Laser system & 1 & 825 \\ \hline
Classical computer & 1 & 150 \\ \hline
Image system   &    1    &    100    \\   \hline
Magnetic field generation &   1 &  1     \\ \hline
Active control system   &  $N_{\rm{control}}$ &  100   \\ \hline
Passive control system & 1 & 200   \\ \hline

Compressor*  & 1  & 7500   \\  \hline

Chiller*        & 1   & 2000    \\\hline
\end{tabular}
\caption{Main components found in a trapped ions based quantum computer. The power consumption values were directly observed on the trapped ion device in the NQCC in the UK. The components marked with a star (*) are not mandatory in QCCD trapped ions devices, and are commonly omitted. Although present in the NQCC laboratory, we removed them from the main analysis.}
\label{tab:trapped_ions_components}
\end{table}

Table~\ref{tab:trapped_ions_optime} shows the reset time between shots, clock time, measurement time and shuttling time of ions between storage and gate zones, for devices with 1, 5 and 10 gate zones. The reset time accounts for transport back to memory (or storage) zone, Doppler and sideband cooling. The clock time accounts for the average time to perform the two-qubit gates, plus any transport that might happen between gate zones. This consequently leads to a longer clock time for the multi-zone computers, due to extra transport needed to arrange the qubits. To estimate the transport time of the multi-zone devices, we assumed that gate zones are connected in a linear geometry, and set the upper bound to the time required to populate the furthest gate zone. Assuming that these devices can move ions safely at $3~\mathrm{m/s}$, and the distance between gate zones is $750~\mathrm{\mu m}$~\cite{Ransford2025}, we expect the clock time of multi-zone devices to be $t_{\rm{clock}}^{\rm{multi-zone}}\approx 250~\mathrm{\mu s}\times (N_{\rm{zones}}^{\rm{gate}}-1)$, to move the ions between the farthest apart gate zones. For the 5- and 10-gate-zones, this movement is the most time-consuming operation in a single circuit layer, an is thus considered to be the clock time. In contrast, the clock time of the single-gate-zone device corresponds to the additional transport to bring a pair of qubits to the same potential well ($100~\mathrm{\mu s}$) and the time to perform a gate itself ($70~\mathrm{\mu s}$). This time is based on laser gates, but the fastest microwave gates would be performed in 300us. Since two qubit gates require more transport, and transport induces heating in the ions, cooling is normally performed in between gates~\cite{Shu2014}. For the sake of simplicity, we exclude this cooling time out of our model, as it would only represent a quantitative difference and not the overall conclusions of the model.

\begin{table}[!h]
\centering
\begin{tabular}{|l|c|c|c|} 
\hline
          & $\mathbf{N_{\rm{\textbf{zones}}}^{\rm{\textbf{gate}}}=1}$ &  $\mathbf{N_{\rm{\textbf{zones}}}^{\rm{\textbf{gate}}}=5}$ &  $\mathbf{N_{\rm{\textbf{zones}}}^{\rm{\textbf{gate}}}=10}$ \\ \hline\hline
$N_{\rm{control}}$  & $1$   &  $2$ & 2\\\hline\hline
$t_{\rm{reset}}$ (ms) & $150$ &  $150$ & $150$\\
$t_{\rm{clock}}$ ($\mathrm{\mu s}$) & $170$ &  $1000$ & $2250$ \\
$t_{\rm{trans}}$ ($\mathrm{ms}$) & $50$ &  $50$ & $50$\\
$t_{\rm{meas}}$ (ms) & $0.5$ &  $0.5$ & 0.5\\\hline
\end{tabular}
\caption{Number of active control systems and operational times in trapped ion quantum computers with 1, 5 and 10 gate zones.}
\label{tab:trapped_ions_optime}
\end{table}

The components detailed in this section correspond to a basic QCCD trapped-ions device build with a storage zone and several gate zones connected linearly. There are multiple modifications that can be made to further improve the performance and efficiency of such device. For instance, one could modify the distribution of the traps to define cache zones, or improve the connectivity between gate zones, which would optimize the transports needed during the execution of a circuit. These modifications, however would need to be combined with a powerful compiler to take advantage of the device's geometry (this is the case of Quantinuum's Helios computer~\cite{Ransford2025}). In that sense, we expect state-of-the-art and near-future devices to display higher power demands, but within the same order of magnitude as our estimations. However, if this resources are used properly, the computing time of such computers would be significantly optimized and could overcompensate the extra power consumption, delivering higher energy efficiency.

\subsection{Neutral Atoms}
\label{app:neutral_atoms}

This appendix presents the architectures used in the main text to compute estimations on the energy efficiency of neutral atoms quantum computers. More specifically, we base our estimation on the rudbium-87 neutral atom processor hosted in Harvard, where digital entangling operations are performed via the $n=53$ Rydberg state. The system itself supports up to a thousand individually trapped atoms with coherent transport performed in parallel on over 100 atoms at a time. In this system, atoms can be re-used and lost atoms can be refilled from a reservoir to decrease cycle time through state-selective loss-detecting lattice readout. The continuous array system that we include in our analysis is based on Ref.~\cite{chiu_continuous_2025} where atoms are reloaded with states up to 30,000 qubits per second.
Table~\ref{tab:neutral_atoms_components} describes a simplified reference component list, implementations that can host up to a thousand traps and $N_{\rm q} = 400$ qubits. The table include hardware components for the baseline neutral-atoms architecture based on the Harvard computer, and the extra components required to implement continuous reloading.

\begin{table}[!h]
\centering
\begin{tabular}{|l|c|c|c|} 
\hline
\textbf{Component}  & \textbf{Number} & \textbf{Power consumption per unit (W) }& \textbf{Ref}. \\ \hline
Trap lasers  &   1    &  1500  &   \cite{noauthor_852nm_nodate}     \\  \hline
Lasers for Rydberg gates    &  1     &  1900 & \cite{photonics_cw_nodate, noauthor_verdi_nodate}\\  \hline
Other lasers   & 2      &  900  & \\  \hline
               & 7      &  150  &     \\   \hline
Magnetic field generation &   1    &  1050  &      \\  \hline
Camera      &   1    &     155 & \cite{noauthor_qcmos_nodate} \\  \hline
Vacuum pump  &  1     &   100  & \\  \hline
Classical computers &    1   &  150   & \\  \hline
HVAC  &   1    &    7500  & \\  \hline
Other electronics &   1    &   2000   & \\  \hline
Laser Chiller (small) &  5     & 600  & \cite{noauthor_t257p-20_nodate} \\  \hline
Laser Chiller (large)  &  2     & 1000  &\cite{noauthor_3u_nodate}\\ \hline\hline
\multicolumn{4}{|c|}{\textbf{Extra components for continuous reloading}}\\\hline
Transport laser  &  1     &   350  & \\  \hline
Laser Chiller &    1   &  1500   & \\  \hline
Additional Vacuum pumps &   1    &    700  & \\  \hline

\end{tabular}
\caption{Main components found in a neutral-atoms quantum computer. The values without reference were directly observed on the experimental setup found at the Harvard laboratory~\cite{bluvstein_logical_2024}.}

\label{tab:neutral_atoms_components}
\end{table}

Table~\ref{tab:neutral_optimes} shows the relevant operational times of the studied neutral-atoms computer. These estimations are based on the processor operating with qubit reuse as presented in~\cite{chiu_continuous_2025}. Here, atoms are loaded and cooled in a magneto-optical trap and then filled into tweezers, consisting of the computational array as well as a atom reservoir. This process is the most time intensive and required approximately 500ms. In order to increase cycle rates, the atoms are not discarded after an experimental shot, but instead, lost atoms are refilled from a reservoir and qubits are reinitialized. In contrast to  MOT reloading, this process takes on the order of 10 ms for each shot, which corresponds to the $t_{\rm{reset}}$ considered in this work. Atoms can be refilled from a reservoir until it is depleted, therefore, it is highly dependent on the array and reservoir size until atoms must be reloaded from the MOT. For the purposes of this calculation, we estimate reloading every 2400 gate layers. For the sake of simplicity, we assume that reloading can be performed \emph{during} a circuit execution, not only between samples. In the case of a unitary circuit that would require to copy the state of the qubits before reloading, which we do not account for. Transport and reconfiguration times are also given as broad estimates that are highly dependent on the circuit structure. Moving atoms in optical tweezers is generally limited by atom loss and heating to approximately $1~\mathrm{ms}$ \cite{bluvstein_quantum_2022}. Generally, the arrays are around $200~\mathrm{\mu m}$ in extent and therefore, considering trap transfer times as well, we estimate that both reconfiguration of the atoms during successive gates as well as moves from the storage region take on the order of $500~\mathrm{\mu s}$. Considering that the reconfiguration of the atoms in the interaction is a process that takes place before most of the layers of the circuit, and that single- and two-qubit gates take the scale of $10~\mathrm{\mu s}$ and $150~\mathrm{\mu s}$, respectively, we assume $t_{\rm{clock}}= t_{\rm{reconfig}}$. On the contrary, transport from the storage to the interaction-zone is a less common operation, characterized by the $\beta_{\rm{trans}}$ parameter. Finally, we have chosen the measurement time to be $\approx10~\mathrm{ms}$. Again this is highly dependent on readout method, and new schemes have been shows readout times on the microsecond scale \cite{muzi_falconi_microsecond-scale_2025}.

\begin{table}[!h]
\centering
\begin{tabular}{|l|c|c|}
\hline
                                     & \textbf{Periodic Reload}    & \textbf{Continuous Reload}  \\ \hline
$t_{\rm{reset}}$ (ms)                & $10$              & -                                         \\\hline 
$t_{\rm{clock}}$ ($\mathrm{\mu s}$)  & $500$              & $500$                                    \\\hline
$t_{\rm{trans}}$ ($\mathrm{\mu s}$) & $500$               & $500$                      \\\hline
$t_{\rm{meas}}$ (ms)                 & $10$              & $10$                                        \\\hline
$t_{\rm{reload}}$ ($\mathrm{m s}$) & $500$               & \multirow{2}{*}{-}     \\\cline{1-2}         
Reload freq. ($\#$ of layers)                & $2400$              &                                         \\ \hline
\end{tabular}
\caption{Extra hardware components and operational times of two neutral-atoms devices, with and without the ability to perform continuous reload of the register during computations.}
\label{tab:neutral_optimes}
\end{table}

\subsection{Photonic architectures}
\label{app:hardwarephoton}

This appendix presents the architectures used in the main text to compute estimations on the energy efficiency of photonic quantum computers. Table~\ref{tab:photonic_component} describes a simplified reference component list, based on real life implementations of chip-based photonic computation, in particular the one from Quandela~\cite{Qandela}. It assumes cryostats that can host up to one quantum dot source and 25 superconducting nanowire single photon detectors (SNSPDs) as well as demultiplexer able to send at most 12 photons on the chip per source. Table~\ref{tab:photonparam1} and \ref{tab:photonparam2} show the efficiencies of components that are used to compute the end-to-end transmission of devices using different chip technologies. More information about the chip technologies can be found in Appendix~\ref{app:chiptech}.

\begin{table}[!h]

\centering
\begin{tabular}{|p{6cm}|c|c|c|} 
\hline
\textbf{Component}      & \textbf{Number} & \textbf{Power consumption (W)} & \textbf{Ref.} \\ \hline
Laser & 1      & 4  & \\ \hline
Cryogenic system for the Quantum dot and the SNSPDs & $\left\lceil\frac{2N_q}{25}\right\rceil$   & 1500 &   \\ \hline
Peltier cooling of the chip           &   1     &        200-300       &        \\ \hline
Demultiplexer system (DMX-12)        &  $\left\lceil \frac{N_{\rm q}}{12}\right\rceil $      &      $\sim$50 per DMX   &              \\ \hline
Classical control electronics (FPGA, DAC)       &  $\left\lceil \frac{4N_{\rm q}^2}{16}\right\rceil $      &      $\sim$15 W   &         \cite{Crampton2025}     \\ \hline
Classical computers         &        &  150          &           \\ \hline
\end{tabular}
\caption{Main components found in a single photon source photonic quantum computer with $N_{\rm q}$ qubits. }
\label{tab:photonic_component}
\end{table}

\begin{table}[!h]

\centering
\begin{tabular}{|p{6cm}|l|l|} 
\hline
\textbf{Name}      & \textbf{Symbol} & \textbf{Value}  \\ \hline
Source efficiency &  $\eta_{\rm of}$    &   71.2\% \\ \hline
Detector efficiency & $\eta_{\rm d}$   &  98\%   \\ \hline
Demultiplexer efficiency   &   $\eta_{\rm dmx}$     &     83\%     \\ \hline
\end{tabular}

\caption{Component efficiencies in photonic computers.}
\label{tab:photonparam1}
\end{table}

\begin{table}[!h]

\centering
\begin{tabular}{|p{6cm}|l|l|l|c|} 
\hline
\textbf{Chip technology}      & \textbf{Glass mesh} & \textbf{EO-LN} & \textbf{EO-BTO} \\ \hline
Power consumption of the chip elements &   50-100 mW    & 60 $\mu$W   & 121 nW \\ \hline
Time to reconfigure the phase shifters &  1 $\mu$s  &  44 ps &  145 ps \\ \hline
Coupling loss at in/output of the chip    &   0.5 dB/Facet   &  3.4 dB/Facet  & .127 dB/Facet \\ \hline
Mach-Zehnder interferometer loss       &  0.1 dB/MZI  &  0.15 dB/MZI   & .21 dB/MZI   \\ \hline
Propagation loss & 0.3 dB/cm   &  0.025 dB/cm  &  0.016 dB/cm \\ \hline
Ref.                             &   \cite{Barzaghi2025}      &   \cite{Sund2023,Zheng2023}    & \cite{Alexander2024,Chrysostomidis2025}  \\ \hline
\end{tabular}
\caption{Differences between the chip technologies considered in photonic computation.}
\label{tab:photonparam2}
\end{table}

\subsection{Cooling systems}
\label{app:cool}
Cryogenics are a very important part of many quantum computing architectures. Since the quantum systems in which information is encoded need to be protected from decoherence and outside influences, it is sometimes necessary to cool them down to temperatures below 20 mK. Moreover, some technologies such as single photon detectors have a higher efficiency at low temperature (around 4K). Pulse tube, water cooler and gas handling systems are a crucial parts of superconducting, spin and photonic quantum computers. Unsurprisingly, they represent one of the main source of energy consumption in the total energy balance of these platforms. 

In a typical dilution fridge used for quantum computing uses different Helium isotopes and cooling happens in two stages. A first phase of pre-cooling down uses a so-called pulse tube to cool down a volume up to $\approx$ 4.2K. This is the maximum achievable due to physical constraints: below this temperature gas density drops. To reach a lower temperature, a dilution cycle of $^3$He/$^4$He starts, using an secondary pulse tube. This allows systems to reach temperatures $<10$ mK. To understand the energy consumption of this technology, we use, as a reference, a cryogenic refrigerator used in our lab in ICFO, Spain~\cite{CryoCooler}. We note, however, that there are few companies that produce this technology and the values that we show in the different tables of this appendix are compatible with all current state-of-the-art dilution fridges.

We observed that most of the energy consumption stems from the primary pulse tube. For example, measurements performed with a voltmeter on the cryogenic refrigerator~\cite{CryoCooler} revealed that the consumption of the secondary mechanism, used to perform the $^3$He/$^4$He cycle, has a power consumption of 200 W while the primary pulse tube has a power consumption around 8000 W. Moreover, the energy spent by the fridge depends on two main factors: the heat production of the system inside the fridge and the volume to cool down. It is challenging, and out of the scope of this study, to derive a general relation between dissipated heat, volume and power consumption, as done for example in~\cite{Fellous-Asiani2023}. In this work, we use the power consumption observed in real-life implementations of solid-state quantum computers using dilution fridges. 

Since the volume and amount of control electronics should increase significantly as the number of qubits grows, the energy spent in cooling can only increase. Our observations show however that the energy spent in the electronic components required to address the qubits might outgrow the cooling energy in the future.

\section{Compilation of algorithms}
\label{app:compil}
To run on a quantum computer, a quantum algorithm needs to be compiled into a sequence of gates that can be run efficiently on the specific hardware used. The gates are scheduled in time steps, where the gates that do not act on the same qubits can be performed in parallel. The number of time steps defines the depth of the circuit. The circuit depth of an algorithm is central to our study, since it defines how long an algorithm will take to be ran on a quantum computer.

\subsection{Solid-state qubits}
\label{app:compilsolid}

Given the qubit connectivity in a solid-state platform, this compilation process can result in very different depths for the same algorithm. In this section, we discuss the influence of the compilation method and the qubit connectivity of a platform on the circuit depth of an algorithm.

\begin{figure}[!h]
    \centering
    \begin{quantikz}[row sep={2em,between origins}, column sep=2em]
       \lstick{{$v_0$}} & \ctrl{4}   & \\
       \lstick{{$v_1$}} &            & \\
       \lstick{{$v_2$}} &            & \\
       \lstick{{$v_3$}} &            & \\
       \lstick{{$v_4$}} &  \targ{-4} & \\
\end{quantikz} 
= \begin{quantikz}[row sep={2em,between origins}, column sep=2em]
       \lstick{{$v_0$}} & \swap{1} &          &          &          &  \swap{1} & \\
       \lstick{{$v_1$}} & \targX{} & \swap{1} &          & \swap{1} &  \targX{} & \\
       \lstick{{$v_2$}} &          & \targX{} & \ctrl{1} & \targX{} &           & \\
       \lstick{{$v_3$}} & \swap{1} &          & \targ{-1}&          &  \swap{1} & \\
       \lstick{{$v_4$}} & \targX{} &          &          &          &  \targX{} & \\
\end{quantikz} 
=
\begin{quantikz}[row sep={2em,between origins}, column sep=2em]
       \lstick{{$v_0$}} & \swap{1} &          &          & \rstick{{$v_1$}}\\
       \lstick{{$v_1$}} & \targX{} & \swap{1} &          &  \rstick{{$v_2$}}\\
       \lstick{{$v_2$}} &          & \targX{} & \ctrl{1} &  \rstick{{$v_0$}}\\
       \lstick{{$v_3$}} & \swap{1} &          & \targ{-1}&  \rstick{{$v_4$}}\\
       \lstick{{$v_4$}} & \targX{} &          &          &  \rstick{{$v_3$}}\\
\end{quantikz} 
    
   \caption{Implementation of a CNOT gate between $v_0$ and $v_4$, on a quantum system with only line connectivity (with qubit connectivity between $v_i$ and $v_{i+1}$). The last two layers of the middle circuit are optional and can be replaced by qubit relabeling.}
   \label{fig:compilationCNOT}
\end{figure}
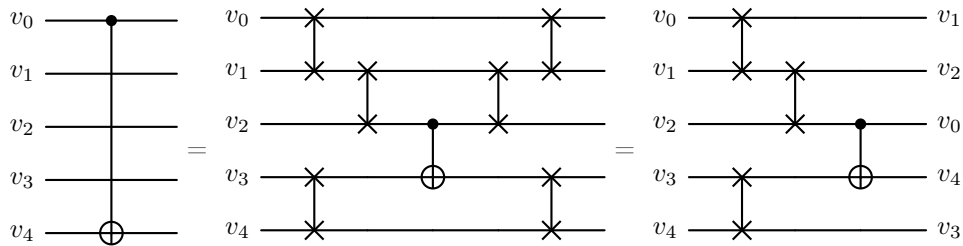

In order to characterize the depth overhead of an algorithm executed on a solid-state processor with given connectivity, we need to account for the distance between qubits that interact together. The connectivity between qubits is represented by a graph $G=(V, E)$ where each vertex $v \in V$ corresponds to a qubit and, if two vertices are connected by an edge $(i,j) \in E$, we can perform a physical two-qubit gate between them. In order to solve this issue, one has to include SWAP gates that place qubit states in adjacent vertices connected with an edge. This process is called routing. For instance in a linear graph, a CNOT gate between adjacent qubits does not result in any overhead. On the other hand, a CNOT gate between qubits that are 4 nodes apart, will add 2 extra layers of SWAP gates and relabeling (see Figure~\ref{fig:compilationCNOT}). 

In a general graph $G$, we can define the Manhattan distance $d(v,v')$ which corresponds to the length of the shortest path between two vertices in $G$. We can then replace a CNOT between two distant qubits $v,v'$ by $d(v,v')-1$ SWAP gates and 1 CNOT gate. Considering that each SWAP gate is done by adding 3 layers of CNOT gates between adjacent qubits, this implies a total overhead of $3\times \left\lceil\frac{d(v,v') - 1}{2}\right\rceil$ layers. In order to generalize this expression for a given graph, we use the average shortest path between all pairs of nodes of a graph $G$:
\begin{equation}
    \overline{d(G)} = \frac{2}{|V|(|V|-1)}\sum_{v\neq v' \in V} d(v,v').
\end{equation}
Therefore, we estimate that the average depth overhead needed to connect any pair of qubits in a processor with connectivity graph $G$ is $3\cdot\left\lceil\frac{ \overline{d(G)}-1}{2}\right\rceil$.

Furthermore, we assume that the gate overhead that connect two different pairs of qubits can also be parallelized. Therefore, we assume that a layer with only one  two-qubit gates will produce the same overhead as a layer with many two-qubit gates. With this in mind, the depth overhead of a circuit depends on the number of layers that have (any number of) two-qubit gates $D^{\rm{2q}}$, as $D^{\rm{2q}}\cdot3\cdot\left\lceil\frac{\overline{d(G)}-1}{2}\right\rceil$.

Let $\alpha_{\rm{2q}} = D^{\mathrm{2q}} / D$ be the fraction of layers which include a two-qubit gate in the transpiled circuit. The circuit depth after routing is given by:
\begin{equation}
    D\approx D_{0}\left( 1+\alpha_{\rm{2q}}\cdot3\cdot \left\lceil \frac{\overline{d(G)}-1}{2} \right\rceil \right)
\end{equation}
An upper bound of this value can be obtained by substituting the average shortest path length $\overline{d(G)}$ with the diameter of the circuit $d(G)$, which represents the \emph{longest} shortest path length between nodes. While this model can be improved in future studies, we believe this allows us to provide a fair order of magnitude of the energy efficiency of solid-state platform. Our compilation model has the benefit of being simple and easily integrable into the larger model to benchmark the energetic cost of quantum computing.

\begin{figure}[!h]
    \centering
    \begin{subfigure}[b]{.5\textwidth}
        \centering
        \includegraphics[width = \textwidth]{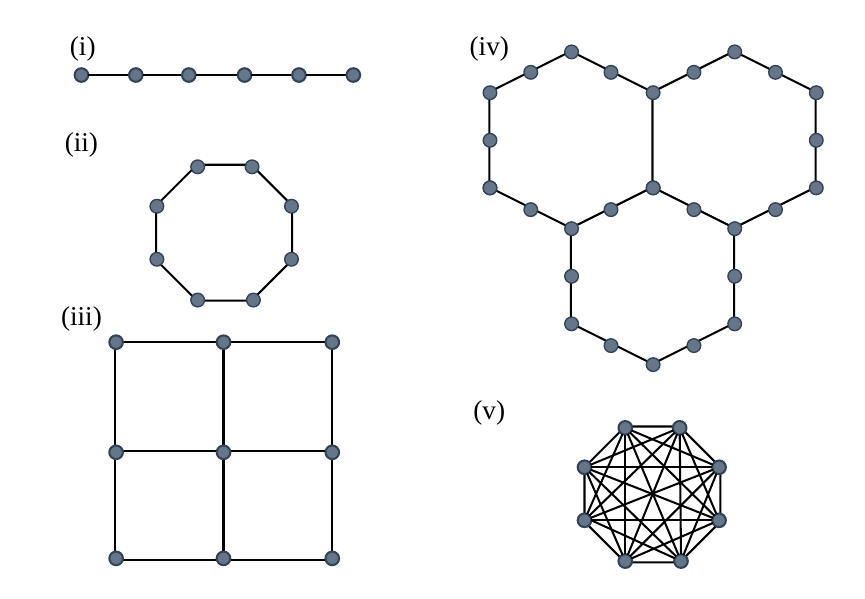}
        \caption{}
    \end{subfigure}
    \begin{subfigure}[b]{.42\textwidth}
        \centering
        \includegraphics[width =\textwidth]{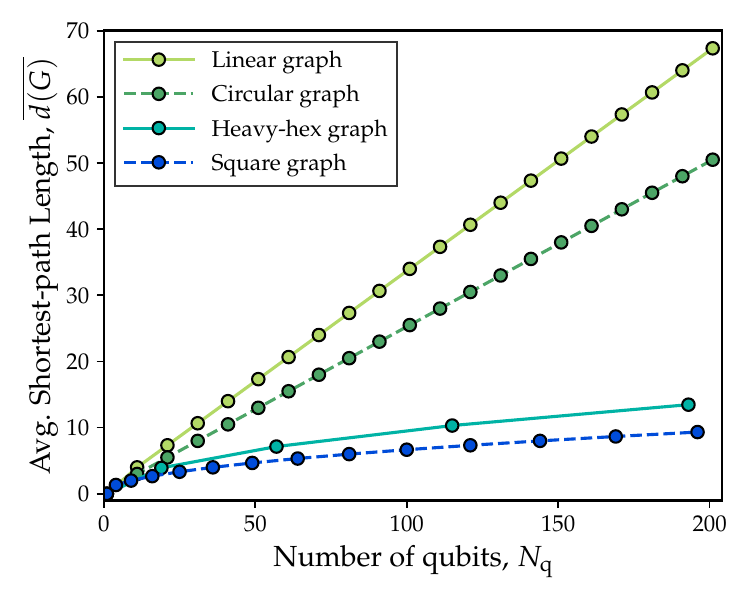}
        \caption{}
    \end{subfigure}
    \caption{Different types of connectivity considered in this work for solid-state-qubits computers. Panel (a) displays the shape of the following connectivity graphs: i) Linear, ii) Circular, iii) Square-lattice, iv) Heavy-hex and v) Fully-connected. Panel (b) shows how the average shortest-path length $\overline{d(G)}$ of these graphs scale with the number of qubits (excluding the fully-connected one, for which $\overline{d(G)}=0$).}
    \label{fig:connectivity_graphs_avg_spl}
\end{figure}

In this work, we used the \texttt{rustworkx} python library to represent different connectivity graphs, and compute $\overline{d(G)}$ exactly for each of them. The graphs that we consider the most relevant for solid-state computing are:

\begin{itemize}
    \item Fully Connected Graphs: In these graphs, all pairs of nodes are connected with an edge. Rather than an actual implementation for solid-state computers, fully connected graphs correspond to an ideal-case scenario in which there is no connectivity constrains at all. This serves us to compare the overhead produced by other constrained graphs with respect to the ideal case, where the circuit does not require any gates. For fully connected graphs, $\overline{d(G)}=d(G)=1$.
    \item Linear Graphs: These graphs correspond to a one-dimensional string of qubits (nodes) connected to their nearest neighbor. This is the most basic form of connectivity that allows all qubits to be able to interact between them, since there is always a path that connects any pair. The shortest paths between nodes of a linear graph can be reduced by connecting the two ends of the graph. This would correspond to a Circular Graph, that we also include in Figure~\ref{fig:connectivity_graphs}. For linear graphs, $\overline{d(G)},d(G)\propto N_{\rm{q}}$.
    \item 2D Lattice Graphs: In general, 2D Lattice Graphs, are the current state-of-the-art approach for scaling solid-state quantum computers. There are several sub-groups of 2D graphs, based on the geometry of the lattice cells. The most common geometries, and the ones that we include in this work, are square lattices and heavy-hex lattices (graphs iii and iv displayed in Panel (b) of Figure~\ref{fig:connectivity_graphs_avg_spl}). These graphs are often optimized for quantum error correction codes~\cite{Het_nyi_2024,vezvaee2025surfacecodescalingheavyhex}.

\end{itemize}

\subsection{Atom-based qubits}
\label{app:coldcomp}

After the transpilation step of an algorithm into a quantum circuit of depth $D_0$, the routing step of the compilation process ensures that gates can be applied to the appropriate qubits. Although it is usually said that atom-based devices have all-to-all connectivity, they still have major parallelization constrains. In general, in an atom-based device the gates can only be performed in a limited number of interaction zones, $N_{\rm{zones}}$, which limits the number of gates that can be executed in parallel. In this work, we propose a simple characterization of circuit compilation for atom-based computers, in order to estimate the extra time taken by devices that have different $N_{\rm{zones}}$.

\begin{figure}[!h]
    \centering
    \includegraphics[width=0.8\linewidth]{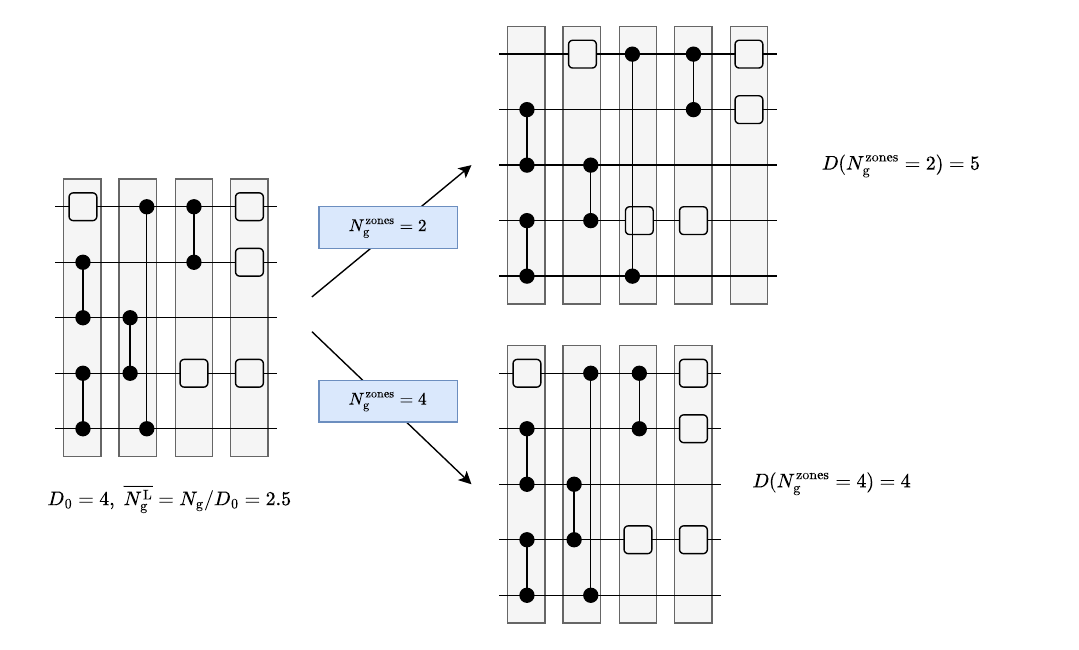}
    \caption{Diagram of a circuit being compiled into two atom-based devices with 2 and 4 gate zones. The horizontal lines represent different qubits, the small squares represent single-qubit gates, and the vertical lines represent two-qubit gates. The layers of the circuits are highlighted with gray vertical rectangles.}
    \label{fig:atom-based_gatezones_diagram}
\end{figure}

In order to execute a quantum circuit on a given device, the number of gates in each layers of the circuit is limited by the number of gate zones. Therefore, all the layers containing more than $N_{\textrm{zones}}$ gates have to be split into more layers, which introduces a depth overhead.  Assuming that large circuits tend to have an even distribution of gates per layer, we can approximate the number of times this happens by using the average number of gates per layer:

\begin{equation}
    \overline{N_{\rm{g}}^{\rm{L}}}=\frac{N_{\rm{g}}}{D_{0}},
\end{equation} 

where $N_{\rm{g}}$ is the total number of gates in the circuit. If the number of gates in the layers is lower than the number of gate zones, the circuit can be executed in the device without adding any extra layers. On the contrary, if there are more gates per layer than the available gate zones, the layers of the circuit will be split into smaller layers with less gates. Figure~\ref{fig:atom-based_gatezones_diagram} shows a diagram of a circuit being compiled into a 2- and a 4-gate-zones device. With this reasoning in mind, the final depth of the circuit reads:
\begin{equation}
    D = 
\left\lceil \frac{\overline{N_{\rm{g}}^{\rm{L}}}}{N_{\rm{zones}}}\right\rceil\cdot D_{0}.
\end{equation}

It is worth mentioning that the final depth of a real circuit could be further optimized for a particular device. For instance, let's say one wants to execute a circuit that has 7 gates per layer on a device with 5 gate zones. According to our description, each layer will be split in two. However, on many cases, the 2 gates that exceed the number of gate zones in some of the layers could be put together to compress the excess overhead. In addition, some circuits might have a large variance in gates per layer, in which case the depth overhead would be uneven through the circuit. However, as the number of gates zones increase, $\overline{N_{\rm{g}}^{\rm{L}}}$ variations also become less relevant, \emph{e.g.} a layer with 21 gates and 39 gates will be both split in two for a device with $N_{\rm{zones}}=20$. For the sake of generality, we assume that the final depth of the circuit will not change in orders of magnitude from our estimations, but we want to stress that architecture-aware compilation processes can provide a certain degree of optimization on the execution time of circuits.

The computing time of a trapped-ions based quantum computer is highly influenced by the shuttling time of ions from zones in the device. In particular, we assume that between all layers, the device has to reconfigure the location of the ions in the traps gate zones, and after a certain number of layers, the ions in the traps have to be exchanged for other ions held in the storage zone. The frequency of storage-to-gate-zone transport depends on the structure of the circuit, the number of qubits being used and the number of gate zones in the device. For instance, a circuit that has several consecutive layers with gates on the same set of qubits, will need less transport than a sparse circuit where most qubits are used sequentially. To characterize the ratio of layers that require ion transport, we define the parameter $\beta_{\rm{trans}}=N_{\rm{trans}}/D$, where $N_{\rm{trans}}$ is the total number of storage-to-gate-zone transportation. Therefore, the time to execute a circuit in an atom-based computer can be written as follows:
\begin{equation}
    t_{\mathcal{\rm circuit}}=t_{\rm{reset}}+D\cdot t_{\rm{clock}}+ \beta_{\rm{trans}}\cdot D \cdot t_{\rm{{transport}}} + t_{\rm{meas}}.
\end{equation}

According to our previous reasoning, $\beta_{\textrm{transport}}$ depends on the structure of the circuit, the number of qubits, and the number of gate zones of the device. For a given circuit, we expect that increasing the number of gate zones by a certain proportion, reduces the number of transports with, at least, the same proportion. For instance, a circuit that has to perform 100 transports in a device with 1 gate zone, will need, at most, 10 transports in a 10-gate-zones device. In that case, the ratio $\beta_{\textrm{transport}}$ will be 10 times smaller for the 10-gate-zones device. However, the relation between  $N_{\rm{trans}}$, $\beta_{\textrm{transport}}$ and the number of gate zones is not always so clear. Let's take a circuit that executes a large number of gates on a set of 10 qubits, and then performs another large number of gates on a different set of 10 qubits. In a 10-gate-zones device, this circuit needs only one transport and, for a big enough depth, has $\beta_{\textrm{transport}}\sim 0$.

Finally, there is a clear lower bound of $\beta_{\textrm{transport}}$ for any circuit with average gates per layer $\overline{N_{\rm{g}}^{\rm{L}}}$, executed on a device with $N_{\rm{zones}}$ gate zones. In a single layer of the original circuit, all the gates are applied to different qubits. Therefore, if $N_{\rm{zones}}<\overline{N_{\rm{g}}^{\rm{L}}}$, the original layers will be split into new layers that use different qubits, with no other option than having to transport them from the storage to the gate zones and vice versa. The minimum ratio of transports per layer is then:
\begin{equation}
    \beta_{\rm{min}}(\overline{N_{\rm{g}}^{\rm{L}}},N_{\rm{zones}})=1-\frac{1}{\left\lceil \frac{\overline{N_{\rm{g}}^{\rm{L}}}}{N_{\rm{zones}}} \right\rceil}
    \label{eq:min_beta}.
\end{equation}

\section{Chip technologies for photonic computing}
\label{app:chiptech}
Programmable photonic chips form the backbone of photonic quantum computing platforms by providing the medium through which gates can be applied to quantum states. Such chips require interconnected waveguides for confining and directing photons to provide a set of orthogonal spatial modes as well as methods of dynamically tuning the coupling strength between these modes. Generally this is accomplished by means of Mach-Zehnder interferometer meshes where mode coupling is tuned by means of electronically controlled phase shifting elements. Different combinations of waveguide and coupling technologies have been mixed and matched to combine the advantages and mitigate the shortcomings of each. In this review we compare examples of these platforms limiting our scope to the technologies which dominate the current literature and industrial efforts. For the purposes of our analysis in Figure \ref{fig:photonic_EE_vs_chip} we compare only those examples which have been explicitly integrated into or designed for programmable interferometer meshes.

A fair comparison between platforms must include electrical power to operate the phase shifters as well as coupling/propagation losses that change the time required to obtain samples, and switching speed, which sets gate initialization overhead. The phase-shifter power is the sum of static and dynamic components where static power maintains a setting and dynamic power updates it. When sampling time is long compared to initialization, the dynamic contribution can be negligible; in large meshes and long sampling windows, static power can dominate, so we focus on it.

Glass-based meshes are compelling for quantum photonics: excess loss as low as 4.35 dB fiber-to-fiber in a 24-mode mesh, fiber coupling losses ~.5 dB/facet, propagation loss below 0.3 dB/cm, ~0.1 dB per MZI unit, polarization invariance, and quasi-3D waveguide fabrication \cite{Barzaghi2025}. The key limitation is that Silicon lacks native optical activity, so phase control relies on either electrically-driven plasma dispersion effects or Joule microheaters using the thermo-optic effect. Plasma dispersion phase shifters induce large and phase-dependent propagation losses making them unsuitable for quantum computing \cite{Rahim2021}. This constrains switching bandwidth to \(< \)1 kHz and creates thermal cross-talk that can require active stabilization, leading to ~2 s algorithm update time. Typical heaters consume ~10–100 mW static power \cite{Barzaghi2025,WANG2024} . Multiplying by the quadratic scaling of tuners in a Clements mesh and integrating over long sampling times can yield significant static energy.

Electro optic (EO) platforms have been the subject of intense development largely driven by telecoms. They use the Pockels effect to induce phase shifts linearly proportional to an applied electric field. The leading EO materials are lithium niobate (LN) and barium titanate (BTO). Both offer high index and wide transparency $n\sim 2.2$, $0.35–5 \mu$m for LN and $n\sim 2.4$, 0.35–5 \(\mu\)m for BTO \cite{Guo2023}. Their EO coefficients -- $r_{33}\approx 26.9$ pm/V for LN and 125 pm/V for BTO -- enable compact, efficient phase shifters \cite{Chelladurai2025}. These phase shifters can be either implemented monolithically or by evanescent coupling with thin films of EO material deposited onto low-loss Silicon-based waveguides \cite{Zhu2021}. Such heterogeneous integration allows the advantages of multiple materials to be combined such as the ultra-low loss of Silicon nitride (SiN) waveguides with the lower energy switching of BTO \cite{Alexander2024}. Device loss, bandwidth, and voltage depend on intricately coupled waveguide/electrode design beyond the scope of this discussion but two dominant device types are traveling-wave (TW) and lumped-element (LE).

In TW devices, electrodes form a transmission line so the RF wave co-propagates with the optical carrier, maximizing overlap and achieving modulation bandwidths \(>\) 100GHz. \cite{Xu2022,Yan2025,Eltes2019}. However, holding a setting requires continuous RF delivery. Even efficient TW designs can incur tens of mW, making them unattractive when sampling times greatly exceed switching times. In LE designs, the device is effectively a capacitor (EO layer as dielectric). Energy is mainly expended dynamically during updates; static consumption is near-zero aside from leakage (often ~100s of nW). State-of-the-art LE LN reports ~15 GHz switching \cite{Yang2024,Wang2018}. At 1 kHz sampling, typical per-device dynamic power can be ~1 \(\mu\)W.

EO phase-shifters are promising for scalable photonic QC, with the main barrier being industrially scalable heterogeneous integration processes; LN is currently most mature, while BTO is advancing rapidly \cite{Alexander2024,Catalalahoz2026,Chrysostomidis2025}. EO platforms may pay higher insertion-loss penalties: LE meshes in monolithic LN report ~0.15 dB per MZI and 0.025 dB/cm propagation, with ~7 dB/facet edge coupling (reduced to 3.4 dB via grating couplers). A 4-mode, mesh dissipated 1.5 mW across 24 phase shifters with ~600 MHz switching \cite{Zheng2023,Sund2023}. Industry BTO-on-SiN reports ~0.1 dB per phase element, 0.127 dB/facet coupling, 0.016 dB/cm propagation, and 6.9 GHz switching, though achievable mode scaling is unclear \cite{Alexander2024}. Another BTO platform reports a 4-mode mesh with 560 nW static per element at 12.5 MHz; scaled to a 4-layer Clements topology, ~14 \(\mu\)W static is expected, but current 1.48 dB/element loss is impractical for quantum computing applications \cite{Catalalahoz2026}.

\bibliographystyle{unsrt}
\bibliography{bibliography}

@article{1st_VQE,
  title={A variational eigenvalue solver on a photonic quantum processor},
  author={Peruzzo, Alberto and McClean, Jarrod and Shadbolt, Peter and Yung, Man-Hong and Zhou, Xiao-Qi and Love, Peter J and Aspuru-Guzik, Al{\'a}n and O’brien, Jeremy L},
  journal={Nat. Commun.},
  volume={5},
  number={1},
  pages={4213},
  year={2014},
   url={https://doi.org/10.1038/ncomms5213},
 doi={10.1038/ncomms5213},
  publisher={Springer Science and Business Media LLC}
}

@article{All_vqe,
   title={Variational quantum algorithms},
   volume={3},
   ISSN={2522-5820},
   url={http://dx.doi.org/10.1038/s42254-021-00348-9},
   DOI={10.1038/s42254-021-00348-9},
   number={9},
   journal={Nat. Rev. Phys.},
   publisher={Springer Science and Business Media LLC},
   author={Cerezo, M. and Arrasmith, Andrew and Babbush, Ryan and Benjamin, Simon C. and Endo, Suguru and Fujii, Keisuke and McClean, Jarrod R. and Mitarai, Kosuke and Yuan, Xiao and Cincio, Lukasz and Coles, Patrick J.},
   year={2021},
   month=aug, pages={625–644} }

@article{Review_VQE,
title = {The Variational Quantum Eigensolver: A review of methods and best practices},
journal = {Phys. Rep.},
volume = {986},
pages = {1-128},
year = {2022},
issn = {0370-1573},
doi = {/10.1016/j.physrep.2022.08.003},
url = {https://www.sciencedirect.com/science/article/pii/S0370157322003118},
author = {Jules Tilly and Hongxiang Chen and Shuxiang Cao and Dario Picozzi and Kanav Setia and Ying Li and Edward Grant and Leonard Wossnig and Ivan Rungger and George H. Booth and Jonathan Tennyson},
}

@article{Ciacco2025,
  title = {Review of Quantum Algorithms for Medicine, Finance and Logistics},
  author = {Ciacco, Alessia and Guerriero, Francesca and Macrina, Giusy},
  year = 2025,
  month = feb,
  journal = {Soft Comput.},
  volume = {29},
  number = {4},
  pages = {2129--2170},
  issn = {1433-7479},
  doi = {10.1007/s00500-025-10540-z},
  urldate = {2026-08-31},
  langid = {english}
}

@misc{Phillipson2025,
  title = {Quantum {{Computing}} in {{Logistics}} and {{Supply Chain Management}} an {{Overview}}},
  author = {Phillipson, Frank},
  year = 2025,
  month = apr,
  number = {arXiv:2402.17520},
  eprint = {2402.17520},
  primaryclass = {quant-ph},
  publisher = {arXiv},
  doi = {10.48550/arXiv.2402.17520},
  urldate = {2026-08-31},
  archiveprefix = {arXiv}
}

@book{Nielsen2010,
  title = {Quantum {{Computation}} and {{Quantum Information}}: 10th {{Anniversary Edition}}},
  shorttitle = {Quantum {{Computation}} and {{Quantum Information}}},
  author = {Nielsen, Michael A. and Chuang, Isaac L.},
  year = 2010,
  month = dec,
  publisher = {Cambridge University Press},
  doi = {10.1017/CBO9780511976667},
  urldate = {2025-10-06},
  isbn = {978-0-511-97666-7},
  langid = {english}
}

@article{ramelow2013,
author = {Sven Ramelow and Alexandra Mech and Marissa Giustina and Simon Gr\"{o}blacher and Witlef Wieczorek and J\"{o}rn Beyer and Adriana Lita and Brice Calkins and Thomas Gerrits and Sae Woo Nam and Anton Zeilinger and Rupert Ursin},
journal = {Opt. Express},
number = {6},
pages = {6707--6717},
publisher = {Optica Publishing Group},
title = {Highly efficient heralding of entangled single photons},
volume = {21},
month = {Mar},
year = {2013},
url = {https://opg.optica.org/oe/abstract.cfm?URI=oe-21-6-6707},
doi = {10.1364/OE.21.006707},
}

@article{xanadu2019source,
  title = {Scalable Squeezed-Light Source for Continuous-Variable Quantum Sampling},
  author = {Vernon, Z. and Quesada, N. and Liscidini, M. and Morrison, B. and Menotti, M. and Tan, K. and Sipe, J.E.},
  journal = {Phys. Rev. Appl.},
  volume = {12},
  issue = {6},
  pages = {064024},
  numpages = {12},
  year = {2019},
  month = {Dec},
  publisher = {American Physical Society},
  doi = {10.1103/PhysRevApplied.12.064024},
  url = {https://link.aps.org/doi/10.1103/PhysRevApplied.12.064024}
}

@article{xanadu2019archi,
  title = {Hybrid spatiotemporal architectures for universal linear optics},
  author = {Su, Daiqin and Dhand, Ish and Helt, Lukas G. and Vernon, Zachary and Br\'adler, Kamil},
  journal = {Phys. Rev. A},
  volume = {99},
  issue = {6},
  pages = {062301},
  numpages = {7},
  year = {2019},
  month = {Jun},
  publisher = {American Physical Society},
  doi = {10.1103/PhysRevA.99.062301},
  url = {https://link.aps.org/doi/10.1103/PhysRevA.99.062301}
}

@misc {milburn2025photonic-qc,
      author = "Jacquiline Romero and Gerard Milburn",
      title = "Photonic Quantum Computing",
      year = "2025",
      month = "07",
      publisher = "Oxford University Press",
      doi = "10.1093/acrefore/9780190871994.013.84",
      url = "https://oxfordre.com/physics/view/10.1093/acrefore/9780190871994.001.0001/acrefore-9780190871994-e-84"
}

@article{klm2001klm,
  title = {A scheme for efficient quantum computation with linear optics},
  author = {Knill, E. and Laflamme, R. and Milburn, G.},
  journal = {Nature},
  volume = {409},
  issue = { },
  pages = {46–52},
  year = {2001},
  doi = {10.1038/35051009},
  url = {https://doi.org/10.1038/35051009}
}

@article{senellart2021bright,
  title = {Bright Polarized Single-Photon Source Based on a Linear Dipole},
  author = {Thomas, S. E. and Billard, M. and Coste, N. and Wein, S. C. and Priya and Ollivier, H. and Krebs, O. and Taza\"{\i}rt, L. and Harouri, A. and Lemaitre, A. and Sagnes, I. and Anton, C. and Lanco, L. and Somaschi, N. and Loredo, J. C. and Senellart, P.},
  journal = {Phys. Rev. Lett.},
  volume = {126},
  issue = {23},
  pages = {233601},
  numpages = {6},
  year = {2021},
  month = {Jun},
  publisher = {American Physical Society},
  doi = {10.1103/PhysRevLett.126.233601},
  url = {https://link.aps.org/doi/10.1103/PhysRevLett.126.233601}
}

@article{francisjones2016,
author = {Robert J. A. Francis-Jones and Rowan A. Hoggarth and Peter J. Mosley},
journal = {Optica},
number = {11},
pages = {1270--1273},
publisher = {Optica Publishing Group},
title = {All-fiber multiplexed source of high-purity single photons},
volume = {3},
month = {Nov},
year = {2016},
url = {https://opg.optica.org/optica/abstract.cfm?URI=optica-3-11-1270},
doi = {10.1364/OPTICA.3.001270},
}

@article{spring2017,
author = {Justin B. Spring and Paolo L. Mennea and Benjamin J. Metcalf and Peter C. Humphreys and James C. Gates and Helen L. Rogers and Christoph S\"{o}ller and Brian J. Smith and W. Steven Kolthammer and Peter G. R. Smith and Ian A. Walmsley},
journal = {Optica},
number = {1},
pages = {90--96},
publisher = {Optica Publishing Group},
title = {Chip-based array of near-identical, pure, heralded single-photon sources},
volume = {4},
month = {Jan},
year = {2017},
url = {https://opg.optica.org/optica/abstract.cfm?URI=optica-4-1-90},
doi = {10.1364/OPTICA.4.000090},
}

@article{kaneda2016,
author = {Fumihiro Kaneda and Karina Garay-Palmett and Alfred B. U'Ren and Paul G. Kwiat},
journal = {Opt. Express},
number = {10},
pages = {10733--10747},
publisher = {Optica Publishing Group},
title = {Heralded single-photon source utilizing highly nondegenerate, spectrally factorable spontaneous parametric downconversion},
volume = {24},
month = {May},
year = {2016},
url = {https://opg.optica.org/oe/abstract.cfm?URI=oe-24-10-10733},
doi = {10.1364/OE.24.010733},
}

@article{weston2016,
author = {Morgan M. Weston and Helen M. Chrzanowski and Sabine Wollmann and Allen Boston and Joseph Ho and Lynden K. Shalm and Varun B. Verma and Michael S. Allman and Sae Woo Nam and Raj B. Patel and Sergei Slussarenko and Geoff J. Pryde},
journal = {Opt. Express},
number = {10},
pages = {10869--10879},
publisher = {Optica Publishing Group},
title = {Efficient and pure femtosecond-pulse-length source of polarization-entangled photons},
volume = {24},
month = {May},
year = {2016},
url = {https://opg.optica.org/oe/abstract.cfm?URI=oe-24-10-10869},
doi = {10.1364/OE.24.010869},
}

@article{Acharya2025,
  title = {Quantum Error Correction below the Surface Code Threshold},
  author = {Acharya, Rajeev and Abanin, Dmitry A. and {Aghababaie-Beni}, Laleh and Aleiner, Igor and Andersen, Trond I. and Ansmann, Markus and others},
  year = {2025},
  month = feb,
  journal = {Nature},
  volume = {638},
  number = {8052},
  pages = {920--926},
  publisher = {Nature Publishing Group},
  issn = {1476-4687},
  doi = {10.1038/s41586-024-08449-y},
  urldate = {2025-06-05},
  copyright = {2024 The Author(s)},
  langid = {english}
}

@article{Huang2020,
  title = {Superconducting {{Quantum Computing}}: {{A Review}}},
  shorttitle = {Superconducting {{Quantum Computing}}},
  author = {Huang, He-Liang and Wu, Dachao and Fan, Daojin and Zhu, Xiaobo},
  year = 2020,
  month = aug,
  journal = {Sci. China Inf. Sci.},
  volume = {63},
  number = {8},
  eprint = {2006.10433},
  primaryclass = {quant-ph},
  pages = {180501},
  issn = {1674-733X, 1869-1919},
  doi = {10.1007/s11432-020-2881-9},
  urldate = {2025-02-26},
  archiveprefix = {arXiv}
}

@article{Blais2004,
  title = {Cavity Quantum Electrodynamics for Superconducting Electrical Circuits: {{An}} Architecture for Quantum Computation},
  shorttitle = {Cavity Quantum Electrodynamics for Superconducting Electrical Circuits},
  author = {Blais, Alexandre and Huang, Ren-Shou and Wallraff, Andreas and Girvin, S. M. and Schoelkopf, R. J.},
  year = {2004},
  month = jun,
  journal = {Phys. Rev. A},
  volume = {69},
  number = {6},
  pages = {062320},
  publisher = {American Physical Society},
  doi = {10.1103/PhysRevA.69.062320},
  urldate = {2025-06-25}
}

@article{Dewes2012,
  title = {Characterization of a {{Two-Transmon Processor}} with {{Individual Single-Shot Qubit Readout}}},
  author = {Dewes, A. and Ong, F. R. and Schmitt, V. and Lauro, R. and Boulant, N. and Bertet, P. and Vion, D. and Esteve, D.},
  year = {2012},
  month = feb,
  journal = {Phys. Rev. Lett.},
  volume = {108},
  number = {5},
  pages = {057002},
  publisher = {American Physical Society},
  doi = {10.1103/PhysRevLett.108.057002},
  urldate = {2025-06-25}
}

@article{Ezratty2023,
  title = {Perspective on Superconducting Qubit Quantum Computing},
  author = {Ezratty, Olivier},
  year = {2023},
  month = may,
  journal = {Eur. Phys. J. A},
  volume = {59},
  number = {5},
  pages = {1--18},
  publisher = {Springer Berlin Heidelberg},
  issn = {1434-601X},
  doi = {10.1140/epja/s10050-023-01006-7},
  urldate = {2025-06-03},
  copyright = {2023 The Author(s), under exclusive licence to Societ{\`a} Italiana di Fisica and Springer-Verlag GmbH Germany, part of Springer Nature},
  langid = {english}
}

@article{Girvin2009,
  title = {Circuit {{QED}} and Engineering Charge-Based Superconducting Qubits},
  author = {Girvin, S M and Devoret, M H and Schoelkopf, R J},
  year = {2009},
  month = dec,
  journal = {Phys. Scr.},
  volume = {2009},
  number = {T137},
  pages = {014012},
  issn = {1402-4896},
  doi = {10.1088/0031-8949/2009/T137/014012},
  urldate = {2025-06-25},
  langid = {english}
}

@article{Bouchiat1998,
  title = {Quantum Coherence with a Single {{Cooper}} Pair},
  author = {Bouchiat, V. and Vion, D. and Joyez, P. and Esteve, D. and Devoret, M. H.},
  year = 1998,
  month = jan,
  journal = {Phys. Scr.},
  volume = {1998},
  number = {T76},
  pages = {165},
  publisher = {IOP Publishing},
  issn = {1402-4896},
  doi = {10.1238/Physica.Topical.076a00165},
  urldate = {2026-07-21},
  langid = {english}
}

@article{Nakamura1999,
  title = {Coherent Control of Macroscopic Quantum States in a Single-{{Cooper-pair}} Box},
  author = {Nakamura, Y. and Pashkin, Yu A. and Tsai, J. S.},
  year = 1999,
  month = apr,
  journal = {Nature},
  volume = {398},
  number = {6730},
  pages = {786--788},
  publisher = {Nature Publishing Group},
  issn = {1476-4687},
  doi = {10.1038/19718},
  urldate = {2026-07-21},
  copyright = {1999 Macmillan Magazines Ltd.},
  langid = {english}
}

@misc{QMIO,
      title={QMIO: A tightly integrated hybrid HPCQC system}, 
      author={Javier Cacheiro and Álvaro C Sánchez and Russell Rundle and George B Long and Gavin Dold and Jamie Friel and Andrés Gómez},
      year={2025},
      eprint={2505.19267},
      archivePrefix={arXiv},
      primaryClass={quant-ph},
      url={https://arxiv.org/abs/2505.19267}, 
}

@article{Jin2015,
  title = {Thermal and {{Residual Excited-State Population}} in a {{3D Transmon Qubit}}},
  author = {Jin, X. Y. and Kamal, A. and Sears, A. P. and Gudmundsen, T. and Hover, D. and Miloshi, J. and Slattery, R. and Yan, F. and Yoder, J. and Orlando, T. P. and Gustavsson, S. and Oliver, W. D.},
  year = {2015},
  month = jun,
  journal = {Phys. Rev. Lett.},
  volume = {114},
  number = {24},
  pages = {240501},
  issn = {0031-9007, 1079-7114},
  doi = {10.1103/PhysRevLett.114.240501},
  urldate = {2025-02-27},
  copyright = {http://link.aps.org/licenses/aps-default-license},
  langid = {english}
}

@article{Josephson1974,
  title = {The Discovery of Tunnelling Supercurrents},
  author = {Josephson, B. D.},
  year = {1974},
  month = apr,
  journal = {Rev. Mod. Phys.},
  volume = {46},
  number = {2},
  pages = {251--254},
  publisher = {American Physical Society},
  doi = {10.1103/RevModPhys.46.251},
  urldate = {2025-06-05}
}

@article{Kim2023,
  title = {Evidence for the Utility of Quantum Computing before Fault Tolerance},
  author = {Kim, Youngseok and Eddins, Andrew and Anand, Sajant and Wei, Ken Xuan and {van den Berg}, Ewout and Rosenblatt, Sami and Nayfeh, Hasan and Wu, Yantao and Zaletel, Michael and Temme, Kristan and Kandala, Abhinav},
  year = {2023},
  month = jun,
  journal = {Nature},
  volume = {618},
  number = {7965},
  pages = {500--505},
  publisher = {Nature Publishing Group},
  issn = {1476-4687},
  doi = {10.1038/s41586-023-06096-3},
  urldate = {2025-06-05},
  copyright = {2023 The Author(s)},
  langid = {english}
}

@article{King2025,
  title = {Beyond-Classical Computation in Quantum Simulation},
  author = {King, Andrew D. and Nocera, Alberto and Rams, Marek M. and Dziarmaga, Jacek and Wiersema, Roeland and others},
  year = {2025},
  month = apr,
  journal = {Science},
  volume = {388},
  number = {6743},
  pages = {199--204},
  issn = {0036-8075, 1095-9203},
  doi = {10.1126/science.ado6285},
  urldate = {2025-06-25},
  langid = {english}
}

@article{Kjaergaard2020,
  title = {Superconducting {{Qubits}}: {{Current State}} of {{Play}}},
  shorttitle = {Superconducting {{Qubits}}},
  author = {Kjaergaard, Morten and Schwartz, Mollie E. and Braum{\"u}ller, Jochen and Krantz, Philip and Wang, Joel I.-Jan and Gustavsson, Simon and Oliver, William D.},
  year = {2020},
  month = mar,
  journal = {Annu. Rev. Condens. Matter Phys.},
  volume = {11},
  number = {1},
  eprint = {1905.13641},
  primaryclass = {quant-ph},
  pages = {369--395},
  issn = {1947-5454, 1947-5462},
  doi = {10.1146/annurev-conmatphys-031119-050605},
  urldate = {2025-02-26},
  archiveprefix = {arXiv}
}

@article{Zu2022,
  title = {Development of Dilution Refrigerators---{{A}} Review},
  author = {Zu, H. and Dai, W. and De Waele, A.T.A.M.},
  year = {2022},
  month = jan,
  journal = {Cryogenics},
  volume = {121},
  pages = {103390},
  issn = {00112275},
  doi = {10.1016/j.cryogenics.2021.103390},
  urldate = {2025-03-18},
  langid = {english}
}

@article{postquantum_sc_qubits,
  title     = {Quantum Computing Modalities: Superconducting Qubits},
  journal   = {PostQuantum},
  year      = {2025},
  url       = {https://postquantum.com/quantum-modalities/superconducting-qubits/},
  note      = {Accessed March 2026},
  author    = {Ivezic, Marin}
}

@article{transmon_qubit,
   title={Transmon qubit readout fidelity at the threshold for quantum error correction without a quantum-limited amplifier},
   volume={9},
   ISSN={2056-6387},
   url={http://dx.doi.org/10.1038/s41534-023-00689-6},
   DOI={10.1038/s41534-023-00689-6},
   number={1},
   journal={npj Quantum Information},
   publisher={Springer Science and Business Media LLC},
   author={Chen, Liangyu and Li, Hang-Xi and Lu, Yong and Warren, Christopher W. and Križan, Christian J. and Kosen, Sandoko and Rommel, Marcus and Ahmed, Shahnawaz and Osman, Amr and Biznárová, Janka and Fadavi Roudsari, Anita and Lienhard, Benjamin and Caputo, Marco and Grigoras, Kestutis and Grönberg, Leif and Govenius, Joonas and Kockum, Anton Frisk and Delsing, Per and Bylander, Jonas and Tancredi, Giovanna},
   year={2023},
   month=mar }

@misc{qblox2025activereset,
  title     = {Speeding up your experiment with active reset of your qubit},
  author    = {Qblox},
  year      = {2025},
  url       = {https://qblox.com/newsroom/speeding-up-your-experiment-with-active-reset-of-your-qubit},
  note      = {Accessed March 2026}
}

@misc{bluefors_ghs2,
  author = {{Bluefors}},
  title  = {Gas Handling System Generation 2},
  year   = {2024},
  url    = {https://bluefors.com/products/dilution-refrigerator-measurement-systems/gas-handling-system/},
  note   = {Accessed March 2026}
}

@inproceedings{cha2020hemt,
    author = {Cha, Eunjung and Wadefalk, Niklas and Moschetti, Giuseppe and Pourkabirian, Arsalan and Stenarson, Jörgen and Grahn, Jenny},
    year = {2020},
    month = {08},
    pages = {1299-1302},
    title = {A 300-µW Cryogenic HEMT LNA for Quantum Computing},
    doi = {10.1109/IMS30576.2020.9223865}
    }

@article{Fluxonium_1,
   title={Fluxonium: Single Cooper-Pair Circuit Free of Charge Offsets},
   volume={326},
   ISSN={1095-9203},
   url={http://dx.doi.org/10.1126/science.1175552},
   DOI={10.1126/science.1175552},
   number={5949},
   journal={Science},
   publisher={American Association for the Advancement of Science (AAAS)},
   author={Manucharyan, Vladimir E. and Koch, Jens and Glazman, Leonid I. and Devoret, Michel H.},
   year={2009},
   month=Oct, pages={113–116} }

@article{Fluxonium_2,
   title={Fluxonium: An Alternative Qubit Platform for High-Fidelity Operations},
   volume={129},
   ISSN={1079-7114},
   url={http://dx.doi.org/10.1103/PhysRevLett.129.010502},
   DOI={10.1103/physrevlett.129.010502},
   number={1},
   journal={Physical Review Letters},
   publisher={American Physical Society (APS)},
   author={Bao, Feng and Deng, Hao and Ding, Dawei and Gao, Ran and Gao, Xun and Huang, Cupjin and Jiang, Xun and Ku, Hsiang-Sheng and Li, Zhisheng and Ma, Xizheng and Ni, Xiaotong and Qin, Jin and Song, Zhijun and Sun, Hantao and Tang, Chengchun and Wang, Tenghui and Wu, Feng and Xia, Tian and Yu, Wenlong and Zhang, Fang and Zhang, Gengyan and Zhang, Xiaohang and Zhou, Jingwei and Zhu, Xing and Shi, Yaoyun and Chen, Jianxin and Zhao, Hui-Hai and Deng, Chunqing},
   year={2022},
   month=June }

@article{Elzerman2003,
  title = {Few-Electron Quantum Dot Circuit with Integrated Charge Read Out},
  author = {Elzerman, J. M. and Hanson, R. and Greidanus, J. S. and Willems Van Beveren, L. H. and De Franceschi, S. and Vandersypen, L. M. K. and Tarucha, S. and Kouwenhoven, L. P.},
  year = 2003,
  month = apr,
  journal = {Phys. Rev. B},
  volume = {67},
  number = {16},
  pages = {161308},
  issn = {0163-1829, 1095-3795},
  doi = {10.1103/PhysRevB.67.161308},
  urldate = {2026-07-21},
  copyright = {http://link.aps.org/licenses/aps-default-license},
  langid = {english}
}

@article{Ha2022,
  title = {A Flexible Design Platform for {{Si}}/{{SiGe}} Exchange-Only Qubits with Low Disorder},
  author = {Ha, Wonill and Ha, Sieu D. and Choi, Maxwell D. and Tang, Yan and Schmitz, Adele E. and Levendorf, Mark P. and Lee, Kangmu and Chappell, James M. and Adams, Tower S. and Hulbert, Daniel R. and Acuna, Edwin and Noah, Ramsey S. and Matten, Justine W. and Jura, Michael P. and Wright, Jeffrey A. and Rakher, Matthew T. and Borselli, Matthew G.},
  year = 2022,
  month = feb,
  journal = {Nano Lett.},
  volume = {22},
  number = {3},
  eprint = {2107.10916},
  primaryclass = {cond-mat.mes-hall},
  pages = {1443--1448},
  issn = {1530-6984, 1530-6992},
  doi = {10.1021/acs.nanolett.1c03026},
  urldate = {2026-07-21},
  archiveprefix = {arXiv}
}

@article{Volk2019,
  title = {Loading a Quantum-Dot Based ``{{Qubyte}}'' Register},
  author = {Volk, C. and Zwerver, A. M. J. and Mukhopadhyay, U. and Eendebak, P. T. and {van Diepen}, C. J. and Dehollain, J. P. and Hensgens, T. and Fujita, T. and Reichl, C. and Wegscheider, W. and Vandersypen, L. M. K.},
  year = 2019,
  month = apr,
  journal = {Npj Quantum Inf.},
  volume = {5},
  number = {1},
  pages = {29},
  publisher = {Nature Publishing Group},
  issn = {2056-6387},
  doi = {10.1038/s41534-019-0146-y},
  urldate = {2026-07-21},
  copyright = {2019 The Author(s)},
  langid = {english}
}

@article{Zajac2016,
  title = {Scalable Gate Architecture for Densely Packed Semiconductor Spin Qubits},
  author = {Zajac, D. M. and Hazard, T. M. and Mi, X. and Nielsen, E. and Petta, J. R.},
  year = 2016,
  month = nov,
  journal = {Phys. Rev. Appl.},
  volume = {6},
  number = {5},
  eprint = {1607.07025},
  primaryclass = {cond-mat.mes-hall},
  pages = {054013},
  issn = {2331-7019},
  doi = {10.1103/PhysRevApplied.6.054013},
  urldate = {2026-07-21},
  archiveprefix = {arXiv}
}

@article{Loss1998,
  title = {Quantum computation with quantum dots},
  author = {Loss, Daniel and DiVincenzo, David P.},
  journal = {Phys. Rev. A},
  volume = {57},
  issue = {1},
  pages = {120--126},
  numpages = {0},
  year = {1998},
  month = {Jan},
  publisher = {American Physical Society},
  doi = {10.1103/PhysRevA.57.120},
  url = {https://link.aps.org/doi/10.1103/PhysRevA.57.120}
}

@article{kane1998,
	title = {A silicon-based nuclear spin quantum computer},
	volume = {393},
	copyright = {1998 Macmillan Magazines Ltd.},
	issn = {1476-4687},
	url = {https://www.nature.com/articles/30156},
	doi = {10.1038/30156},
	language = {en},
	number = {6681},
	urldate = {2025-01-08},
	journal = {Nature},
	author = {Kane, B. E.},
	month = may,
	year = {1998},
	note = {Publisher: Nature Publishing Group},
	pages = {133--137},
}

@article{stano2022,
	title = {Review of performance metrics of spin qubits in gated semiconducting nanostructures},
	volume = {4},
	copyright = {2022 Springer Nature Limited},
	issn = {2522-5820},
	url = {https://www.nature.com/articles/s42254-022-00484-w},
	doi = {10.1038/s42254-022-00484-w},
	language = {en},
	number = {10},
	urldate = {2025-01-02},
	journal = {Nature Reviews Physics},
	author = {Stano, Peter and Loss, Daniel},
	month = oct,
	year = {2022},
	note = {Publisher: Nature Publishing Group},
	pages = {672--688},
}

@article{levy2002,
	title = {Universal {Quantum} {Computation} with {Spin}-\$1/2\$ {Pairs} and {Heisenberg} {Exchange}},
	volume = {89},
	url = {https://link.aps.org/doi/10.1103/PhysRevLett.89.147902},
	doi = {10.1103/PhysRevLett.89.147902},
	number = {14},
	urldate = {2025-01-08},
	journal = {Physical Review Letters},
	author = {Levy, Jeremy},
	month = sep,
	year = {2002},
	note = {Publisher: American Physical Society},
	pages = {147902},
}

@article{divincenzo2000,
	title = {Universal quantum computation with the exchange interaction},
	volume = {408},
	copyright = {2000 Macmillan Magazines Ltd.},
	issn = {1476-4687},
	url = {https://www.nature.com/articles/35042541},
	doi = {10.1038/35042541},
	language = {en},
	number = {6810},
	urldate = {2025-01-08},
	journal = {Nature},
	author = {DiVincenzo, D. P. and Bacon, D. and Kempe, J. and Burkard, G. and Whaley, K. B.},
	month = nov,
	year = {2000},
	note = {Publisher: Nature Publishing Group},
	pages = {339--342},
}

@article{maurand2016,
	title = {A {CMOS} silicon spin qubit},
	volume = {7},
	copyright = {2016 The Author(s)},
	issn = {2041-1723},
	url = {https://www.nature.com/articles/ncomms13575},
	doi = {10.1038/ncomms13575},
	language = {en},
	number = {1},
	urldate = {2022-06-21},
	journal = {Nature Communications},
	author = {Maurand, R. and Jehl, X. and Kotekar-Patil, D. and Corna, A. and Bohuslavskyi, H. and Laviéville, R. and Hutin, L. and Barraud, S. and Vinet, M. and Sanquer, M. and De Franceschi, S.},
	month = nov,
	year = {2016},
	note = {Number: 1
Publisher: Nature Publishing Group},
	pages = {13575},
}

@misc{Undseth2026,
  title = {Weight-Four Parity Checks with Silicon Spin Qubits},
  author = {Undseth, Brennan and Meggiato, Nicola and Wu, Yi-Hsien and {Katiraee-Far}, Sam R. and Tryputen, Larysa and de Snoo, Sander L. and Esposti, Davide Degli and Scappucci, Giordano and Greplov{\'a}, Eli{\v s}ka and Vandersypen, Lieven M. K.},
  year = 2026,
  month = jan,
  number = {arXiv:2601.23267},
  eprint = {2601.23267},
  primaryclass = {cond-mat},
  publisher = {arXiv},
  doi = {10.48550/arXiv.2601.23267},
  urldate = {2026-04-30},
  archiveprefix = {arXiv}
}

@article{pioro-ladriere2008,
	title = {Electrically driven single-electron spin resonance in a slanting {Zeeman} field},
	volume = {4},
	copyright = {2008 Springer Nature Limited},
	issn = {1745-2481},
	url = {https://www.nature.com/articles/nphys1053},
	doi = {10.1038/nphys1053},
	language = {en},
	number = {10},
	urldate = {2024-10-29},
	journal = {Nature Physics},
	author = {Pioro-Ladrière, M. and Obata, T. and Tokura, Y. and Shin, Y.-S. and Kubo, T. and Yoshida, K. and Taniyama, T. and Tarucha, S.},
	month = oct,
	year = {2008},
	note = {Publisher: Nature Publishing Group},
	pages = {776--779},
}

@article{veldhorst2015,
	title = {A two-qubit logic gate in silicon},
	volume = {526},
	doi = {10.1038/nature15263},
	number = {7573},
	journal = {Nature},
	author = {Veldhorst, M. and Yang, C. H. and Hwang, J. C. C. and Huang, W. and Dehollain, J. P. and Muhonen, J. T. and Simmons, S. and Laucht, A. and Hudson, F. E. and Itoh, K. M. and Morello, A. and Dzurak, A. S.},
	month = oct,
	year = {2015},
	note = {Publisher: Springer Science and Business Media LLC},
	pages = {410--414},
}

@article{koppens2006,
	title = {Driven coherent oscillations of a single electron spin in a quantum dot},
	volume = {442},
	issn = {1476-4687},
	url = {https://doi.org/10.1038/nature05065},
	doi = {10.1038/nature05065},
	number = {7104},
	journal = {Nature},
	author = {Koppens, F. H. L. and Buizert, C. and Tielrooij, K. J. and Vink, I. T. and Nowack, K. C. and Meunier, T. and Kouwenhoven, L. P. and Vandersypen, L. M. K.},
	year = {2006},
	pages = {766--771},
}

@article{dijkema2025,
	title = {Cavity-mediated {iSWAP} oscillations between distant spins},
	volume = {21},
	copyright = {2024 The Author(s)},
	issn = {1745-2481},
	url = {https://www.nature.com/articles/s41567-024-02694-8},
	doi = {10.1038/s41567-024-02694-8},
	language = {en},
	number = {1},
	urldate = {2025-01-23},
	journal = {Nature Physics},
	author = {Dijkema, Jurgen and Xue, Xiao and Harvey-Collard, Patrick and Rimbach-Russ, Maximilian and de Snoo, Sander L. and Zheng, Guoji and Sammak, Amir and Scappucci, Giordano and Vandersypen, Lieven M. K.},
	month = jan,
	year = {2025},
	note = {Publisher: Nature Publishing Group},
	pages = {168--174},
}

@misc{hamonic2025,
	title = {A foundry-fabricated spin qubit unit cell with in-situ dispersive readout},
	url = {http://arxiv.org/abs/2504.20572},
	doi = {10.48550/arXiv.2504.20572},
	urldate = {2025-10-10},
	publisher = {arXiv},
	author = {Hamonic, Pierre and Toubeix, Mathieu and Haas, Guillermo and Nath, Jayshankar and Dartiailh, Matthieu C. and Martinez, Biel and Bertrand, Benoit and Niebojewski, Heimanu and Vinet, Maud and Bäuerle, Christopher and Balestro, Franck and Meunier, Tristan and Urdampilleta, Matias},
	month = apr,
	year = {2025},
	note = {arXiv:2504.20572 [cond-mat]},
}

@article{george2025,
	title = {12-{Spin}-{Qubit} {Arrays} {Fabricated} on a 300 mm {Semiconductor} {Manufacturing} {Line}},
	volume = {25},
	issn = {1530-6984},
	url = {https://doi.org/10.1021/acs.nanolett.4c05205},
	doi = {10.1021/acs.nanolett.4c05205},
	number = {2},
	urldate = {2025-05-06},
	journal = {Nano Letters},
	author = {George, Hubert C. and Mądzik, Mateusz T. and Henry, Eric M. and Wagner, Andrew J. and Islam, Mohammad M. and Borjans, Felix and Connors, Elliot J. and Corrigan, J. and Curry, Matthew and Harper, Michael K. and Keith, Daniel and Lampert, Lester and Luthi, Florian and Mohiyaddin, Fahd A. and Murcia, Sandra and Nair, Rohit and Nahm, Rambert and Nethwewala, Aditi and Neyens, Samuel and Patra, Bishnu and Raharjo, Roy D. and Rogan, Carly and Savytskyy, Rostyslav and Watson, Thomas F. and Ziegler, Josh and Zietz, Otto K. and Pellerano, Stefano and Pillarisetty, Ravi and Bishop, Nathaniel C. and Bojarski, Stephanie A. and Roberts, Jeanette and Clarke, James S.},
	month = jan,
	year = {2025},
	note = {Publisher: American Chemical Society},
	pages = {793--799},
}

@article{veldhorst2014,
	title = {An addressable quantum dot qubit with fault-tolerant control-fidelity},
	volume = {9},
	issn = {1748-3395},
	url = {https://doi.org/10.1038/nnano.2014.216},
	doi = {10.1038/nnano.2014.216},
	number = {12},
	journal = {Nature Nanotechnology},
	author = {Veldhorst, M. and Hwang, J. C. C. and Yang, C. H. and Leenstra, A. W. and de Ronde, B. and Dehollain, J. P. and Muhonen, J. T. and Hudson, F. E. and Itoh, K. M. and Morello, A. and Dzurak, A. S.},
	year = {2014},
	pages = {981--985},
}

@article{xue2022,
	title = {Quantum logic with spin qubits crossing the surface code threshold},
	volume = {601},
	copyright = {2022 The Author(s)},
	issn = {1476-4687},
	url = {https://www.nature.com/articles/s41586-021-04273-w},
	doi = {10.1038/s41586-021-04273-w},
	language = {en},
	number = {7893},
	urldate = {2025-01-23},
	journal = {Nature},
	author = {Xue, Xiao and Russ, Maximilian and Samkharadze, Nodar and Undseth, Brennan and Sammak, Amir and Scappucci, Giordano and Vandersypen, Lieven M. K.},
	month = jan,
	year = {2022},
	note = {Publisher: Nature Publishing Group},
	pages = {343--347},
}

@article{noiri2022,
	title = {Fast universal quantum gate above the fault-tolerance threshold in silicon},
	volume = {601},
	copyright = {2022 The Author(s), under exclusive licence to Springer Nature Limited},
	issn = {1476-4687},
	url = {https://www.nature.com/articles/s41586-021-04182-y},
	doi = {10.1038/s41586-021-04182-y},
	language = {en},
	number = {7893},
	urldate = {2025-01-23},
	journal = {Nature},
	author = {Noiri, Akito and Takeda, Kenta and Nakajima, Takashi and Kobayashi, Takashi and Sammak, Amir and Scappucci, Giordano and Tarucha, Seigo},
	month = jan,
	year = {2022},
	note = {Publisher: Nature Publishing Group},
	pages = {338--342},
}

@article{mills2022,
	title = {Two-qubit silicon quantum processor with operation fidelity exceeding 99\%},
	volume = {8},
	url = {https://www.science.org/doi/10.1126/sciadv.abn5130},
	doi = {10.1126/sciadv.abn5130},
	number = {14},
	urldate = {2025-01-23},
	journal = {Science Advances},
	author = {Mills, Adam R. and Guinn, Charles R. and Gullans, Michael J. and Sigillito, Anthony J. and Feldman, Mayer M. and Nielsen, Erik and Petta, Jason R.},
	month = apr,
	year = {2022},
	note = {Publisher: American Association for the Advancement of Science},
	pages = {eabn5130},
}

@article{yoneda2018,
	title = {A quantum-dot spin qubit with coherence limited by charge noise and fidelity higher than 99.9\%},
	volume = {13},
	copyright = {2017 The Author(s)},
	issn = {1748-3395},
	url = {https://www.nature.com/articles/s41565-017-0014-x},
	doi = {10.1038/s41565-017-0014-x},
	language = {en},
	number = {2},
	urldate = {2024-09-16},
	journal = {Nature Nanotechnology},
	author = {Yoneda, Jun and Takeda, Kenta and Otsuka, Tomohiro and Nakajima, Takashi and Delbecq, Matthieu R. and Allison, Giles and Honda, Takumu and Kodera, Tetsuo and Oda, Shunri and Hoshi, Yusuke and Usami, Noritaka and Itoh, Kohei M. and Tarucha, Seigo},
	month = feb,
	year = {2018},
	note = {Publisher: Nature Publishing Group},
	pages = {102--106},
}

@article{john2025,
	title = {A two-dimensional 10-qubit array in germanium with robust and localised qubit control},
	url = {http://arxiv.org/abs/2412.16044},
	doi = {10.48550/arXiv.2412.16044},
	urldate = {2025-09-05},
	author = {John, Valentin and Yu, Cécile X. and Straaten, Barnaby van and Rodríguez-Mena, Esteban A. and Rodríguez, Mauricio and Oosterhout, Stefan and Stehouwer, Lucas E. A. and Scappucci, Giordano and Bosco, Stefano and Rimbach-Russ, Maximilian and Niquet, Yann-Michel and Borsoi, Francesco and Veldhorst, Menno},
	month = feb,
	year = {2025},
	note = {arXiv:2412.16044 [cond-mat]},
}

@article{camenzind2022,
	title = {A hole spin qubit in a fin field-effect transistor above 4 kelvin},
	volume = {5},
	copyright = {2022 The Author(s), under exclusive licence to Springer Nature Limited},
	issn = {2520-1131},
	url = {https://www.nature.com/articles/s41928-022-00722-0},
	doi = {10.1038/s41928-022-00722-0},
	language = {en},
	number = {3},
	urldate = {2023-05-15},
	journal = {Nature Electronics},
	author = {Camenzind, Leon C. and Geyer, Simon and Fuhrer, Andreas and Warburton, Richard J. and Zumbühl, Dominik M. and Kuhlmann, Andreas V.},
	month = mar,
	year = {2022},
	note = {Number: 3
Publisher: Nature Publishing Group},
	pages = {178--183},
}

@article{petersson2012,
	title = {Circuit quantum electrodynamics with a spin qubit},
	volume = {490},
	copyright = {2012 Springer Nature Limited},
	issn = {1476-4687},
	url = {https://www.nature.com/articles/nature11559},
	doi = {10.1038/nature11559},
	language = {en},
	number = {7420},
	urldate = {2023-12-14},
	journal = {Nature},
	author = {Petersson, K. D. and McFaul, L. W. and Schroer, M. D. and Jung, M. and Taylor, J. M. and Houck, A. A. and Petta, J. R.},
	month = oct,
	year = {2012},
	note = {Number: 7420
Publisher: Nature Publishing Group},
	pages = {380--383},
}

@article{li2018,
author = {Ruoyu Li  and Luca Petit  and David P. Franke  and Juan Pablo Dehollain  and Jonas Helsen  and Mark Steudtner  and Nicole K. Thomas  and Zachary R. Yoscovits  and Kanwal J. Singh  and Stephanie Wehner  and Lieven M. K. Vandersypen  and James S. Clarke  and Menno Veldhorst },
title = {A crossbar network for silicon quantum dot qubits},
journal = {Science Advances},
volume = {4},
number = {7},
pages = {eaar3960},
year = {2018},
doi = {10.1126/sciadv.aar3960},
URL = {https://www.science.org/doi/abs/10.1126/sciadv.aar3960},
eprint = {https://www.science.org/doi/pdf/10.1126/sciadv.aar3960}}

@article{Qandela,
    author = {Maring, N. and Fyrillas, A. and Pont, M.  et al. },
    title = {A versatile single-photon-based quantum computing platform} ,
    journal = {Nature Photonics},
    year = {2024},
    doi = {https://doi.org/10.1038/s41566-024-01403-4}
}

@article{FBQC,
  title = {Fusion-Based Quantum Computation},
  author = {Bartolucci, Sara and Birchall, Patrick and Bombín, Hector and Cable, Hugo and Dawson, Chris and Gimeno-Segovia, Mercedes and Johnston, Eric and Kieling, Konrad and Nickerson, Naomi and Pant, Mihir and Pastawski, Fernando and Rudolph, Terry and Sparrow, Chris},
  date = {2023-02-17},
  journaltitle = {Nature Communications},
  shortjournal = {Nat Commun},
  volume = {14},
  number = {1},
  pages = {912},
  publisher = {Nature Publishing Group},
  issn = {2041-1723},
  doi = {10.1038/s41467-023-36493-1},
  url = {https://www.nature.com/articles/s41467-023-36493-1},
  urldate = {2025-06-25}
}

@article{garcia2023photonicqrc,
  title = {Scalable Photonic Platform for Real-Time Quantum Reservoir Computing},
  author = {Garc\'{\i}a-Beni, Jorge and Giorgi, Gian Luca and Soriano, Miguel C. and Zambrini, Roberta},
  journal = {Phys. Rev. Appl.},
  volume = {20},
  issue = {1},
  pages = {014051},
  numpages = {20},
  year = {2023},
  month = {Jul},
  publisher = {American Physical Society},
  doi = {10.1103/PhysRevApplied.20.014051},
  url = {https://link.aps.org/doi/10.1103/PhysRevApplied.20.014051}
}

@article{Monbroussou2025photonicqcnn,
   title={Photonic quantum convolutional neural networks with adaptive state injection},
   volume={7},
   ISSN={2577-5421},
   url={http://dx.doi.org/10.1117/1.AP.7.6.066012},
   DOI={10.1117/1.ap.7.6.066012},
   number={06},
   journal={Advanced Photonics},
   publisher={SPIE-Intl Soc Optical Eng},
   author={Monbroussou, Léo and Polacchi, Beatrice and Yacoub, Verena and Caruccio, Eugenio and Rodari, Giovanni and Hoch, Francesco and Carvacho, Gonzalo and Spagnolo, Nicolò and Giordani, Taira and Bossi, Mattia and Rajan, Abhiram and Di Giano, Niki and Albiero, Riccardo and Ceccarelli, Francesco and Osellame, Roberto and Kashefi, Elham and Sciarrino, Fabio},
   year={2025},
   month=nov }

@misc{rambach2025photonicquantumacceleratedmachinelearning,
      title={Photonic Quantum-Accelerated Machine Learning}, 
      author={Markus Rambach and Abhishek Roy and Alexei Gilchrist and Akitada Sakurai and William J. Munro and Kae Nemoto and Andrew G. White},
      year={2025},
      eprint={2512.08318},
      archivePrefix={arXiv},
      primaryClass={quant-ph},
      url={https://arxiv.org/abs/2512.08318}, 
}

@article{mills2026mitigatingphoton,
  doi = {10.22331/q-2026-03-16-2030},
  url = {https://doi.org/10.22331/q-2026-03-16-2030},
  title = {Mitigating photon loss in linear optical quantum circuits},
  author = {Mills, James and Mezher, Rawad},
  journal = {{Quantum}},
  issn = {2521-327X},
  publisher = {{Verein zur F{\"{o}}rderung des Open Access Publizierens in den Quantenwissenschaften}},
  volume = {10},
  pages = {2030},
  month = mar,
  year = {2026}
}

@misc{anguita2025experimentaldemonstrationbosonsampling,
      title={Experimental demonstration of boson sampling as a hardware accelerator for monte carlo integration}, 
      author={Malaquias Correa Anguita and Teun Roelink and Sara Marzban and Wim Briels and Claudia Filippi and Jelmer Renema},
      year={2025},
      eprint={2509.25404},
      archivePrefix={arXiv},
      primaryClass={quant-ph},
      url={https://arxiv.org/abs/2509.25404}, 
}

@article{sakurai2025qorc,
author = {Akitada Sakurai and Aoi Hayashi and William John Munro and Kae Nemoto},
journal = {Optica Quantum},
number = {3},
pages = {238--245},
publisher = {Optica Publishing Group},
title = {Quantum optical reservoir computing powered by boson sampling},
volume = {3},
month = {Jun},
year = {2025},
url = {https://opg.optica.org/opticaq/abstract.cfm?URI=opticaq-3-3-238},
doi = {10.1364/OPTICAQ.541432}}

@misc{wang2026qcircnn,
      title={QCirCNN: a photonic-native quantum circular convolution network based on circulant matrices }, 
      author={Wang, Daphn\'e and Walsh, Anthony and Fruchet, Hugo and Soret, Ariane and Emeriau, Pierre-Emmanuel },
      year={2026},
      journal={In preparation},
}

@article{QEImanifesto,
  title = {Quantum Technologies Need a Quantum Energy Initiative},
  author = {Auff\`eves, Alexia},
  journal = {PRX Quantum},
  volume = {3},
  issue = {2},
  pages = {020101},
  numpages = {12},
  year = {2022},
  month = {Jun},
  publisher = {American Physical Society},
  doi = {10.1103/PRXQuantum.3.020101},
  url = {https://link.aps.org/doi/10.1103/PRXQuantum.3.020101}
}

@misc{eisert2025mindgapsfraughtroad,
      title={Mind the gaps: The fraught road to quantum advantage}, 
      author={Jens Eisert and John Preskill},
      year={2025},
      eprint={2510.19928},
      archivePrefix={arXiv},
      primaryClass={quant-ph},
      url={https://arxiv.org/abs/2510.19928}, 
}

@article{Shu2014,
  title = {Heating Rates and Ion-Motion Control in a \$\textbackslash mathsf\textbraceleft{{Y}}\textbraceright\$-Junction Surface-Electrode Trap},
  author = {Shu, G. and Vittorini, G. and Buikema, A. and Nichols, C. S. and Volin, C. and Stick, D. and Brown, Kenneth R.},
  year = 2014,
  month = jun,
  journal = {Phys. Rev. A},
  volume = {89},
  number = {6},
  pages = {062308},
  publisher = {American Physical Society},
  doi = {10.1103/PhysRevA.89.062308},
  urldate = {2026-05-13}
}

@article{Itano1995,
  title = {Cooling Methods in Ion Traps},
  author = {Itano, Wayne M and Bergquist, J C and Bollinger, J J and Wineland, D J},
  year = 1995,
  month = jan,
  journal = {Phys. Scr.},
  volume = {T59},
  pages = {106--120},
  issn = {0031-8949, 1402-4896},
  doi = {10.1088/0031-8949/1995/T59/013},
  urldate = {2026-05-13},
  langid = {english}
}

@book{Metcalf1999,
  title = {Laser {{Cooling}} and {{Trapping}}},
  author = {Metcalf, Harold J. and Van Der Straten, Peter},
  editor = {Berry, R. Stephen and Birman, Joseph L. and Lynn, Jeffrey W. and Silverman, Mark P. and Stanley, H. Eugene and Voloshin, Mikhail},
  year = 1999,
  series = {Graduate {{Texts}} in {{Contemporary Physics}}},
  publisher = {Springer},
  address = {New York, NY},
  doi = {10.1007/978-1-4612-1470-0},
  urldate = {2026-05-13},
  copyright = {http://www.springer.com/tdm},
  isbn = {978-0-387-98728-6 978-1-4612-1470-0}
}

@article{Todaro2021,
  title = {State {{Readout}} of a {{Trapped Ion Qubit Using}} a {{Trap-Integrated Superconducting Photon Detector}}},
  author = {Todaro, S. L. and Verma, V. B. and McCormick, K. C. and Allcock, D. T. C. and Mirin, R. P. and Wineland, D. J. and Nam, S. W. and Wilson, A. C. and Leibfried, D. and Slichter, D. H.},
  year = 2021,
  month = jan,
  journal = {Phys. Rev. Lett.},
  volume = {126},
  number = {1},
  pages = {010501},
  issn = {0031-9007, 1079-7114},
  doi = {10.1103/PhysRevLett.126.010501},
  urldate = {2026-05-13},
  langid = {english}
}

@article{Ballance2016,
  title = {High-{{Fidelity Quantum Logic Gates Using Trapped-Ion Hyperfine Qubits}}},
  author = {Ballance, C. J. and Harty, T. P. and Linke, N. M. and Sepiol, M. A. and Lucas, D. M.},
  year = 2016,
  month = aug,
  journal = {Phys. Rev. Lett.},
  volume = {117},
  number = {6},
  pages = {060504},
  issn = {0031-9007, 1079-7114},
  doi = {10.1103/PhysRevLett.117.060504},
  urldate = {2026-05-13},
  copyright = {http://link.aps.org/licenses/aps-default-license},
  langid = {english}
}

@article{Brown2011,
  title = {Single-Qubit-Gate Error below 10 - 4 in a Trapped Ion},
  author = {Brown, K. R. and Wilson, A. C. and Colombe, Y. and Ospelkaus, C. and Meier, A. M. and Knill, E. and Leibfried, D. and Wineland, D. J.},
  year = 2011,
  month = sep,
  journal = {Phys. Rev. A},
  volume = {84},
  number = {3},
  pages = {030303},
  issn = {1050-2947, 1094-1622},
  doi = {10.1103/PhysRevA.84.030303},
  urldate = {2026-05-13},
  copyright = {http://link.aps.org/licenses/aps-default-license},
  langid = {english}
}

@misc{Hughes2025,
  title = {Trapped-Ion Two-Qubit Gates with {$>$}99.99\% Fidelity without Ground-State Cooling},
  author = {Hughes, A. C. and Srinivas, R. and L{\"o}schnauer, C. M. and Knaack, H. M. and Matt, R. and Ballance, C. J. and Malinowski, M. and Harty, T. P. and Sutherland, R. T.},
  year = 2025,
  month = oct,
  number = {arXiv:2510.17286},
  eprint = {2510.17286},
  primaryclass = {quant-ph},
  publisher = {arXiv},
  doi = {10.48550/arXiv.2510.17286},
  urldate = {2026-05-13},
  archiveprefix = {arXiv}
}

@article{Bruzewicz2019,
  title = {Trapped-{{Ion Quantum Computing}}: {{Progress}} and {{Challenges}}},
  shorttitle = {Trapped-{{Ion Quantum Computing}}},
  author = {Bruzewicz, Colin D. and Chiaverini, John and McConnell, Robert and Sage, Jeremy M.},
  year = 2019,
  month = jun,
  journal = {Applied Physics Reviews},
  volume = {6},
  number = {2},
  eprint = {1904.04178},
  primaryclass = {quant-ph},
  pages = {021314},
  issn = {1931-9401},
  doi = {10.1063/1.5088164},
  urldate = {2026-05-13},
  archiveprefix = {arXiv}
}

@article{Leibfried2003,
  title = {Quantum Dynamics of Single Trapped Ions},
  author = {Leibfried, D. and Blatt, R. and Monroe, C. and Wineland, D.},
  year = 2003,
  month = mar,
  journal = {Rev. Mod. Phys.},
  volume = {75},
  number = {1},
  pages = {281--324},
  issn = {0034-6861, 1539-0756},
  doi = {10.1103/RevModPhys.75.281},
  urldate = {2026-05-13},
  copyright = {http://link.aps.org/licenses/aps-default-license},
  langid = {english}
}

@article{Turchette1998,
  title = {Deterministic {{Entanglement}} of {{Two Trapped Ions}}},
  author = {Turchette, Q. A. and Wood, C. S. and King, B. E. and Myatt, C. J. and Leibfried, D. and Itano, W. M. and Monroe, C. and Wineland, D. J.},
  year = 1998,
  month = oct,
  journal = {Phys. Rev. Lett.},
  volume = {81},
  number = {17},
  pages = {3631--3634},
  issn = {0031-9007, 1079-7114},
  doi = {10.1103/PhysRevLett.81.3631},
  urldate = {2026-05-13},
  copyright = {http://link.aps.org/licenses/aps-default-license},
  langid = {english}
}

@article{Pino2021a,
  title = {Demonstration of the Trapped-Ion Quantum {{CCD}} Computer Architecture},
  author = {Pino, J. M. and Dreiling, J. M. and Figgatt, C. and Gaebler, J. P. and Moses, S. A. and Allman, M. S. and Baldwin, C. H. and {Foss-Feig}, M. and Hayes, D. and Mayer, K. and {Ryan-Anderson}, C. and Neyenhuis, B.},
  year = 2021,
  month = apr,
  journal = {Nature},
  volume = {592},
  number = {7853},
  pages = {209--213},
  publisher = {Nature Publishing Group},
  issn = {1476-4687},
  doi = {10.1038/s41586-021-03318-4},
  urldate = {2026-04-01},
  copyright = {2021 The Author(s), under exclusive licence to Springer Nature Limited part of Springer Nature},
  langid = {english}
}

@misc{Ransford2025,
  title = {Helios: {{A}} 98-Qubit Trapped-Ion Quantum Computer},
  shorttitle = {Helios},
  author = {Ransford, Anthony and Allman, M. S. and Arkinstall, Jake and Campora, J. P. and Cooper, Samuel F. and Delaney, Robert D. and Dreiling, Joan M. and Estey, Brian and Figgatt, Caroline and Hall, Alex and Husain, Ali A. and Isanaka, Akhil and Kennedy, Colin J. and Kotibhaskar, Nikhil and Madjarov, Ivaylo S. and Mayer, Karl and Milne, Alistair R. and Park, Annie J. and Reed, Adam P. and Ancona, Riley and Andersen, Molly P. and {Andres-Martinez}, Pablo and Angenent, Will and Argueta, Liz and Arkin, Benjamin and Ascarrunz, Leonardo and Baker, William and Barnes, Corey and Bartolotta, John and Berg, Jordan and Besand, Ryan and Bjork, Bryce and Blain, Matt and Blanchard, Paul and {Blume-Kohout}, Robin and Bohn, Matt and Borgna, Agustin and Botamanenko, Daniel Y. and Boutelle, Robert and Brown, Natalie and Buckingham, Grant T. and Burdick, Nathaniel Q. and Burton, William Cody and Carey, Varis and Carron, Christopher J. and Chambers, Joe and Children, John and Colussi, Victor E. and Crepinsek, Steven and Cureton, Andrew and Davies, Joe and Davis, Daniel and DeCross, Matthew and Deen, David and Delaney, Conor and DelVento, Davide and DeSalvo, B. J. and Dominy, Jason and Duncan, Ross and Eccles, Vanya and Edgington, Alec and Erickson, Neal and Erickson, Stephen and Ertsgaard, Christopher T. and Evans, Bruce and Evans, Tyler and Fabrikant, Maya I. and Fischer, Andrew and Foltz, Cameron and {Foss-Feig}, Michael and Francois, David and Freyberg, Brad and Gao, Charles and Garay, Robert and Garvin, Jane and Gaudiosi, David M. and Gilbreth, Christopher N. and Giles, Josh and Glynn, Erin and Graves, Jeff and Hansen, Azure and Hayes, David and Heidemann, Lukas and Higashi, Bob and Hilbun, Tyler and Hines, Jordan and Hlavaty, Ariana and Hoffman, Kyle and Hoffman, Ian M. and Holliman, Craig and Hooper, Isobel and Horning, Bob and Hostetter, James and Hothem, Daniel and Houlton, Jack and Hout, Jared and Hutson, Ross and Jacobs, Ryan T. and Jacobs, Trent and Johannsen, Melf and Johansen, Jacob and Jones, Loren and Julian, Sydney and Jung, Ryan and Keay, Aidan and Klein, Todd and Koch, Mark and Kondo, Ryo and Kong, Chang and Kosto, Asa and Lawrence, Alan and Liefer, David and Lollie, Michelle and Lucchetti, Dominic and Lysne, Nathan K. and Lytle, Christian and MacPherson, Callum and Malm, Andrew and Mather, Spencer and Mathewson, Brian and Maxwell, Daniel and McCaffrey, Lauren and McDougall, Hannah and Mendoza, Robin and Mills, Michael and Morrison, Richard and Narmour, Louis and Nguyen, Nhung and Nugent, Lora and Olson, Scott and Ouellette, Daniel and Parks, Jeremy and Peters, Zach and Petricka, Jessie and Pino, Juan M. and Polito, Frank and Preidl, Matthias and Price, Gabriel and Proctor, Timothy and Pugh, McKinley and Ratcliff, Noah and Raymondson, Daisy and Rhodes, Peter and Roman, Conrad and Roy, Craig and {Ryan-Anderson}, Ciaran and Sanchez, Fernando Betanzo and Sangiolo, George and Sawadski, Tatiana and Schaffer, Andrew and Schow, Peter and Sedlacek, Jon and Semenenko, Henry and Shevchuk, Peter and Shore, Susan and Siegfried, Peter and Singhal, Kartik and Sivarajah, Seyon and Skripka, Thomas and Sletten, Lucas and Spaun, Ben and Sprenkle, R. Tucker and Stoufer, Paul and Tader, Mariel and Taylor, Stephen F. and Thompson, Travis H. and Tobey, Raanan and Tran, Anh and Tran, Tam and Vittorini, Grahame and Volin, Curtis and Walker, Jim and White, Sam and Wilson, Douglas and Wolf, Quinn and Wringe, Chester and Young, Kevin and Zheng, Jian and Zuraski, Kristen and Baldwin, Charles H. and Chernoguzov, Alex and Gaebler, John P. and Sanders, Steven J. and Neyenhuis, Brian and Stutz, Russell and Bohnet, Justin G.},
  year = 2025,
  month = nov,
  number = {arXiv:2511.05465},
  eprint = {2511.05465},
  primaryclass = {quant-ph},
  publisher = {arXiv},
  doi = {10.48550/arXiv.2511.05465},
  urldate = {2026-03-06},
  archiveprefix = {arXiv}
}

@article{Allcock2013,
  title = {A Microfabricated Ion Trap with Integrated Microwave Circuitry},
  author = {Allcock, D. T. C. and Harty, T. P. and Ballance, C. J. and Keitch, B. C. and Linke, N. M. and Stacey, D. N. and Lucas, D. M.},
  year = {2013},
  month = jan,
  journal = {Applied Physics Letters},
  volume = {102},
  number = {4},
  eprint = {1210.3272},
  primaryclass = {physics},
  pages = {044103},
  issn = {0003-6951, 1077-3118},
  doi = {10.1063/1.4774299},
  urldate = {2025-06-25},
  archiveprefix = {arXiv}
}

@misc{Bernardini2023,
  title = {Quantum Computing with Trapped Ions: A Beginner's Guide},
  shorttitle = {Quantum Computing with Trapped Ions},
  author = {Bernardini, Francesco and Chakraborty, Abhijit and Ord{\'o}{\~n}ez, Carlos},
  year = {2023},
  month = sep,
  number = {arXiv:2303.16358},
  eprint = {2303.16358},
  primaryclass = {quant-ph},
  publisher = {arXiv},
  doi = {10.48550/arXiv.2303.16358},
  urldate = {2025-03-12},
  archiveprefix = {arXiv}
}

@article{Cirac1995,
  title = {Quantum {{Computations}} with {{Cold Trapped Ions}}},
  author = {Cirac, J. I. and Zoller, P.},
  year = {1995},
  month = may,
  journal = {Phys. Rev. Lett.},
  volume = {74},
  number = {20},
  pages = {4091--4094},
  publisher = {American Physical Society},
  doi = {10.1103/PhysRevLett.74.4091},
  urldate = {2025-03-12}
}

@article{Liu2025,
  title = {Certified Randomness Using a Trapped-Ion Quantum Processor},
  author = {Liu, Minzhao and Shaydulin, Ruslan and Niroula, Pradeep and DeCross, Matthew and Hung, Shih-Han and Kon, Wen Yu and {Cervero-Mart{\'i}n}, Enrique and Chakraborty, Kaushik and Amer, Omar and Aaronson, Scott and Acharya, Atithi and Alexeev, Yuri and Berg, K. Jordan and Chakrabarti, Shouvanik and Curchod, Florian J. and Dreiling, Joan M. and Erickson, Neal and Foltz, Cameron and {Foss-Feig}, Michael and Hayes, David and Humble, Travis S. and Kumar, Niraj and Larson, Jeffrey and Lykov, Danylo and Mills, Michael and Moses, Steven A. and Neyenhuis, Brian and Eloul, Shaltiel and Siegfried, Peter and Walker, James and Lim, Charles and Pistoia, Marco},
  year = {2025},
  month = apr,
  journal = {Nature},
  volume = {640},
  number = {8058},
  pages = {343--348},
  publisher = {Nature Publishing Group},
  issn = {1476-4687},
  doi = {10.1038/s41586-025-08737-1},
  urldate = {2025-06-25},
  copyright = {2025 The Author(s)},
  langid = {english}
}

@article{Weber2024,
  title = {Cryogenic Ion Trap System for High-Fidelity near-Field Microwave-Driven Quantum Logic},
  author = {Weber, M. A. and L{\"o}schnauer, C. and Wolf, J. and Gely, M. F. and Hanley, R. K. and Goodwin, J. F. and Ballance, C. J. and Harty, T. P. and Lucas, D. M.},
  year = {2024},
  month = jan,
  journal = {Quantum Sci. Technol.},
  volume = {9},
  number = {1},
  eprint = {2207.11364},
  primaryclass = {quant-ph},
  pages = {015007},
  issn = {2058-9565},
  doi = {10.1088/2058-9565/acfba8},
  urldate = {2025-06-20},
  archiveprefix = {arXiv},
  langid = {english}
}

@misc{Ye2025,
  title = {Quantum Error Correction for Long Chains of Trapped Ions},
  author = {Ye, Min and Delfosse, Nicolas},
  year = {2025},
  month = apr,
  number = {arXiv:2503.22071},
  eprint = {2503.22071},
  primaryclass = {quant-ph},
  publisher = {arXiv},
  doi = {10.48550/arXiv.2503.22071},
  urldate = {2025-06-25},
  archiveprefix = {arXiv}
}

@article{Ebadi2021,
  title = {Quantum Phases of Matter on a 256-Atom Programmable Quantum Simulator},
  author = {Ebadi, Sepehr and Wang, Tout T. and Levine, Harry and Keesling, Alexander and Semeghini, Giulia and Omran, Ahmed and Bluvstein, Dolev and Samajdar, Rhine and Pichler, Hannes and Ho, Wen Wei and Choi, Soonwon and Sachdev, Subir and Greiner, Markus and Vuleti{\'c}, Vladan and Lukin, Mikhail D.},
  year = 2021,
  month = jul,
  journal = {Nature},
  volume = {595},
  number = {7866},
  pages = {227--232},
  publisher = {Nature Publishing Group},
  issn = {1476-4687},
  doi = {10.1038/s41586-021-03582-4},
  urldate = {2026-07-30},
  copyright = {2021 The Author(s), under exclusive licence to Springer Nature Limited},
  langid = {english}
}

@article{Bluvstein2024,
  title = {Logical Quantum Processor Based on Reconfigurable Atom Arrays},
  author = {Bluvstein, Dolev and Evered, Simon J. and Geim, Alexandra A. and Li, Sophie H. and Zhou, Hengyun and Manovitz, Tom and Ebadi, Sepehr and Cain, Madelyn and Kalinowski, Marcin and Hangleiter, Dominik and Bonilla Ataides, J. Pablo and Maskara, Nishad and Cong, Iris and Gao, Xun and Sales Rodriguez, Pedro and Karolyshyn, Thomas and Semeghini, Giulia and Gullans, Michael J. and Greiner, Markus and Vuleti{\'c}, Vladan and Lukin, Mikhail D.},
  year = {2024},
  month = feb,
  journal = {Nature},
  volume = {626},
  number = {7997},
  pages = {58--65},
  publisher = {Nature Publishing Group},
  issn = {1476-4687},
  doi = {10.1038/s41586-023-06927-3},
  urldate = {2025-05-14},
  copyright = {2023 The Author(s)},
  langid = {english}
}

@misc{soret2026quantumenergeticadvantagecomputational,
      title={Quantum Energetic Advantage before Computational Advantage in Boson Sampling}, 
      author={Ariane Soret and Nessim Dridi and Stephen C. Wein and Valérian Giesz and Shane Mansfield and Pierre-Emmanuel Emeriau},
      year={2026},
      eprint={2601.08068},
      archivePrefix={arXiv},
      primaryClass={quant-ph},
      url={https://arxiv.org/abs/2601.08068}, 
}

@article{Fellous-Asiani2023,
	title = {Optimizing {Resource} {Efficiencies} for {Scalable} {Full}-{Stack} {Quantum} {Computers}},
	volume = {4},
	issn = {2691-3399},
	url = {https://link.aps.org/doi/10.1103/PRXQuantum.4.040319},
	doi = {10.1103/PRXQuantum.4.040319},
	language = {en},
	number = {4},
	urldate = {2025-01-08},
	journal = {PRX Quantum},
	author = {Fellous-Asiani, Marco and Chai, Jing Hao and Thonnart, Yvain and Ng, Hui Khoon and Whitney, Robert S. and Auffèves, Alexia},
	month = oct,
	year = {2023},
	pages = {040319},
}

@misc{Yehia2025,
	title = {Energetic {Analysis} of {Emerging} {Quantum} {Communication} {Protocols}},
	url = {http://arxiv.org/abs/2410.10661},
	doi = {10.48550/arXiv.2410.10661},
	urldate = {2025-03-28},
	publisher = {arXiv},
	author = {Yehia, Raja and Piétri, Yoann and Pascual-García, Carlos and Lefebvre, Pascal and Centrone, Federico},
	month = mar,
	year = {2025},
	note = {arXiv:2410.10661 [quant-ph]},
}

@misc{ShorDepth1,
      author = {Junpei Yamaguchi and Masafumi Yamazaki and Akihiro Tabuchi and Takumi Honda and Tetsuya Izu and Noboru Kunihiro},
      title = {Estimation of Shor's Circuit for 2048-bit Integers based on Quantum Simulator},
      howpublished = {Cryptology {ePrint} Archive, Paper 2023/092},
      year = {2023},
      url = {https://eprint.iacr.org/2023/092}
}

@article{ShorDepth2,
  title = {How to Factor 2048 Bit {{RSA}} Integers in 8 Hours Using 20 Million Noisy Qubits},
  author = {Gidney, Craig and Eker{\aa}, Martin},
  year = {2021},
  month = apr,
  journal = {Quantum},
  volume = {5},
  eprint = {1905.09749},
  primaryclass = {quant-ph},
  pages = {433},
  issn = {2521-327X},
  doi = {10.22331/q-2021-04-15-433},
  urldate = {2025-06-17},
  archiveprefix = {arXiv},

}

@article{DepthChemistry1,
  title = {Elucidating {{Reaction Mechanisms}} on {{Quantum Computers}}},
  author = {Reiher, Markus and Wiebe, Nathan and Svore, Krysta M. and Wecker, Dave and Troyer, Matthias},
  year = {2017},
  month = jul,
  journal = {Proceedings of the National Academy of Sciences},
  volume = {114},
  number = {29},
  eprint = {1605.03590},
  primaryclass = {quant-ph},
  pages = {7555--7560},
  issn = {0027-8424, 1091-6490},
  doi = {10.1073/pnas.1619152114},
  urldate = {2025-06-17},
  archiveprefix = {arXiv},
}

@article{DepthChemistry2,
  title = {Even More Efficient Quantum Computations of Chemistry through Tensor Hypercontraction},
  author = {Lee, Joonho and Berry, Dominic W. and Gidney, Craig and Huggins, William J. and McClean, Jarrod R. and Wiebe, Nathan and Babbush, Ryan},
  year = {2021},
  month = jul,
  journal = {PRX Quantum},
  volume = {2},
  number = {3},
  eprint = {2011.03494},
  primaryclass = {quant-ph},
  pages = {030305},
  issn = {2691-3399},
  doi = {10.1103/PRXQuantum.2.030305},
  urldate = {2025-06-17},
  archiveprefix = {arXiv},
}

@misc{FourierDepth1,
  title = {Approximate {{Quantum Fourier Transform}} in {{Logarithmic Depth}} on a {{Line}}},
  author = {B{\"a}umer, Elisa and Sutter, David and Woerner, Stefan},
  year = {2025},
  month = apr,
  number = {arXiv:2504.20832},
  eprint = {2504.20832},
  primaryclass = {quant-ph},
  publisher = {arXiv},
  doi = {10.48550/arXiv.2504.20832},
  urldate = {2025-06-17},
  archiveprefix = {arXiv},
  langid = {english},
}

@misc{FourierDepth2,
  title = {Fast Parallel Circuits for the Quantum {{Fourier}} Transform},
  author = {Cleve, Richard and Watrous, John},
  year = {2000},
  month = jun,
  number = {arXiv:quant-ph/0006004},
  eprint = {quant-ph/0006004},
  publisher = {arXiv},
  doi = {10.48550/arXiv.quant-ph/0006004},
  urldate = {2025-06-17},
  archiveprefix = {arXiv},

}

@article{VQEdepth,
  title = {Variational Quantum Eigensolver with Reduced Circuit Complexity},
  author = {Zhang, Yu and Cincio, Lukasz and Negre, Christian and Czarnik, Piotr and Coles, Patrick and Anisimov, Petr and Mniszewski, Susan and Tretiak, Sergei and Dub, Pavel},
  year = {2022},
  month = aug,
  journal = {npj Quantum Information},
  volume = {8},
  doi = {10.1038/s41534-022-00599-z},
}

@misc{CDKMadder,
      title={A new quantum ripple-carry addition circuit}, 
      author={Steven A. Cuccaro and Thomas G. Draper and Samuel A. Kutin and David Petrie Moulton},
      year={2004},
      eprint={quant-ph/0410184},
      archivePrefix={arXiv},
      primaryClass={quant-ph},
      url={https://arxiv.org/abs/quant-ph/0410184}, 
}

@misc{CryoCooler, 
title = {Cryomech Cryogenic Refrigerator Model PT410, Technical instruction manual},
url = {https://files.wmich.edu/s3fs-public/attachments/u1045/2019/PT410_CP2880_Technical_Manual-9139.pdf}
}

@article{woitzikEnergyEstimationBenchmark2024,
  title = {An {{Energy Estimation Benchmark}} for {{Quantum Computers}}},
  author = {Woitzik, Andreas J. C. and Hoffmann, Lukas and Buchleitner, Andreas and Carnio, Edoardo G.},
  year = {2024},
  month = mar,
  journal = {Open Systems \& Information Dynamics},
  volume = {31},
  number = {01},
  eprint = {2302.04144},
  primaryclass = {quant-ph},
  pages = {2450006},
  issn = {1230-1612, 1793-7191},
  doi = {10.1142/S1230161224500069},
  urldate = {2025-08-04},
  archiveprefix = {arXiv}
}

@article{Krinner2019,
  title = {Engineering Cryogenic Setups for 100-Qubit Scale Superconducting Circuit Systems},
  author = {Krinner, Sebastian and Storz, Simon and Kurpiers, Philipp and Magnard, Paul and Heinsoo, Johannes and Keller, Raphael and Luetolf, Janis and Eichler, Christopher and Wallraff, Andreas},
  year = 2019,
  month = dec,
  journal = {EPJ Quantum Technol.},
  volume = {6},
  number = {1},
  eprint = {1806.07862},
  primaryclass = {quant-ph},
  pages = {2},
  issn = {2662-4400, 2196-0763},
  doi = {10.1140/epjqt/s40507-019-0072-0},
  urldate = {2024-12-30},
  archiveprefix = {arXiv}
}

@article{Gambetta2017,
  title = {Building Logical Qubits in a Superconducting Quantum Computing System},
  author = {Gambetta, Jay M. and Chow, Jerry M. and Steffen, Matthias},
  year = 2017,
  month = jan,
  journal = {Npj Quantum Inf.},
  volume = {3},
  number = {1},
  pages = {2},
  publisher = {Nature Publishing Group},
  issn = {2056-6387},
  doi = {10.1038/s41534-016-0004-0},
  urldate = {2026-07-30},
  copyright = {2017 The Author(s)},
  langid = {english}
}

@article{Roth2023,
  title = {An {{Introduction}} to the {{Transmon Qubit}} for {{Electromagnetic Engineers}}},
  author = {Roth, Thomas E. and Ma, Ruichao and Chew, Weng C.},
  year = 2023,
  month = apr,
  journal = {IEEE Antennas Propag. Mag.},
  volume = {65},
  number = {2},
  eprint = {2106.11352},
  primaryclass = {quant-ph},
  pages = {8--20},
  issn = {1045-9243, 1558-4143},
  doi = {10.1109/MAP.2022.3176593},
  urldate = {2025-06-04},
  archiveprefix = {arXiv}
}

@article{horsman2012surface,
  title={Surface code quantum computing by lattice surgery},
  author={Horsman, Dominic and Fowler, Austin G and Devitt, Simon and Van Meter, Rodney},
  journal={New Journal of Physics},
  volume={14},
  number={12},
  pages={123011},
  year={2012},
  publisher={IOP Publishing}
}

@article{bravyi2005universal,
  title={Universal quantum computation with ideal Clifford gates and noisy ancillas},
  author={Bravyi, Sergey and Kitaev, Alexei},
  journal={Physical Review A—Atomic, Molecular, and Optical Physics},
  volume={71},
  number={2},
  pages={022316},
  year={2005},
  publisher={APS}
}

@article{gidney2025factor,
  title={How to factor 2048 bit RSA integers with less than a million noisy qubits},
  author={Gidney, Craig},
  journal={arXiv preprint arXiv:2505.15917},
  year={2025}
}

@article{breuckmann2021quantum,
  title={Quantum low-density parity-check codes},
  author={Breuckmann, Nikolas P and Eberhardt, Jens Niklas},
  journal={PRX quantum},
  volume={2},
  number={4},
  pages={040101},
  year={2021},
  publisher={APS}
}

@misc{DARPA,
title ={Quantum Benchmarking initiative: https://www.darpa.mil/research/programs/quantum-benchmarking-initiative},
 url={https://www.darpa.mil/research/programs/quantum-benchmarking-initiative},

}

@misc{QEIWG,
title={{Quantum Energy Initiative IEEE} working group, https://sagroups.ieee.org/qei/},
url={https://sagroups.ieee.org/qei/},

}

@article{Martinis2009,
  title = {Superconducting Phase Qubits},
  author = {Martinis, John M.},
  year = 2009,
  month = jun,
  journal = {Quantum Inf. Process.},
  volume = {8},
  number = {2},
  pages = {81--103},
  issn = {1573-1332},
  doi = {10.1007/s11128-009-0105-1},
  urldate = {2026-07-21},
  langid = {english}
}

@article{Mooij1999,
  title = {Josephson {{Persistent-Current Qubit}}},
  author = {Mooij, J. E. and Orlando, T. P. and Levitov, L. and Tian, Lin and {van der Wal}, Caspar H. and Lloyd, Seth},
  year = 1999,
  month = aug,
  journal = {Science},
  volume = {285},
  number = {5430},
  pages = {1036--1039},
  publisher = {American Association for the Advancement of Science},
  doi = {10.1126/science.285.5430.1036},
  urldate = {2026-07-21}
}

@article{ReviewSuperconducting,
  title = {A {{Review}} of {{Design Concerns}} in {{Superconducting Quantum Circuits}}},
  author = {{Levenson-Falk}, Eli M. and Shanto, Sadman Ahmed},
  year = 2025,
  month = jun,
  journal = {Materials for Quantum Technology},
  volume = {5},
  number = {2},
  eprint = {2411.16967},
  primaryclass = {quant-ph},
  pages = {022003},
  issn = {2633-4356},
  doi = {10.1088/2633-4356/ade10d},
  urldate = {2025-11-21},
  archiveprefix = {arXiv},
  langid = {english}
}

@misc{QbloxCurrentControler,  
title = {{ Qblox Current Source Module, \url{https://cdn.prod.website-files.com/653289e64ff83c71222f6bf2/67124ca0c409d9f6f7d46784_QBLOX_PRODUCTSHEET_SPI-S4G_V1_7.pdf} }},
url = {https://cdn.prod.website-files.com/653289e64ff83c71222f6bf2/67124ca0c409d9f6f7d46784_QBLOX_PRODUCTSHEET_SPI-S4G_V1_7.pdf}
}

@misc{QbloxBasebandControler,  
title = {{ Qblox Qubit Control Module, \url{https://cdn.prod.website-files.com/6532891b0dd891b360d705ff/67ab39ad8b515abf846fc54d_Qblox%20Productsheet%20-%20QCM%20v1.7.pdf} }},
url = {https://cdn.prod.website-files.com/6532891b0dd891b360d705ff/67ab39ad8b515abf846fc54d_Qblox%20Productsheet%20-%20QCM%20v1.7.pdf}
}

@misc{QbloxMWControler,  
title = {{Qblox MW Module, \url{https://cdn.prod.website-files.com/6532891b0dd891b360d705ff/682c534c283d724c27e82786_Qblox%20Productsheet%20-%20QCM-RF%20II%20v1.8.9.pdf}}},
url = {https://cdn.prod.website-files.com/6532891b0dd891b360d705ff/682c534c283d724c27e82786_Qblox%20Productsheet%20-%20QCM-RF%20II%20v1.8.9.pdf}
}

@misc{QbloxReadOut,  
title = {{Qblox Readout Module, \url{https://cdn.prod.website-files.com/6532891b0dd891b360d705ff/682c534c95c510787d786c8d_Qblox%20Productsheet%20-%20QRM-RF%20v1.7.pdf}}},
url = {https://cdn.prod.website-files.com/6532891b0dd891b360d705ff/682c534c95c510787d786c8d_Qblox%20Productsheet%20-%20QRM-RF%20v1.7.pdf}
}

@misc{LNFlownoise,  
title = {{LNF-LNC0.3\_14B ultra-low noise cryogenic amplifier, \url{https://lownoisefactory.com/product/lnf-lnc0-3_14b/}}},
url = {https://lownoisefactory.com/product/lnf-lnc0-3_14b/}
}

@misc{RoomTempAmpl,  
title = {{Low Noise Amplifier ZKL-33ULN-S+, \url{https://www.minicircuits.com/pdfs/ZKL-33ULN-S+.pdf}}},
url = {https://www.minicircuits.com/pdfs/ZKL-33ULN-S+.pdf}
}

@article{Kielpinski2002ArchitectureFA,
  title={Architecture for a large-scale ion-trap quantum computer},
  author={David Kielpinski and Christopher R. Monroe and David J. Wineland},
  journal={Nature},
  year={2002},
  volume={417},
  pages={709-711},
  url={https://api.semanticscholar.org/CorpusID:4347109}
}

@article{eastin2009,
  title={Restrictions on transversal encoded quantum gate sets},
  author={Eastin, Bryan and Knill, Emanuel},
  journal={Physical review letters},
  volume={102},
  number={11},
  pages={110502},
  year={2009},
  publisher={APS}
}

@inproceedings{Barzaghi2025,
author = {Andrea Barzaghi and Ma{\"e}lle B{\'e}n{\'e}fice and Francesco Ceccarelli and Giacomo Corrielli and Valerio Galli and Marco Gardina and Vittorio Grimaldi and Jakub Kaczorowski and Francesco Malaspina and Roberto Osellame and Ciro Pentangelo and Andrea Rocchetto and Alessandro Rudi},
title = {{A 24-mode laser-written universal photonic processor in a glass-based platform}},
volume = {13391},
booktitle = {Quantum Computing, Communication, and Simulation V},
editor = {Philip R. Hemmer and Alan L. Migdall},
organization = {International Society for Optics and Photonics},
publisher = {SPIE},
pages = {1339112},
year = {2025},
doi = {10.1117/12.3043829},
URL = {https://doi.org/10.1117/12.3043829}
}

@article{WANG2024,
title = {Optimization and comprehensive comparison of thermo-optic phase shifter with folded waveguide on SiN and SOI platforms},
journal = {Optics Communications},
volume = {555},
pages = {130242},
year = {2024},
issn = {0030-4018},
doi = {https://doi.org/10.1016/j.optcom.2023.130242},
url = {https://www.sciencedirect.com/science/article/pii/S0030401823009902},
author = {Jin Wang and Wei Cheng and Wanghua Zhu and Mengjia Lu and Yifei Chen and Shangqing Shi and Chen Guo and Guohua Hu and Yiping Cui and Binfeng Yun},
}

@article{Guo2023,
url = {https://doi.org/10.1515/nanoph-2022-0575},
title = {Ultra-wideband integrated photonic devices on silicon platform: from visible to mid-IR},
title = {},
author = {Xuhan Guo and Xingchen Ji and Baicheng Yao and Teng Tan and Allen Chu and Ohad Westreich and Avik Dutt and Cheewei Wong and Yikai Su},
pages = {167--196},
volume = {12},
number = {2},
journal = {Nanophotonics},
doi = {doi:10.1515/nanoph-2022-0575},
year = {2023},
lastchecked = {2026-02-10}
}

@article{Chelladurai2025,
   author = {Daniel Chelladurai and Manuel Kohli and Joel Winiger and David Moor and Andreas Messner and Yuriy Fedoryshyn and Mohammed Eleraky and Yuqi Liu and Hua Wang and Juerg Leuthold},
   doi = {10.1038/s41563-025-02158-1},
   issn = {14764660},
   issue = {6},
   journal = {Nature Materials},
   month = {6},
   pages = {868-875},
   pmid = {40097599},
   publisher = {Nature Research},
   title = {Barium titanate and lithium niobate permittivity and Pockels coefficients from megahertz to sub-terahertz frequencies},
   volume = {24},
   year = {2025}
}

@inproceedings{Reck1994,
author = {Michael Reck and Anton Zeilinger and Herbert J. Bernstein and Philip Bertani},
booktitle = {European Quantum Electronics Conference},
journal = {European Quantum Electronics Conference},
pages = {QTuC6},
publisher = {Optica Publishing Group},
title = {How to build any discrete unitary operator in your laboratory},
year = {1994},
url = {https://opg.optica.org/abstract.cfm?URI=EQEC-1994-QTuC6},
}

@article{Nerenberg2025,
author = {Sam Nerenberg and Oliver D. Neill and Giulia Marcucci and Daniele Faccio},
journal = {Optica Quantum},
number = {2},
pages = {201--210},
publisher = {Optica Publishing Group},
title = {Photon number-resolving quantum reservoir computing},
volume = {3},
month = {Apr},
year = {2025},
url = {https://opg.optica.org/opticaq/abstract.cfm?URI=opticaq-3-2-201},
doi = {10.1364/OPTICAQ.553294},
}

@INPROCEEDINGS{Perez2016,
  author={Pérez, Daniel and Gasulla, Ivana and Capmany, José and Soref, Richard A.},
  booktitle={2016 IEEE Photonics Conference (IPC)}, 
  title={Hexagonal waveguide mesh design for universal multiport interferometers}, 
  year={2016},
  volume={},
  number={},
  pages={285-286},
  doi={10.1109/IPCon.2016.7831099}}

@article{Zhu2021,
   author = {Di Zhu and Linbo Shao and Mengjie Yu and Rebecca Cheng and Boris Desiatov and C. J. Xin and Yaowen Hu and Jeffrey Holzgrafe and Soumya Ghosh and Amirhassan Shams-Ansari and Eric Puma and Neil Sinclair and Christian Reimer and Mian Zhang and Marko Lončar},
   doi = {10.1364/aop.411024},
   issn = {19438206},
   issue = {2},
   journal = {Advances in Optics and Photonics},
   month = {6},
   pages = {242},
   publisher = {Optica Publishing Group},
   title = {Integrated photonics on thin-film lithium niobate},
   volume = {13},
   year = {2021}
}

@misc{Rahim2021,
   author = {Abdul Rahim and Artur Hermans and Benjamin Wohlfeil and Despoina Petousi and Bart Kuyken and Dries Van Thourhout and Roel Baets},
   doi = {10.1117/1.AP.3.2.024003},
   issn = {25775421},
   issue = {2},
   journal = {Advanced Photonics},
   month = {3},
   publisher = {SPIE},
   title = {Taking silicon photonics modulators to a higher performance level: State-of-the-art and a review of new technologies},
   volume = {3},
   year = {2021}
}

@article{Xu2022,
author = {Mengyue Xu and Yuntao Zhu and Fabio Pittal\`{a} and Jin Tang and Mingbo He and Wing Chau Ng and Jingyi Wang and Ziliang Ruan and Xuefeng Tang and Maxim Kuschnerov and Liu Liu and Siyuan Yu and Bofang Zheng and Xinlun Cai},
journal = {Optica},
number = {1},
pages = {61--62},
publisher = {Optica Publishing Group},
title = {Dual-polarization thin-film lithium niobate in-phase quadrature modulators for terabit-per-second transmission},
volume = {9},
month = {Jan},
year = {2022},
url = {https://opg.optica.org/optica/abstract.cfm?URI=optica-9-1-61},
doi = {10.1364/OPTICA.449691},
}

@article{Yan2025,
author = {Yan, Yixin and Zhang, Haoran and Liu, Xiaolei and Peng, Liuxing and Zhang, Qian and Yu, Guangbin and Wu, Qing and Li, Haitao},
title = {Recent Progress in Electro-Optic Modulators: Physical Phenomenon, Structures Properties, and Integration Strategy},
journal = {Laser \& Photonics Reviews},
volume = {19},
number = {3},
pages = {2400624},
doi = {https://doi.org/10.1002/lpor.202400624},
url = {https://onlinelibrary.wiley.com/doi/abs/10.1002/lpor.202400624},
eprint = {https://onlinelibrary.wiley.com/doi/pdf/10.1002/lpor.202400624},
year = {2025}
}

@article{Eltes2019,
   title={A BaTiO3-Based Electro-Optic Pockels Modulator Monolithically Integrated on an Advanced Silicon Photonics Platform},
   volume={37},
   ISSN={1558-2213},
   url={http://dx.doi.org/10.1109/JLT.2019.2893500},
   DOI={10.1109/jlt.2019.2893500},
   number={5},
   journal={Journal of Lightwave Technology},
   publisher={Institute of Electrical and Electronics Engineers (IEEE)},
   author={Eltes, Felix and Mai, Christian and Caimi, Daniele and Kroh, Marcel and Popoff, Youri and Winzer, Georg and Petousi, Despoina and Lischke, Stefan and Ortmann, J. Elliott and Czornomaz, Lukas and Zimmermann, Lars and Fompeyrine, Jean and Abel, Stefan},
   year={2019},
   month=mar, pages={1456–1462} }

@article{Wang2018,
author = {Cheng Wang and Mian Zhang and Brian Stern and Michal Lipson and Marko Lon\v{c}ar},
journal = {Opt. Express},
number = {2},
pages = {1547--1555},
publisher = {Optica Publishing Group},
title = {Nanophotonic lithium niobate electro-optic modulators},
volume = {26},
month = {Jan},
year = {2018},
url = {https://opg.optica.org/oe/abstract.cfm?URI=oe-26-2-1547},
doi = {10.1364/OE.26.001547}
}

@Article{Yang2024,
AUTHOR = {Yang, Peng and Sun, Siwei and Zhang, Yuqiang and Cao, Rui and He, Huimin and Xue, Haiyun and Liu, Fengman},
TITLE = {High-Bandwidth Lumped Mach-Zehnder Modulators Based on Thin-Film Lithium Niobate},
JOURNAL = {Photonics},
VOLUME = {11},
YEAR = {2024},
NUMBER = {5},
ARTICLE-NUMBER = {399},
URL = {https://www.mdpi.com/2304-6732/11/5/399},
ISSN = {2304-6732},
DOI = {10.3390/photonics11050399}
}

@article{Alexander2024,
   author = {Koen Alexander and Andrea Bahgat and Avishai Benyamini and Dylan Black and Damien Bonneau and Stanley Burgos and Ben Burridge and Geoff Campbell and Gabriel Catalano and Alex Ceballos and Chia-Ming Chang and CJ Chung and Fariba Danesh and Tom Dauer and Michael Davis and Eric Dudley and Ping Er-Xuan and Josep Fargas and Alessandro Farsi and Colleen Fenrich and Jonathan Frazer and Masaya Fukami and Yogeeswaran Ganesan and Gary Gibson and Mercedes Gimeno-Segovia and Sebastian Goeldi and Patrick Goley and Ryan Haislmaier and Sami Halimi and Paul Hansen and Sam Hardy and Jason Horng and Matthew House and Hong Hu and Mehdi Jadidi and Henrik Johansson and Thomas Jones and Vimal Kamineni and Nicholas Kelez and Ravi Koustuban and George Kovall and Peter Krogen and Nikhil Kumar and Yong Liang and Nicholas LiCausi and Dan Llewellyn and Kimberly Lokovic and Michael Lovelady and Vitor Manfrinato and Ann Melnichuk and Mario Souza and Gabriel Mendoza and Brad Moores and Shaunak Mukherjee and Joseph Munns and Francois-Xavier Musalem and Faraz Najafi and Jeremy L. O'Brien and J. Elliott Ortmann and Sunil Pai and Bryan Park and Hsuan-Tung Peng and Nicholas Penthorn and Brennan Peterson and Matt Poush and Geoff J. Pryde and Tarun Ramprasad and Gareth Ray and Angelita Rodriguez and Brian Roxworthy and Terry Rudolph and Dylan J. Saunders and Pete Shadbolt and Deesha Shah and Hyungki Shin and Jake Smith and Ben Sohn and Young-Ik Sohn and Gyeongho Son and Chris Sparrow and Matteo Staffaroni and Camille Stavrakas and Vijay Sukumaran and Davide Tamborini and Mark G. Thompson and Khanh Tran and Mark Triplet and Maryann Tung and Alexey Vert and Mihai D. Vidrighin and Ilya Vorobeichik and Peter Weigel and Mathhew Wingert and Jamie Wooding and Xinran Zhou},
   month = {4},
   title = {A manufacturable platform for photonic quantum computing},
   url = {http://arxiv.org/abs/2404.17570},
   year = {2024}
}

@misc{Catalalahoz2026,
      title={High-Speed Non-Volatile Barium Titanate Field Programmable Photonic Gate Array}, 
      author={Cristina Catalá-Lahoz and Jose Roberto Rausell-Campo and Daniel Pérez-López and Lucas Güniat and Clarissa Convertino and Felix Eltes and Jean Fompeyrine and Charis Mesaritakis and Adonis Bogris and Quentin Wilmart and Jonathan Faugier-Tovar and Benoit Charbonnier and José Capmany},
      year={2026},
      eprint={2601.07456},
      archivePrefix={arXiv},
      primaryClass={physics.optics},
      url={https://arxiv.org/abs/2601.07456}, 
}

@article{Zheng2023,
  title = {Electro-optically programmable photonic circuits enabled by wafer-scale integration on thin-film lithium niobate},
  author = {Zheng, Yong and Zhong, Haozong and Zhang, Haisu and Song, Lvbin and Liu, Jian and Liang, Youting and Liu, Zhaoxiang and Chen, Jinming and Zhou, Junxia and Fang, Zhiwei and Wang, Min and Li, Lin and Wu, Rongbo and Cheng, Ya},
  journal = {Phys. Rev. Res.},
  volume = {5},
  issue = {3},
  pages = {033206},
  numpages = {9},
  year = {2023},
  month = {Sep},
  publisher = {American Physical Society},
  doi = {10.1103/PhysRevResearch.5.033206},
  url = {https://link.aps.org/doi/10.1103/PhysRevResearch.5.033206}
}

@article{
Sund2023,
author = {Patrik I. Sund  and Emma Lomonte  and Stefano Paesani  and Ying Wang  and Jacques Carolan  and Nikolai Bart  and Andreas D. Wieck  and Arne Ludwig  and Leonardo Midolo  and Wolfram H. P. Pernice  and Peter Lodahl  and Francesco Lenzini },
title = {High-speed thin-film lithium niobate quantum processor driven by a solid-state quantum emitter},
journal = {Science Advances},
volume = {9},
number = {19},
pages = {eadg7268},
year = {2023},
doi = {10.1126/sciadv.adg7268},
URL = {https://www.science.org/doi/abs/10.1126/sciadv.adg7268},
eprint = {https://www.science.org/doi/pdf/10.1126/sciadv.adg7268},
}

@article{Chrysostomidis2025,
   author = {T. Chrysostomidis and D. Chatzitheocharis and F. Eltes and C. Convertino and T. Buriakova and M. Zervas and N. Pleros and K. Vyrsokinos},
   doi = {10.1109/JLT.2025.3537286},
   issn = {15582213},
   issue = {9},
   journal = {Journal of Lightwave Technology},
   pages = {4404-4415},
   publisher = {Institute of Electrical and Electronics Engineers Inc.},
   title = {Ultra-Efficient Si3N4 MZIs With BaTiO3 as Weight Elements for Neuromorphic Photonics},
   volume = {43},
   year = {2025}
}

@article{Georgescu2014,
  title = {Quantum simulation},
  author = {Georgescu, I. M. and Ashhab, S. and Nori, Franco},
  journal = {Rev. Mod. Phys.},
  volume = {86},
  issue = {1},
  pages = {153--185},
  numpages = {33},
  year = {2014},
  month = {Mar},
  publisher = {American Physical Society},
  doi = {10.1103/RevModPhys.86.153},
  url = {https://link.aps.org/doi/10.1103/RevModPhys.86.153}
}

@article{Jaksch2000,
  title = {Fast Quantum Gates for Neutral Atoms},
  author = {Jaksch, D. and Cirac, J. I. and Zoller, P. and Rolston, S. L. and C\^ot\'e, R. and Lukin, M. D.},
  journal = {Phys. Rev. Lett.},
  volume = {85},
  issue = {10},
  pages = {2208--2211},
  numpages = {0},
  year = {2000},
  month = {Sep},
  publisher = {American Physical Society},
  doi = {10.1103/PhysRevLett.85.2208},
  url = {https://link.aps.org/doi/10.1103/PhysRevLett.85.2208}
}

@article{Crampton2025,
   author = {Oliver M. Crampton and Toby J. Dowling and Thomas Roger and Peter R. Smith and James F. Dynes and Matthew S. Winnel and Davide G. Marangon and Mirko Sanzaro and Ravinder Singh and Chithrabhanu Perumangatt and Joseph A. Dolphin and Taofiq K. Paraiso and Andrew J. Shields},
   doi = {10.1038/s41534-025-01100-2},
   issn = {20566387},
   issue = {1},
   journal = {npj Quantum Information},
   month = {12},
   publisher = {Nature Research},
   title = {A 2-Gbps low-SWaP quantum random number generator with photonic integrated circuits for satellite applications},
   volume = {11},
   year = {2025}
}

@article{shaw_benchmarking_2024,
	title = {Benchmarking highly entangled states on a 60-atom analogue quantum simulator},
	volume = {628},
	issn = {1476-4687},
	url = {https://doi.org/10.1038/s41586-024-07173-x},
	doi = {10.1038/s41586-024-07173-x},
	number = {8006},
	journal = {Nature},
	author = {Shaw, Adam L. and Chen, Zhuo and Choi, Joonhee and Mark, Daniel K. and Scholl, Pascal and Finkelstein, Ran and Elben, Andreas and Choi, Soonwon and Endres, Manuel},
	month = apr,
	year = {2024},
	pages = {71--77},
}

@article{evered_high-fidelity_2023,
	title = {High-fidelity parallel entangling gates on a neutral-atom quantum computer},
	volume = {622},
	copyright = {2023 The Author(s)},
	issn = {1476-4687},
	url = {https://www.nature.com/articles/s41586-023-06481-y},
	doi = {10.1038/s41586-023-06481-y},
	language = {en},
	number = {7982},
	urldate = {2026-02-20},
	journal = {Nature},
	publisher = {Nature Publishing Group},
	author = {Evered, Simon J. and Bluvstein, Dolev and Kalinowski, Marcin and Ebadi, Sepehr and Manovitz, Tom and Zhou, Hengyun and Li, Sophie H. and Geim, Alexandra A. and Wang, Tout T. and Maskara, Nishad and Levine, Harry and Semeghini, Giulia and Greiner, Markus and Vuletić, Vladan and Lukin, Mikhail D.},
	month = oct,
	year = {2023},
	pages = {268--272},
}

@article{bluvstein_logical_2024,
	title = {Logical quantum processor based on reconfigurable atom arrays},
	volume = {626},
	copyright = {2023 The Author(s)},
	issn = {1476-4687},
	url = {https://www.nature.com/articles/s41586-023-06927-3},
	doi = {10.1038/s41586-023-06927-3},
	language = {en},
	number = {7997},
	urldate = {2026-02-20},
	journal = {Nature},
	publisher = {Nature Publishing Group},
	author = {Bluvstein, Dolev and Evered, Simon J. and Geim, Alexandra A. and Li, Sophie H. and Zhou, Hengyun and Manovitz, Tom and Ebadi, Sepehr and Cain, Madelyn and Kalinowski, Marcin and Hangleiter, Dominik and Bonilla Ataides, J. Pablo and Maskara, Nishad and Cong, Iris and Gao, Xun and Sales Rodriguez, Pedro and Karolyshyn, Thomas and Semeghini, Giulia and Gullans, Michael J. and Greiner, Markus and Vuletić, Vladan and Lukin, Mikhail D.},
	month = feb,
	year = {2024},
	pages = {58--65},

}

@article{jenkins_ytterbium_2022,
	title = {Ytterbium {Nuclear}-{Spin} {Qubits} in an {Optical} {Tweezer} {Array}},
	volume = {12},
	url = {https://link.aps.org/doi/10.1103/PhysRevX.12.021027},
	doi = {10.1103/PhysRevX.12.021027},
	number = {2},
	urldate = {2026-02-20},
	journal = {Physical Review X},
	publisher = {American Physical Society},
	author = {Jenkins, Alec and Lis, Joanna W. and Senoo, Aruku and McGrew, William F. and Kaufman, Adam M.},
	month = may,
	year = {2022},
	pages = {021027},
}

@article{ma_universal_2022,
	title = {Universal {Gate} {Operations} on {Nuclear} {Spin} {Qubits} in an {Optical} {Tweezer} {Array} of \${\textasciicircum}\{171\}{\textbackslash}mathrm\{{Yb}\}\$ {Atoms}},
	volume = {12},
	url = {https://link.aps.org/doi/10.1103/PhysRevX.12.021028},
	doi = {10.1103/PhysRevX.12.021028},
	number = {2},
	urldate = {2026-02-20},
	journal = {Physical Review X},
	publisher = {American Physical Society},
	author = {Ma, Shuo and Burgers, Alex P. and Liu, Genyue and Wilson, Jack and Zhang, Bichen and Thompson, Jeff D.},
	month = may,
	year = {2022},
	pages = {021028},
}

@article{chiu_continuous_2025,
	title = {Continuous operation of a coherent 3,000-qubit system},
	volume = {646},
	copyright = {2025 The Author(s)},
	issn = {1476-4687},
	url = {https://www.nature.com/articles/s41586-025-09596-6},
	doi = {10.1038/s41586-025-09596-6},
	language = {en},
	number = {8087},
	urldate = {2026-02-20},
	journal = {Nature},
	publisher = {Nature Publishing Group},
	author = {Chiu, Neng-Chun and Trapp, Elias C. and Guo, Jinen and Abobeih, Mohamed H. and Stewart, Luke M. and Hollerith, Simon and Stroganov, Pavel L. and Kalinowski, Marcin and Geim, Alexandra A. and Evered, Simon J. and Li, Sophie H. and Lyu, Xingjian and Peters, Lisa M. and Bluvstein, Dolev and Wang, Tout T. and Greiner, Markus and Vuletić, Vladan and Lukin, Mikhail D.},
	month = oct,
	year = {2025},
	pages = {1075--1080},
}

@article{manetsch_tweezer_2025,
	title = {A tweezer array with 6,100 highly coherent atomic qubits},
	volume = {647},
	copyright = {2025 The Author(s)},
	issn = {1476-4687},
	url = {https://www.nature.com/articles/s41586-025-09641-4},
	doi = {10.1038/s41586-025-09641-4},
	language = {en},
	number = {8088},
	urldate = {2026-02-20},
	journal = {Nature},
	publisher = {Nature Publishing Group},
	author = {Manetsch, Hannah J. and Nomura, Gyohei and Bataille, Elie and Lv, Xudong and Leung, Kon H. and Endres, Manuel},
	month = nov,
	year = {2025},
	pages = {60--67},
}

@misc{noauthor_852nm_nodate,
	title = {852nm Narrow linewidth laser, \url{https://www.precilasers.com/en/page-75-235.html}},
	url = {https://www.precilasers.com/en/page-75-235.html},
	urldate = {2026-02-20},
}

@misc{noauthor_t257p-20_nodate,
	title = {{T257P}-20 {\textbar} {Product} {\textbar} {ThermoTek}},
	url = {https://www.thermotekusa.com/product/t257p-20},
	urldate = {2026-02-20},
}

@misc{noauthor_verdi_nodate,
	title = {{Verdi - High Performance CW Green Lasers, Coherent, \url{https://www.coherent.com/lasers/cw-solid-state/verdi}}},
	url = {https://www.coherent.com/lasers/cw-solid-state/verdi},
	language = {en-US},
	urldate = {2026-02-20},
}

@misc{photonics_cw_nodate,
	title = {{CW Fiber Amplifiers, \url{https://www.ipgphotonics.com/products/lasers/specialized-lasers/cw-amplifiers}}},
	url = {https://www.ipgphotonics.com/products/lasers/specialized-lasers/cw-amplifiers},
	language = {en},
	urldate = {2026-02-20},
	journal = {IPG Photonics Corporation},
	author = {Photonics, I. P. G.},
}

@misc{noauthor_3u_nodate,
	title = {{3U Rack Mount  Product,  ThermoTek, \url{https://www.thermotekusa.com/product/3u-rack}}},
	url = {https://www.thermotekusa.com/product/3u-rack},
	urldate = {2026-02-20},
}

@misc{noauthor_qcmos_nodate,
	title = {{qCMOS cameras, Hamamatsu Photonics, \url{https://www.hamamatsu.com/jp/en/product/cameras/qcmos-cameras.html}}},
	url = {https://www.hamamatsu.com/jp/en/product/cameras/qcmos-cameras.html},
	language = {en},
	urldate = {2026-02-20},
}

@misc{Qiskit,
    title ={{Qiskit Documentation from IBM:} https://quantum.cloud.ibm.com/docs/en/api/qiskit },
    url = {https://quantum.cloud.ibm.com/docs/en/api/qiskit } }

@article{levine_parallel_2019,
	title = {Parallel {Implementation} of {High}-{Fidelity} {Multiqubit} {Gates} with {Neutral} {Atoms}},
	volume = {123},
	url = {https://link.aps.org/doi/10.1103/PhysRevLett.123.170503},
	doi = {10.1103/PhysRevLett.123.170503},
	number = {17},
	urldate = {2026-02-20},
	journal = {Physical Review Letters},
	publisher = {American Physical Society},
	author = {Levine, Harry and Keesling, Alexander and Semeghini, Giulia and Omran, Ahmed and Wang, Tout T. and Ebadi, Sepehr and Bernien, Hannes and Greiner, Markus and Vuletić, Vladan and Pichler, Hannes and Lukin, Mikhail D.},
	month = oct,
	year = {2019},
	pages = {170503},
}

@article{gyger_continuous_2024,
	title = {Continuous operation of large-scale atom arrays in optical lattices},
	volume = {6},
	url = {https://link.aps.org/doi/10.1103/PhysRevResearch.6.033104},
	doi = {10.1103/PhysRevResearch.6.033104},
	number = {3},
	urldate = {2026-02-20},
	journal = {Physical Review Research},
	publisher = {American Physical Society},
	author = {Gyger, Flavien and Ammenwerth, Maximilian and Tao, Renhao and Timme, Hendrik and Snigirev, Stepan and Bloch, Immanuel and Zeiher, Johannes},
	month = jul,
	year = {2024},
	pages = {033104},
}

@misc{li_fast_2025,
	title = {Fast, continuous and coherent atom replacement in a neutral atom qubit array},
	url = {http://arxiv.org/abs/2506.15633},
	doi = {10.48550/arXiv.2506.15633},
	urldate = {2026-02-20},
	publisher = {arXiv},
	author = {Li, Yiyi and Bao, Yicheng and Peper, Michael and Li, Chenyuan and Thompson, Jeff D.},
	month = jun,
	year = {2025},
	note = {arXiv:2506.15633 [quant-ph]},
}

@article{Het_nyi_2024,
   title={Creating Entangled Logical Qubits in the Heavy-Hex Lattice with Topological Codes},
   volume={5},
   ISSN={2691-3399},
   url={http://dx.doi.org/10.1103/PRXQuantum.5.040334},
   DOI={10.1103/prxquantum.5.040334},
   number={4},
   journal={PRX Quantum},
   publisher={American Physical Society (APS)},
   author={Hetényi, Bence and Wootton, James R.},
   year={2024},
   month=dec }

@misc{vezvaee2025surfacecodescalingheavyhex,
      title={Surface code scaling on heavy-hex superconducting quantum processors}, 
      author={Arian Vezvaee and Cesar Benito and Mario Morford-Oberst and Alejandro Bermudez and Daniel A. Lidar},
      year={2025},
      eprint={2510.18847},
      archivePrefix={arXiv},
      primaryClass={quant-ph},
      url={https://arxiv.org/abs/2510.18847}, 
}

@misc{clements2017,
      title={An Optimal Design for Universal Multiport Interferometers}, 
      author={William R. Clements and Peter C. Humphreys and Benjamin J. Metcalf and W. Steven Kolthammer and Ian A. Walmsley},
      year={2017},
      eprint={1603.08788},
      archivePrefix={arXiv},
      primaryClass={physics.optics},
      url={https://arxiv.org/abs/1603.08788}, 
}

@article{delegated1,
author = {Fitzsimons, Joseph},
year = {2016},
month = {11},
pages = {},
title = {Private quantum computation: An introduction to blind quantum computing and related protocols},
volume = {3},
journal = {npj Quantum Information},
doi = {10.1038/s41534-017-0025-3}
}

@article{LeoSecurity,
   title={Security Limitations of Classical-Client Delegated Quantum Computing},
   ISBN={9783030648343},
   ISSN={1611-3349},
   url={http://dx.doi.org/10.1007/978-3-030-64834-3_23},
   DOI={10.1007/978-3-030-64834-3_23},
   journal={Lecture Notes in Computer Science},
   publisher={Springer International Publishing},
   author={Badertscher, Christian and Cojocaru, Alexandru and Colisson, Léo and Kashefi, Elham and Leichtle, Dominik and Mantri, Atul and Wallden, Petros},
   year={2020},
   pages={667–696}
}

@article{BroadbentDelegated,
   title={Delegating private quantum computations},
   volume={93},
   ISSN={1208-6045},
   url={http://dx.doi.org/10.1139/cjp-2015-0030},
   DOI={10.1139/cjp-2015-0030},
   number={9},
   journal={Canadian Journal of Physics},
   publisher={Canadian Science Publishing},
   author={Broadbent, Anne},
   year={2015},
   month={Sep},
   pages={941–946}
}

@article{Broadbent_2009,
   title={Universal Blind Quantum Computation},
   ISBN={9781424451166},
   url={http://dx.doi.org/10.1109/FOCS.2009.36},
   DOI={10.1109/focs.2009.36},
   journal={2009 50th Annual IEEE Symposium on Foundations of Computer Science},
   publisher={IEEE},
   author={Broadbent, Anne and Fitzsimons, Joseph and Kashefi, Elham},
   year={2009},
   month={Oct}
}

@article{MBQC,
   title={Measurement-based quantum computation},
   volume={5},
   ISSN={1745-2481},
   url={http://dx.doi.org/10.1038/nphys1157},
   DOI={10.1038/nphys1157},
   number={1},
   journal={Nature Physics},
   publisher={Springer Science and Business Media LLC},
   author={Briegel, Hans J. and Browne, Dan E. and Dür, Wolfgang and Raussendorf, Robert and Van den Nest, Marteen},
   year={2009},
   month={Jan},
   pages={19–26}
}

@article{annealing1,
    author = {Rajak, Atanu and Suzuki, Sei and Dutta, Amit and Chakrabarti, Bikas K.},
    title = {Quantum annealing: an overview},
    journal = {Philosophical Transactions of the Royal Society A: Mathematical, Physical and Engineering Sciences},
    volume = {381},
    number = {2241},
    pages = {20210417},
    year = {2022},
    month = {12},
    issn = {1364-503X},
    doi = {10.1098/rsta.2021.0417},
    url = {https://doi.org/10.1098/rsta.2021.0417},
    eprint = {https://royalsocietypublishing.org/rsta/article-pdf/doi/10.1098/rsta.2021.0417/1326707/rsta.2021.0417.pdf},
}

@article{annealing2,
   title={Mathematical foundation of quantum annealing},
   volume={49},
   ISSN={1089-7658},
   url={http://dx.doi.org/10.1063/1.2995837},
   DOI={10.1063/1.2995837},
   number={12},
   journal={Journal of Mathematical Physics},
   publisher={AIP Publishing},
   author={Morita, Satoshi and Nishimori, Hidetoshi},
   year={2008},
   month=dec }

@article{Albash_2018,
   title={Adiabatic quantum computation},
   volume={90},
   ISSN={1539-0756},
   url={http://dx.doi.org/10.1103/RevModPhys.90.015002},
   DOI={10.1103/revmodphys.90.015002},
   number={1},
   journal={Reviews of Modern Physics},
   publisher={American Physical Society (APS)},
   author={Albash, Tameem and Lidar, Daniel A.},
   year={2018},
   month=jan }

@article{Markovic_2020,
   title={Quantum neuromorphic computing},
   volume={117},
   ISSN={1077-3118},
   url={http://dx.doi.org/10.1063/5.0020014},
   DOI={10.1063/5.0020014},
   number={15},
   journal={Applied Physics Letters},
   publisher={AIP Publishing},
   author={Marković, Danijela and Grollier, Julie},
   year={2020},
   month=oct }

@misc{carles2026experimentalquantumreservoircomputing,
      title={Experimental quantum reservoir computing with a circuit quantum electrodynamics system}, 
      author={Baptiste Carles and Julien Dudas and Léo Balembois and Julie Grollier and Danijela Marković},
      year={2026},
      eprint={2506.22016},
      archivePrefix={arXiv},
      primaryClass={quant-ph},
      url={https://arxiv.org/abs/2506.22016}, 
}

@misc{dudas2023quantumreservoirneuralnetwork,
      title={Quantum reservoir neural network implementation on coherently coupled quantum oscillators}, 
      author={Julien Dudas and Baptiste Carles and Erwan Plouet and Alice Mizrahi and Julie Grollier and Danijela Marković},
      year={2023},
      eprint={2209.03221},
      archivePrefix={arXiv},
      primaryClass={quant-ph},
      url={https://arxiv.org/abs/2209.03221}, 
}

@misc{Altaisky2001,
  title = {Quantum Neural Network},
  author = {Altaisky, M. V.},
  year = 2001,
  month = jul,
  number = {arXiv:quant-ph/0107012},
  eprint = {quant-ph/0107012},
  publisher = {arXiv},
  doi = {10.48550/arXiv.quant-ph/0107012},
  urldate = {2026-03-27},
  archiveprefix = {arXiv}
}

@article{Schuld2014,
  title = {The Quest for a {{Quantum Neural Network}}},
  author = {Schuld, Maria and Sinayskiy, Ilya and Petruccione, Francesco},
  year = 2014,
  month = nov,
  journal = {Quantum Inf Process},
  volume = {13},
  number = {11},
  pages = {2567--2586},
  issn = {1573-1332},
  doi = {10.1007/s11128-014-0809-8},
  urldate = {2026-03-27},
  langid = {english},
}

@article{deGliniasty2024spinopticalquantum,
  doi = {10.22331/q-2024-07-24-1423},
  url = {https://doi.org/10.22331/q-2024-07-24-1423},
  title = {A {S}pin-{O}ptical {Q}uantum {C}omputing {A}rchitecture},
  author = {de Gliniasty, Gr{\'{e}}goire and Hilaire, Paul and Emeriau, Pierre-Emmanuel and Wein, Stephen C. and Salavrakos, Alexia and Mansfield, Shane},
  journal = {{Quantum}},
  issn = {2521-327X},
  publisher = {{Verein zur F{\"{o}}rderung des Open Access Publizierens in den Quantenwissenschaften}},
  volume = {8},
  pages = {1423},
  month = jul,
  year = {2024}
}

@misc{gyurik2024exponentialseparationsclassicalquantum,
      title={Exponential separations between classical and quantum learners}, 
      author={Casper Gyurik and Vedran Dunjko},
      year={2024},
      eprint={2306.16028},
      archivePrefix={arXiv},
      primaryClass={quant-ph},
      url={https://arxiv.org/abs/2306.16028}, 
}

@article{salavrakos2025photonnative,
doi = {10.1088/2633-4356/adc531},
url = {https://doi.org/10.1088/2633-4356/adc531},
year = {2025},
month = {apr},
publisher = {IOP Publishing},
volume = {5},
number = {2},
pages = {023001},
author = {Salavrakos, Alexia and Maring, Nicolas and Emeriau, Pierre-Emmanuel and Mansfield, Shane},
title = {Photon-native quantum algorithms},
journal = {Materials for Quantum Technology}}

@misc{mezher2023solvinggraphproblemssinglephotons,
      title={Solving graph problems with single-photons and linear optics}, 
      author={Rawad Mezher and Ana Filipa Carvalho and Shane Mansfield},
      year={2023},
      eprint={2301.09594},
      archivePrefix={arXiv},
      primaryClass={quant-ph},
      url={https://arxiv.org/abs/2301.09594}, 
}

@inproceedings{clifford2018classical,
  title={The classical complexity of {B}oson {S}ampling},
  author={Clifford, Peter and Clifford, Rapha{\"e}l},
  booktitle={Proceedings of the Twenty-Ninth Annual ACM-SIAM Symposium on Discrete Algorithms},
  doi={https://doi.org/10.1137/1.9781611975031.10},
  pages={146--155},
  year={2018},
  organization={SIAM}
}

@article{go2025bs,
  title = {Quantum Computational Advantage of Noisy {B}oson {S}ampling with Partially Distinguishable Photons},
  author = {Go, Byeongseon and Oh, Changhun and Jeong, Hyunseok},
  journal = {PRX Quantum},
  volume = {6},
  issue = {3},
  pages = {030362},
  numpages = {20},
  year = {2025},
  month = {Sep},
  publisher = {American Physical Society},
  doi = {10.1103/rflv-gc66},
  url = {https://link.aps.org/doi/10.1103/rflv-gc66}
}

@article{oszmaniec2025complexitytheoreticfoundationsbosonsamplinglinear,
  author  = {Adam Bouland and Daniel Brod and Ishaun Datta and Bill Fefferman and Daniel Grier and Felipe Hernandez and Michal Oszmaniec},
  title   = {Complexity-theoretic foundations of {B}oson {S}ampling with a linear number of modes},
  journal = {arXiv},
  year    = {2025},
  doi     = {10.48550/arXiv.2312.00286}
}

@article{wu2018benchmark,
    author = {Wu, Junjie and Liu, Yong and Zhang, Baida and Jin, Xianmin and Wang, Yang and Wang, Huiquan and Yang, Xuejun},
    title = {A benchmark test of {B}oson {S}ampling on Tianhe-2 supercomputer},
    journal = {National Science Review},
    volume = {5},
    number = {5},
    pages = {715-720},
    year = {2018},
    month = {07},
    issn = {2095-5138},
    doi = {10.1093/nsr/nwy079},
    url = {https://doi.org/10.1093/nsr/nwy079}
}

@article{wang2019bs-20-photons,
  title = {{B}oson {S}ampling with 20 Input Photons and a 60-Mode Interferometer in a $1{0}^{14}$-Dimensional Hilbert Space},
  author = {Wang, Hui and Qin, Jian and Ding, Xing and Chen, Ming-Cheng and Chen, Si and You, Xiang and He, Yu-Ming and Jiang, Xiao and You, L. and Wang, Z. and Schneider, C. and Renema, Jelmer J. and H\"ofling, Sven and Lu, Chao-Yang and Pan, Jian-Wei},
  journal = {Phys. Rev. Lett.},
  volume = {123},
  issue = {25},
  pages = {250503},
  numpages = {7},
  year = {2019},
  month = {Dec},
  publisher = {American Physical Society},
  doi = {10.1103/PhysRevLett.123.250503},
  url = {https://link.aps.org/doi/10.1103/PhysRevLett.123.250503}
}

@inproceedings{aaronson2011computational,
  title={The computational complexity of linear optics},
  author={Aaronson, Scott and Arkhipov, Alex},
  booktitle={Proceedings of the forty-third annual ACM symposium on Theory of computing},
  pages={333--342},
  doi={https://doi.org/10.1145/1993636.1993682},
  year={2011}
}

@misc{cain2026shorsalgorithmpossible10000,
      title={Shor's algorithm is possible with as few as 10,000 reconfigurable atomic qubits}, 
      author={Madelyn Cain and Qian Xu and Robbie King and Lewis R. B. Picard and Harry Levine and Manuel Endres and John Preskill and Hsin-Yuan Huang and Dolev Bluvstein},
      year={2026},
      eprint={2603.28627},
      archivePrefix={arXiv},
      primaryClass={quant-ph},
      url={https://arxiv.org/abs/2603.28627}, 
}

@article{evered_probing_2025,
	title = {Probing the {Kitaev} honeycomb model on a neutral-atom quantum computer},
	volume = {645},
	copyright = {2025 The Author(s)},
	issn = {1476-4687},
	url = {https://www.nature.com/articles/s41586-025-09475-0},
	doi = {10.1038/s41586-025-09475-0},
	language = {en},
	number = {8080},
	urldate = {2026-05-11},
	journal = {Nature},
	publisher = {Nature Publishing Group},
	author = {Evered, Simon J. and Kalinowski, Marcin and Geim, Alexandra A. and Manovitz, Tom and Bluvstein, Dolev and Li, Sophie H. and Maskara, Nishad and Zhou, Hengyun and Ebadi, Sepehr and Xu, Muqing and Campo, Joseph and Cain, Madelyn and Ostermann, Stefan and Yelin, Susanne F. and Sachdev, Subir and Greiner, Markus and Vuletić, Vladan and Lukin, Mikhail D.},
	month = sep,
	year = {2025},
	pages = {341--347},
}

@article{reichardt_fault-tolerant_2024,
	title = {Fault-tolerant quantum computation with a neutral atom processor},
	url = {https://arxiv.org/abs/2411.11822},
	doi = {10.48550/arXiv.2411.11822},
	language = {en},
	urldate = {2026-05-11},
	journal = {arXiv preprint arXiv:2411.11822},
	author = {Reichardt, Ben W. and Paetznick, Adam and Aasen, David and Basov, Ivan and Bello-Rivas, Juan M. and Bonderson, Parsa and Chao, Rui and van Dam, Wim and Hastings, Matthew B. and Mishmash, Ryan V. and Paz, Andres and da Silva, Marcus P. and Sundaram, Aarthi and Svore, Krysta M. and Vaschillo, Alexander and Wang, Zhenghan and Zanner, Matt and Cairncross, William B. and Chen, Cheng-An and Crow, Daniel and Kim, Hyosub and Kindem, Jonathan M. and King, Jonathan and McDonald, Michael and Norcia, Matthew A. and Ryou, Albert and Stone, Mark and Wadleigh, Laura and Barnes, Katrina and Battaglino, Peter and Bohdanowicz, Thomas C. and Booth, Graham and Brown, Andrew and Brown, Mark O. and Cassella, Kayleigh and Coxe, Robin and Epstein, Jeffrey M. and Feldkamp, Max and Griger, Christopher and Halperin, Eli and Heinz, Andre and Hummel, Frederic and Jaffe, Matthew and Jones, Antonia M. W. and Kapit, Eliot and Kotru, Krish and Lauigan, Joseph and Li, Ming and Marjanovic, Jan and Megidish, Eli and Meredith, Matthew and Morshead, Ryan and Muniz, Juan A. and Narayanaswami, Sandeep and Nishiguchi, Ciro and Paule, Timothy and Pawlak, Kelly A. and Pudenz, Kristen L. and Rodríguez Pérez, David and Simon, Jon and Smull, Aaron and Stack, Daniel and Urbanek, Miroslav and van de Veerdonk, René J. M. and Vendeiro, Zachary and Weverka, Robert T. and Wilkason, Thomas and Wu, Tsung-Yao and Xie, Xin and Zalys-Geller, Evan and Zhang, Xiaogang and Bloom, Benjamin J.},
	month = nov,
	year = {2024},
	eprint = {2411.11822},
	archivePrefix = {arXiv},
	primaryClass = {quant-ph},
}

@article{chinnarasu_variational_2025,
	title = {Variational Simulation of the {Lipkin}-{Meshkov}-{Glick} Model on a Neutral Atom Quantum Computer},
	volume = {6},
	copyright = {2025 American Physical Society},
	issn = {2691-3399},
	url = {https://journals.aps.org/prxquantum/abstract/10.1103/PRXQuantum.6.020350},
	doi = {10.1103/PRXQuantum.6.020350},
	language = {en},
	number = {2},
	urldate = {2026-05-11},
	journal = {PRX Quantum},
	publisher = {American Physical Society},
	author = {Chinnarasu, R. and Poole, C. and Phuttitarn, L. and Noori, A. and Graham, T. M. and Coppersmith, S. N. and Balantekin, A. B. and Saffman, M.},
	month = jun,
	year = {2025},
	pages = {020350},
}

@article{bluvstein_quantum_2022,
	title = {A quantum processor based on coherent transport of entangled atom arrays},
	volume = {604},
	copyright = {2022 The Author(s)},
	issn = {1476-4687},
	url = {https://www.nature.com/articles/s41586-022-04592-6},
	doi = {10.1038/s41586-022-04592-6},
	language = {en},
	number = {7906},
	urldate = {2026-05-11},
	journal = {Nature},
	publisher = {Nature Publishing Group},
	author = {Bluvstein, Dolev and Levine, Harry and Semeghini, Giulia and Wang, Tout T. and Ebadi, Sepehr and Kalinowski, Marcin and Keesling, Alexander and Maskara, Nishad and Pichler, Hannes and Greiner, Markus and Vuletić, Vladan and Lukin, Mikhail D.},
	month = apr,
	year = {2022},
	pages = {451--456},
}

@article{evered_high-fidelity_2026,
	title = {High-fidelity entangling gates and nonlocal circuits with neutral atoms},
	url = {https://arxiv.org/abs/2604.25987},
	doi = {10.48550/arXiv.2604.25987},
	language = {en},
	urldate = {2026-05-11},
	journal = {arXiv preprint arXiv:2604.25987},
	author = {Evered, Simon J. and Xu, Muqing and Li, Sophie H. and Geim, Alexandra A. and Bonilla Ataides, J. Pablo and Kalinowski, Marcin and Bluvstein, Dolev and Maskara, Nishad and Kokail, Christian and Greiner, Markus and Vuletić, Vladan and Lukin, Mikhail D.},
	month = apr,
	year = {2026},
	eprint = {2604.25987},
	archivePrefix = {arXiv},
	primaryClass = {quant-ph},
}

@article{bluvstein_architectural_2025,
	title = {Architectural mechanisms of a universal fault-tolerant quantum computer},
	url = {https://arxiv.org/abs/2506.20661},
	doi = {10.48550/arXiv.2506.20661},
	language = {en},
	urldate = {2026-05-11},
	journal = {arXiv preprint arXiv:2506.20661},
	author = {Bluvstein, Dolev and Geim, Alexandra A. and Li, Sophie H. and Evered, Simon J. and Bonilla Ataides, J. Pablo and Baranes, Gefen and Gu, Andi and Manovitz, Tom and Xu, Muqing and Kalinowski, Marcin and Majidy, Shayan and Kokail, Christian and Maskara, Nishad and Trapp, Elias C. and Stewart, Luke M. and Hollerith, Simon and Zhou, Hengyun and Gullans, Michael J. and Yelin, Susanne F. and Greiner, Markus and Vuletic, Vladan and Cain, Madelyn and Lukin, Mikhail D.},
	month = jun,
	year = {2025},
	eprint = {2506.20661},
	archivePrefix = {arXiv},
	primaryClass = {quant-ph},
}

@article{muzi_falconi_microsecond-scale_2025,
	title = {Microsecond-{Scale} High-{Survival} and Number-{Resolved} Detection of Ytterbium Atom Arrays},
	volume = {135},
	copyright = {2025 American Physical Society},
	issn = {0031-9007, 1079-7114},
	url = {https://link.aps.org/doi/10.1103/n3bg-7yw7},
	doi = {10.1103/n3bg-7yw7},
	language = {en},
	number = {20},
	urldate = {2026-05-11},
	journal = {Physical Review Letters},
	publisher = {American Physical Society},
	author = {Muzi Falconi, Alessandro and Panza, Riccardo and Sbernardori, Sara and Forti, Riccardo and Klemt, Ralf and Abdel Karim, Omar and Marinelli, Matteo and Scazza, Francesco},
	month = nov,
	year = {2025},
	pages = {203402},
}

\end{document}